\documentclass[journal]{IEEEtran}
\usepackage{amsmath,amsfonts}
\usepackage{array}
\usepackage{textcomp}
\usepackage{stfloats}
\usepackage{url}
\usepackage{verbatim}
\usepackage{graphicx}
\usepackage[onehalfspacing]{setspace}
\usepackage{booktabs,colortbl,xcolor}
\usepackage{graphicx,wrapfig}
\usepackage[section]{placeins} 
\usepackage{float} 
\usepackage{enumitem}
\usepackage{subfiles}
\usepackage{scrhack}
\usepackage[%
bookmarksnumbered=true,
pdfborder={0 0 0},
pdfa
]{hyperref}
\usepackage{subcaption}
 
\usepackage{zx-calculus}
\usepackage{tikz}
\usetikzlibrary{backgrounds,fit,decorations.pathreplacing,calc,matrix,arrows.meta,positioning,decorations.pathreplacing,scopes}
\usepackage{nicefrac}
\usepackage{csquotes}
\usepackage{mathtools}
\usepackage[binary-units=true, detect-all=true]{siunitx} %
\usepackage{url}
\usepackage{multirow, makecell}
\usepackage{multicol}
\usepackage{array}
\usepackage[braket, qm]{qcircuit}
\usepackage{marvosym}
\usepackage{verbatim}
\usepackage{nicematrix}
\usepackage{stmaryrd}
\usepackage{soul}
\usepackage{etoolbox}
\usepackage{algorithm}
\usepackage{algorithmic}
\usepackage{balance}
\usepackage{csvsimple}
\usepackage{bbm}
\makeatletter
\patchcmd{\SOUL@ulunderline}{\dimen@}{\SOUL@dimen}{}{}
\patchcmd{\SOUL@ulunderline}{\dimen@}{\SOUL@dimen}{}{}
\patchcmd{\SOUL@ulunderline}{\dimen@}{\SOUL@dimen}{}{}
\newdimen\SOUL@dimen
\makeatother
\mathchardef\mhyphen="2D
\usepackage[style=ieee, maxnames=2, minnames=1, date=year, doi=false,isbn=false,backend=biber, sortcites, dashed=false]{biblatex}
\makeatletter
\let\oldtheequation\theequation
\renewcommand\tagform@[1]{\maketag@@@{\ignorespaces#1\unskip\@@italiccorr}}
\renewcommand\theequation{(\oldtheequation)}
\makeatother

\usepackage{amsthm}

\newtheorem{example}{Example}[]

\newtheorem{theorem}{Theorem}[]

\newcounter{relctr}
\everydisplay\expandafter{\the\everydisplay\setcounter{relctr}{0}}

\AtBeginDocument{}

\begin{document}

	\title{Equivalence Checking of\\Quantum Circuits with the ZX-Calculus}

\author{
  \IEEEauthorblockN{Tom Peham\IEEEauthorrefmark{1}~\IEEEmembership{Graduate Student Member,~IEEE,}\\ Lukas
    Burgholzer\IEEEauthorrefmark{2}~\IEEEmembership{Graduate Student Member,~IEEE,} Robert Wille\IEEEauthorrefmark{1}\IEEEauthorrefmark{3}~\IEEEmembership{Senior Member,~IEEE}\\}\vspace*{1mm}
	\IEEEauthorblockA{\IEEEauthorrefmark{1}Chair for Design Automation, Technical University of Munich, Germany\\}
	\IEEEauthorblockA{\IEEEauthorrefmark{2}Institute for Integrated Circuits, Johannes Kepler University Linz, Austria\\}
	\IEEEauthorblockA{\IEEEauthorrefmark{3}Software Competence Center Hagenberg GmbH (SCCH), Hagenberg, Austria}\vspace*{1mm}
	
	\IEEEauthorblockA{tom.peham@tum.de \hspace{1cm}lukas.burgholzer@jku.at \hspace{1cm}
          robert.wille@tum.de\\}\vspace*{1mm}
        \IEEEauthorblockA{https://www.cda.cit.tum.de/research/quantum/}
}
\maketitle

\begin{abstract}
As state-of-the-art quantum computers are capable of running increasingly complex algorithms, the need for automated
methods to design and test potential applications rises. Equivalence checking of quantum circuits is an important, yet
hardly automated, task in the development of the quantum software stack. Recently, new methods have been proposed that
tackle this problem from widely different perspectives. One of them is based on the ZX-calculus, a graphical rewriting
system for quantum computing. However, the power and capability of this equivalence checking method has barely been
explored. The aim of this work is to
evaluate the ZX-calculus as a tool for equivalence checking of quantum circuits.
To this end, it is demonstrated how the
ZX-calculus based approach for equivalence checking can be expanded in order to verify the results of
compilation flows and optimizations on quantum circuits. It is also shown that the \mbox{ZX-calculus} based method is not
complete---especially for quantum circuits with ancillary qubits. In order to properly evaluate the proposed method, we
conduct a detailed case study by comparing it to two other state-of-the-art methods for equivalence checking: one based on path-sums and another based on decision diagrams. The proposed methods have been integrated into the publicly available QCEC tool~(\mbox{\url{https://github.com/cda-tum/qcec}}) which is part of the Munich Quantum Toolkit (MQT).
\end{abstract}

	\section{Introduction}\label{cha:intro}
	
	Quantum computing~\cite{nielsenQuantumComputationQuantum2010} has had a surge in research endeavors by academia and
	industry in recent years. While quantum computers have not reached a stage of widespread practical usability yet, they
	promise to outperform classical computers in various important tasks, such as unstructured search, integer factorization,
	optimization problems, the simulation of molecules, and more~\cite{groverFastQuantumMechanical1996,shorPolynomialtimeAlgorithmsPrime1997,cerezoVariationalQuantumAlgorithms2020,farhiQuantumApproximateOptimization2014,hermanSurveyQuantumComputing2022,farhiQuantumApproximateOptimization2014,biamonteQuantumMachineLearning2017}. To keep pace with the rapid developments in
	quantum hardware, various tools have been developed that help in designing corresponding applications.
	
	Initially, a quantum computation is described as a sequence of (high-level) quantum gates---somewhat similar to a classical C program.
	However, just like assembly for a classical processor, the actual machine instructions that may be performed on a given quantum processor are generally restricted to a small (low-level) gate-set and might only allow interactions between specific pairs of qubits. 
	Therefore, in order to execute a given circuit on quantum hardware, it needs to be \emph{compiled} to a representation that adheres to all constraints imposed by the targeted device~\cite{sivarajahKetRetargetableCompiler2020,amyStaqFullstackQuantum2020,smithQuantumComputationalCompiler2019,hanerSoftwareMethodologyCompiling2018}.
	Since quantum computers are heavily affected by noise and decoherence, it is paramount to optimize circuits as much as possible in order to maximize the expected fidelity when running the \mbox{circuit~\cite{hattoriQuantumCircuitOptimization2018,sasanianReversibleQuantumCircuit2013,itokoOptimizationQuantumCircuit2020,namAutomatedOptimizationLarge2018,cordierBiologyMedicineLandscape2021}}.
	
	Since the compiled quantum circuit might be altered drastically from its original high-level description, it is of utmost importance that the circuit to be executed on the hardware still implements the same functionality as originally intended.
	Verification of compilation results or, more generally, \emph{equivalence checking of quantum circuits}, turns out to be
	an extremely complex, even \mbox{QMA-complete}\footnote{QMA is the quantum computing analogue to NP. Indeed, NP is a
		subset of QMA}~\cite{janzingNonidentityCheckQMAcomplete2005}, task and is in dire need of automation.
	Although various methods have been
	proposed~\cite{yamashitaFastEquivalencecheckingQuantum2010,berentSATEncodingQuantum2022,hongApproximateEquivalenceChecking2021,viamontesCheckingEquivalenceQuantum2007,amyLargescaleFunctionalVerification2019,
		burgholzerAdvancedEquivalenceChecking2021,wangXQDDbasedVerificationMethod2008,cowtanGenericCompilationStrategy2020} to tackle the equivalence checking
	problem from completely different perspectives, a baseline indicating which paradigm is suited best for which use case
	is yet to be established.
	
	One method for equivalence checking of quantum circuits is based on the
	\emph{ZX}-calculus~\cite{vandeweteringZXcalculusWorkingQuantum2020,cowtanGenericCompilationStrategy2020,duncanGraphtheoreticSimplificationQuantum2019,kissingerReducingTcountZXcalculus2020},
	a graphical calculus used for reasoning about quantum computing. While some results on this equivalence checking
	method exist~\cite{kissingerReducingTcountZXcalculus2020}, it was introduced as more of a \mbox{side-note} in the original work
	than a \mbox{fully-fledged} method. At the time of writing, the only publicly available implementation of this algorithm is written in
	Python~\cite{kissingerPyZXLargeScale2019}. Therefore it is difficult to assess the performance of this method due to the
	inherently slower runtime of the Python interpreter when compared to a compiled language like C++. This makes it somewhat
	problematic to compare the method with other established equivalence checking algorithms. Furthermore, issues unique to
	design automation in quantum computing, like inaccurate representations of complex numbers, different logical-to-physical
	qubit mappings, and ancillary qubits,
	have not been addressed in the ZX-calculus framework. Apart from practical
	considerations, theoretical aspects have also hardly been investigated so far. It is not known for what class of circuits the
	equivalence checking method based on the ZX-calculus is complete, i.e., whether it can actually prove the equivalence of
	any two equivalent quantum circuits. 
	
	Motivated by that, the aim of this work is twofold. Firstly, to establish whether the \mbox{ZX-calculus} provides a solid equivalence checking
	methodology, we review the current state of the art in equivalence checking with the ZX-calculus. To expand on this, we
	discuss how this method can be augmented to handle
	inaccurate representations of complex numbers arising from compilation and optimization processes, deal with alterations of the input and output layout of a circuit that
	happen during compilation, and integrate
	ancillary qubits into the equivalence checking procedure. We provide first results on the completeness of this
	equivalence checking algorithm.
	Secondly, in order to empirically show that the \mbox{ZX-calculus} is a practically relevant method in equivalence checking, we conduct a detailed case study\footnote{A preliminary version of this case study has been published in~\cite{pehamEquivalenceCheckingParadigms2022}.} to establish a baseline for the current state of the art in equivalence checking
	of quantum circuits considering a large range of benchmarks. To this end, we \mbox{re-implemented} the ZX-calculus based
	equivalence checking algorithm in C++ and expanded it with the capabilities mentioned above. We compare this
	implementation---which has been integrated into the publicly available equivalence checking tool \emph{QCEC}~(\mbox{\url{https://github.com/cda-tum/qcec}})---with two other state-of-the-art equivalence checking methods: one based on path-sums~\cite{amyLargescaleFunctionalVerification2019} and another based on quantum decision
	diagrams~\cite{viamontesGatelevelSimulationQuantum2003,wangXQDDbasedVerificationMethod2008,niemannQMDDsEfficientQuantum2016,millerQMDDDecisionDiagram2006,zulehnerHowEfficientlyHandle2019}.
	
	Overall, we show that the ZX-calculus can be adapted to verify the results of compilation flows effectively, that it
	outperforms the path-sum approach by a constant factor, and that it performs on par with the decision diagram based method in many cases.
        However, both the ZX-calculus and the decision diagram based  methods have domains where they
	clearly outperform the other. Since we show that the ZX-calculus based approach
	may fail to prove the equivalence of two equivalent quantum circuits, it cannot be used to prove non-equivalence of
	quantum circuits---only to give an indication of non-equivalence. All in all, we conclude that neither the ZX-calculus nor the decision diagram based method is clearly better
	than the other but that they rather serve as complementary methods that are best used in conjunction.

	The remainder of this work is structured as follows: \autoref{cha:background} provides the necessary background and motivates the necessity of equivalence checking routines in quantum computing. Then,
	\autoref{cha:equ_checking} describes the equivalence checking problem in detail. Based on that, \autoref{cha:zx}
	recapitulates the theory of the ZX-calculus, explains the state-of-the art equivalence checking algorithm based on
	the ZX-calculus in detail, and shows how the method can be expanded to handle more relevant equivalence checking problems
	in quantum circuit compilation and optimization. There we also prove that the ZX-calculus based equivalence
	checking method is complete for Clifford circuits and illustrate that it is not complete in general---failing to prove the equivalence of a
	simple example. In \autoref{cha:experiments} we compare ZX-calculus based equivalence checking with methods
        based on paths-sums and decision diagrams. Finally, \autoref{cha:conclusion} concludes this work.

	\section{Background}\label{cha:background}

        To keep this work as self-contained as possible, this section provides a brief introduction to the concepts of
        quantum computing and quantum circuits relevant for this work.
        
        \subsection{Quantum Computing}
\label{sec:quantum_comp}
In classical computing, information is encoded in classical bits that can be either $0$ or $1$. Analogously, in quantum computing, \emph{quantum bits} (or \emph{qubits} in short) are used which can be either
in the $\ket{0}$ or $\ket{1}$ state (in Dirac notation). Contrary to the classical domain, qubits can also be in \emph{superposition} of multiple
states. Formally, the state~$\ket{\phi}$ of a qubit is written as
$$\ket{\phi} = \alpha_0 \ket{0} + \alpha_1 \ket{1} = \alpha_0 \begin{bmatrix} 1 \\ 0\end{bmatrix} + \alpha_1 \begin{bmatrix} 0 \\
	1\end{bmatrix} = \begin{bmatrix}\alpha_0 \\ \alpha_1 \end{bmatrix}$$
with \emph{amplitudes} $\alpha_0, \alpha_1 \in \mathbb{C}, |\alpha_0|^2 + |\alpha_1|^2 = 1$.

The basis states of multi-qubit systems are obtained as the \emph{tensor product} of single qubit states. So a basis state of a
$3$-qubit system would for example be written as $\ket{1} \otimes \ket{1} \otimes \ket{0} = \ket{110} =: \ket{6}$. In
general, an $n$-qubit state $\ket{\phi}$ is described by a linear combination of basis vectors, i.e.,
$$\sum_{i=0}^{2^n-1}\alpha_i \ket{i} \mbox{ with } \sum_{i=0}^{2^n-1}|\alpha_i|^2 = 1 \mbox{ and } \alpha_i \in \mathbb{C}.$$

Any operation manipulating the state of a quantum system must again yield a valid quantum state. As a consequence, any
such operation $U$ must be \emph{unitary}, i.e., it must obey the equation \mbox{$UU^\dagger = U^\dagger U = I$} where
$U^\dagger$ is the \emph{conjugate transpose} of $U$ and $I$ is the identity transformation.

\begin{example}
	Consider the \emph{Hadamard transform} \mbox{$H = \frac{1}{\sqrt{2}} \left[\begin{smallmatrix} 1 & 1 \\ 1 & -1 \end{smallmatrix}\right]$}.
	It can be easily checked by matrix multiplication that $H$ is a unitary transformation.
	The Hadamard transform maps \emph{Z-basis} states to \emph{X-basis} states, i.e.,
	\begin{align*}
		H \ket{0} &= \frac{1}{\sqrt{2}} \ket{0} + \frac{1}{\sqrt{2}} \ket{1} =: \ket{+} \\
		H \ket{1} &= \frac{1}{\sqrt{2}} \ket{0} - \frac{1}{\sqrt{2}} \ket{1} =: \ket{-}.
	\end{align*}
	\noindent
	An important unitary acting on two qubits is the \emph{controlled not} or CNOT gate. It is defined by the matrix 
	$ \left[ 
	\begin{smallmatrix}
		1&0&0&0\\
		0&1&0&0\\
		0&0&0&1\\
		0&0&1&0\\
	\end{smallmatrix} \right]
	$
	and flips the second qubit (the \emph{target}) when the first qubit (the \emph{control}) is in state $\ket{1}$.
	
\end{example}

A quantum computation is a unitary transformation acting on some initial state (usually the qubits are all prepared to
be $\ket{0}$). Instead of writing the \emph{system matrix} (i.e., the unitary describing the behavior of the whole circuit) explicitly, a common way to describe the \emph{unitary evolution} of a quantum system is
through \emph{quantum circuit} notation~\cite{nielsenQuantumComputationQuantum2010}. There, qubits are represented by wires and operations (called
\emph{gates}) are annotated as boxes and circles on the wires. The evolution of the initial state is read from left to right. Thus, a quantum circuit $G$ is described as a sequence of gates $g_{0}\dots g_{m-1}$.
Due to their unitary nature, quantum circuits are inherently reversible. More specifically, the inverse of a quantum
circuit \mbox{$G = g_{0}\dots g_{m-1}$} is obtained by inverting each gate and reversing the order of operations,
i.e., $G^\dagger = g_{m-1}^\dagger \dots g_{0}^\dagger$.

\begin{figure}[t]
	\centering
	\begin{subfigure}[b]{.25\textwidth}
		\centering\leavevmode
		\mbox{\scriptsize
		 \Qcircuit @C=1.0em @R=0.2em @!R { \\
			\nghost{ {q}_{0} :  } & \lstick{ {q}_{2} :  \ket{0}} & \qw & \qw & \targ \barrier[0em]{2} & \qw & \meter & \cw & \rstick{ {c}_{2} }\\ 
			\nghost{ {q}_{1} :  } & \lstick{ {q}_{1} :  \ket{0}} & \qw & \targ & \qw & \qw & \meter & \cw & \rstick{ {c}_{1} }\\
			\nghost{ {q}_{2} :  } & \lstick{ {q}_{0} :  \ket{0}} & \gate{\mathrm{H}} & \ctrl{-1} & \ctrl{-2} & \qw & \meter & \cw & \rstick{ {c}_{0} }\\ 
		}
		}
		\caption{GHZ state preparation circuit $G$}
		\label{fig:ghz}
	\end{subfigure}  \begin{subfigure}[b]{.2\textwidth}
		\[\resizebox{0.7\linewidth}{!}{$
			\tfrac{1}{\sqrt{2}} 
			\begin{bNiceArray}{CC:CC|CC:CC}
				1   &   &   &    & 1         &           &           &           \\
				& 1 &   &    &           & 1         &           &           \\ \hdottedline
				&   & 1 &    &           &           & 1         &           \\
				&   &   & 1  &           &           &           & 1         \\ \hline
				&   &   & 1  &           &           &           & \text{-}1 \\
				&   & 1 &    &           &           & \text{-}1 &           \\ \hdottedline
				& 1 &   &    &           & \text{-}1 &           &           \\
				1 &   &   &    & \text{-}1 &           &           &           \\
			\end{bNiceArray} 	$}\]
		\caption{System matrix $U$ of $G$}
		\label{fig:ghz_system_matrix}
	\end{subfigure}%
	\caption{GHZ state preparation}\vspace*{-4mm}
\end{figure}
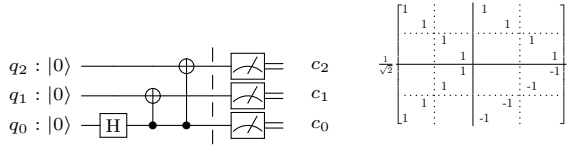

\vspace{3mm}
\begin{example}
	The circuit $G$ in Fig.~\ref{fig:ghz} represents a 3-qubit system. The box annotated with H is a Hadamard transform on
	qubit $q_0$ and the connected circles and dots are CNOT gates with $q_0$ as control and $q_1$ and $q_2$ as
	target qubit, respectively. The circuit maps $\ket{000}$ to $\frac{1}{\sqrt{2}} \ket{000} + \frac{1}{\sqrt{2}} \ket{111}$,
	the well-known GHZ state~\cite{greenbergerGoingBellTheorem2007}. The system matrix describing the unitary this circuit
	realizes is given in Fig.~\ref{fig:ghz_system_matrix}.
\end{example}

 \vspace{-1mm}
	
	\subsection{Quantum Circuit Compilation}\label{sec:quant-circ-comp}
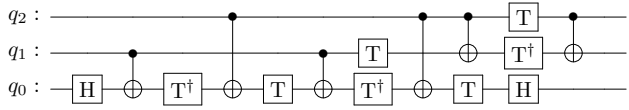
\begin{figure}[t]
		\centering
		\resizebox{\linewidth}{!}{
		\scalebox{1.0}{
			\Qcircuit @C=1.0em @R=0.2em @!R { \\
				\nghost{ {q}_{0} :  } & \lstick{ {q}_{2} :  } & \qw & \qw & \qw & \ctrl{2} & \qw & \qw & \qw & \ctrl{2} & \ctrl{1} & \gate{\mathrm{T}} & \ctrl{1} & \qw & \qw\\ 
				\nghost{ {q}_{1} :  } & \lstick{ {q}_{1} :  } & \qw & \ctrl{1} & \qw & \qw & \qw & \ctrl{1} & \gate{\mathrm{T}} & \qw & \targ & \gate{\mathrm{T^\dagger}} & \targ & \qw & \qw\\ 
				\nghost{ {q}_{2} :  } & \lstick{ {q}_{0} :  } & \gate{\mathrm{H}} & \targ & \gate{\mathrm{T^\dagger}} & \targ & \gate{\mathrm{T}} & \targ & \gate{\mathrm{T^\dagger}} & \targ & \gate{\mathrm{T}} & \gate{\mathrm{H}} & \qw & \qw & \qw\\ 
				\\ }}}
		\caption{Decomposition of the Toffoli gate in Clifford+T}\label{fig:toff_decomp}
	\end{figure}

	Quantum algorithms are typically designed at a rather high abstraction level without considering specific hardware
	restrictions. In order to execute a conceptual quantum algorithm on an actual device, it has to be \emph{compiled} to a
	representation that conforms to all restrictions imposed by the targeted device. Since quantum computers typically only
	support a limited gate-set, every high-level operation has to be \emph{decomposed} into that gate-set~\cite{vidalUniversalQuantumCircuit2004, barencoElementaryGatesQuantum1995, maslovAdvantagesUsingRelative2016}. This
	can sometimes significantly increase the size of the circuit. \autoref{fig:toff_decomp} shows an exemplary decomposition
	of the Toffoli gate in the Clifford+T gate-set.
	
	In addition,
	many architectures (such as those based on superconducting qubits)
	restrict the pairs of qubits that operations may be applied to. Hence, it is necessary to \emph{map} the decomposed
	circuit to the device such that it adheres to the device's coupling constraints~\cite{willeMappingQuantumCircuits2019, muraliNoiseadaptiveCompilerMappings2019, liTacklingQubitMapping2019}. In general, this is accomplished by
	establishing a mapping between the circuit's logical qubits and the device's physical qubits. Since it is generally not
	possible to determine a conforming mapping in a static fashion, SWAP gates are inserted into the circuit that allow to
	dynamically change the logical-to-physical qubit mapping over the course of the compilation.
	
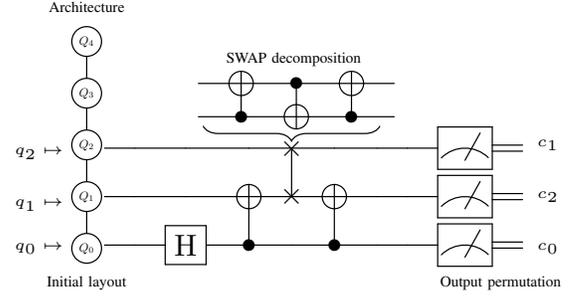
\begin{figure}[t]
		\centering
		\resizebox{0.85\linewidth}{!}{%
                  \begin{tikzpicture}[qnode/.style={circle, draw=black,scale=0.4}]
				\node[qnode] (Q0) at (-2.2,-0.6) {$Q_0$};
				\node[qnode] (Q1) at (-2.2, 0) {$Q_1$} ;
				\node[qnode] (Q2) at (-2.2, 0.6) {$Q_2$};
				\node[qnode] (Q3) at (-2.2, 1.2) {$Q_3$};
				\node[qnode] (Q4) at (-2.2, 1.8) {$Q_4$};
				\draw[] (Q0.north) -- (Q1.south);
				\draw[] (Q1.north) -- (Q2.south);
				\draw[] (Q2.north) -- (Q3.south);
				\draw[] (Q3.north) -- (Q4.south);
				
				\node[] (q0) at (-2.75,-0.6) {\tiny$q_0 \mapsto$};
				\node[] (q1) at (-2.75, -0.1) {\tiny$q_1 \mapsto$} ;
				\node[] (q2) at (-2.75 , 0.5) {\tiny$q_2 \mapsto$};
				\node[] (c0) at (3.15,-0.6) {\tiny$c_0$};
				\node[] (c1) at (3.15, 0) {\tiny$c_2$} ;
				\node[] (c2) at (3.15 , 0.6) {\tiny$c_1$};
				
				\node[]  at (-2.2 , -1) {\tiny Initial layout};
				\node[]  at (0.2, 1.6) {\tiny SWAP decomposition};
				\node[]  at (2.6 , -1) {\tiny Output permutation};
				\node[]  at (-2.2 , 2.2) {\tiny Architecture};
				\node at (0,0){
					\Qcircuit @C=1.0em @R=0.2em @!R { \\
						\nghost{ {q}_{2} :  } & \qw & \qw               & \qw       & \qswap \qwx[1] & \qw       & \qw & \qw & \meter & \cw\\
						\nghost{ {q}_{2} :  } &	 \qw & \qw               & \targ     & \qswap         & \targ     & \qw & \qw & \meter & \cw \\
						\nghost{ {q}_{2} :  } &	 \qw & \gate{\mathrm{H}} & \ctrl{-1} & \qw            & \ctrl{-1} & \qw & \qw & \meter & \cw\\
						\\ }};
				\node at (-0.2,1.12) {
					\Qcircuit @C=1.0em @R=0.2em @!R { \\
						\nghost{ {q}_{2} :  } & \targ     & \ctrl{1}  & \targ & \qw       \\ 
						\nghost{ {q}_{1} :  }   & \ctrl{-1} & \targ & \ctrl{-1} & \qw \\ 
						\\ }};
				
				\draw[decorate, decoration={brace, amplitude=5pt}] (1.2,0.82)--(-0.85,0.82);
		\end{tikzpicture}}\vspace*{-2mm}
		\centering
		\caption{Compilation of GHZ state preparation circuit}\vspace*{-2mm}
		\label{fig:ghz_mapped}
	\end{figure}

	\begin{example}
		\label{ex:ghz_mapping}
		Consider again the GHZ preparation circuit shown in \autoref{fig:ghz} and assume it shall be mapped to the $5$-qubit,
		linear architecture shown on the left-hand side of \autoref{fig:ghz_mapped}. Assume that, initially, logical qubit $q_i$ is mapped to physical qubit
		$Q_i$ for $0 \leq i \leq 2$. Then, the first two operations can be directly applied, while the last operation cannot---due to the fact that $Q_0$ and $Q_2$ are not directly connected on the architecture. Hence, a SWAP operation between
		$Q_2$ and $Q_1$ is introduced, which allows to execute the final gate. At the end of the circuit $q_0$ is measured on $Q_0$, $q_1$ on $Q_2$ and $q_2$ on $Q_1$.
	\end{example}
	
        \vspace*{-3mm}
	\subsection{Quantum Circuit Optimization}\label{sec:opt}
	
	Through decomposition and mapping, even small quantum circuits can significantly increase in size. In classical
	computing, circuits are usually optimized in order to require less space, time, or energy. While time is still
	an important factor in quantum computing, this is because of a different reason. The \emph{coherence time} of a quantum
	mechanical system is the time for which the system remains quantum-mechanically
	coherent~\cite{nielsenQuantumComputationQuantum2010}. When this time is exceeded, the system collapses into some basis
	state and is therefore no longer in superposition and qubits are not entangled anymore. The decoherence time essentially
	puts a limit on the maximum number of operations that can be performed on a quantum system before it collapses. 
	
	Another factor unique to quantum computing is the \emph{gate error rate}. Gates are difficult to realize precisely in
	practice. Every operation performed on qubits potentially introduces some error. This is often quantified using the
	\emph{fidelity} $\mathcal{F}$, which measures the distance between two quantum states. The \emph{gate fidelity}~\cite{nielsenQuantumComputationQuantum2010} is then a
	measure of the fidelity of the quantum state after applying a noisy gate compared to the quantum state if an ideal
	(noiseless) gate was applied.
	The higher the gate fidelity, the better the implementation of the gate. Some gates are easier to realize than
	others. On superconducting architectures, CNOT error is the dominating error factor. At the time of writing, for
	example, the single-qubit Pauli $X$ error on the IBMQ Montreal quantum computer was $\num{2.003e-4}$ on the physical qubit
	$Q0$ compared to the CNOT error between qubit $Q0$ and $Q1$ of $\num{2.276e-3}$.
	\begin{figure}[t]
		\centering
		\begin{subfigure}[b]{\linewidth}
		\resizebox{\linewidth}{!}{
			\begin{tikzpicture}
				\node at (5.5, 0.4) {$\cdots$};
				\node at (5.5, 0.05) {$\cdots$};
				\node at (5.5, -0.3) {$\cdots$};
				\node at (5.5, -0.65) {$\cdots$};
				\node at (-5.5, -1.5) {$\cdots$};
				\node at (-5.5, -1.85) {$\cdots$};
				\node at (-5.5, -2.2) {$\cdots$};
				\node at (-5.5, -2.55) {$\cdots$};
				\node at (-1,0){\scalebox{0.6}{
						\Qcircuit @C=1.0em @R=0.2em @!R { \\
							\nghost{ {q}_{0} :  } & \lstick{ {q}_{3} :  } & \gate{R_Z\,(\frac{\pi}{8})}  & \ctrl{1}                    & \qw                          & \ctrl{1} & \qw                          & \qw                          & \ctrl{2}                    & \qw                         & \qw                          & \qw                          & \ctrl{2} … & \qw      & \qw                          & \qw      & \qw                         & \qw \\
							\nghost{ {q}_{1} :  } & \lstick{ {q}_{2} :  } & \gate{R_Z\,(\frac{\pi}{8})}  & \targ                       & \gate{R_Z\,(-\frac{\pi}{8})} & \targ    & \ctrl{1}                     & \qw                          & \qw                         & \qw                         & \ctrl{1}                     & \qw                          & \qw        & \qw      & \qw                          & \ctrl{2} & \qw                         & \qw \\
							\nghost{ {q}_{2} :  } & \lstick{ {q}_{1} :  } & \gate{R_Z\,(\frac{\pi}{8})}  & \qw                         & \qw                          & \qw      & \targ                        & \gate{R_Z\,(-\frac{\pi}{8})} & \targ                       & \gate{R_Z\,(\frac{\pi}{8})} & \targ                        & \gate{R_Z\,(-\frac{\pi}{8})} & \targ      & \ctrl{1} & \qw                          & \qw      & \qw                         & \qw \\ 
							\nghost{ {q}_{3} :  } & \lstick{ {q}_{0} :  } & \gate{H}                     & \gate{R_Z\,(\frac{\pi}{8})} & \qw                          & \qw      & \qw                          & \qw                          & \qw                         & \qw                         & \qw                          & \qw                          & \qw        & \targ    & \gate{R_Z\,(-\frac{\pi}{8})} & \targ    & \gate{R_Z\,(\frac{\pi}{8})} & \qw \\
				}}};
				\node at (-1, -2) {\scalebox{0.6}{\Qcircuit @C=1.0em @R=0.2em @!R {\\
							\nghost{ {q}_{0} :  } & \qw                   & \qw                          & \ctrl{3}                    & \qw                          & \qw      & \qw                          & \qw                          & \qw                         & \qw                         & \qw                          & \ctrl{3}                     & \qw        & \qw      & \qw                          & \qw      & \qw                               \\ 
							\nghost{ {q}_{1} :  } & \qw                   & \qw                          & \qw                         & \qw                          & \qw      & \qw                          & \ctrl{2}                     & \qw                         & \qw                         & \qw                          & \qw                          & \qw        & \qw      & \qw                          & \qw      & \qw                               \\ 
							\nghost{ {q}_{2} :  } & \ctrl{1}              & \qw                          & \qw                         & \qw                          & \ctrl{1} & \qw                          & \qw                          & \qw                         & \ctrl{1}                    & \qw                          & \qw                          & \qw        & \qw      & \qw                          & \qw      & \qw                               \\ 
							\nghost{ {q}_{3} :  } & \targ                 & \gate{R_Z\,(-\frac{\pi}{8})} & \targ                       & \gate{R_Z\,(\frac{\pi}{8})}  & \targ    & \gate{R_Z\,(-\frac{\pi}{8})} & \targ                        & \gate{R_Z\,(\frac{\pi}{8})} & \targ                       & \gate{R_Z\,(-\frac{\pi}{8})} & \targ                        & \gate{H}   & \qw      & \qw                          & \qw      & \qw                               \\                                                                                                                                                                                                                                                                                                                                                                                                                                                                                                    \\ }}
				};
			\end{tikzpicture}}
			\caption{Multi-controlled Toffoli without ancillary qubits}\label{fig:mcx_orig}
		\end{subfigure}

		\begin{subfigure}[b]{\linewidth}
			\resizebox{\linewidth}{!}{
			\scalebox{0.5}{
				\Qcircuit @C=1.0em @R=0.2em @!R { \\
					\nghost{ {q}_{0} :  } & \lstick{ {q}_{3} :  } & \qw & \qw & \qw & \qw & \ctrl{4} & \qw & \qw & \qw & \qw & \qw & \qw & \qw & \qw & \qw & \qw & \qw & \qw & \qw & \qw & \qw & \qw & \qw & \qw & \qw & \ctrl{4} & \qw & \qw & \qw & \qw & \qw & \qw\\ 
					\nghost{ {q}_{1} :  } & \lstick{ {q}_{2} :  } & \qw & \qw & \ctrl{3} & \qw & \qw & \qw & \ctrl{3} & \qw & \qw & \qw & \qw & \qw & \qw & \qw & \qw & \qw & \qw & \qw & \qw & \qw & \qw & \qw & \ctrl{3} & \qw & \qw & \qw & \ctrl{3} & \qw & \qw & \qw & \qw\\ 
					\nghost{ {q}_{2} :  } & \lstick{ {q}_{1} :  } & \qw & \qw & \qw & \qw & \qw & \qw & \qw & \qw & \qw & \qw & \qw & \ctrl{1} & \qw & \qw & \qw & \ctrl{1} & \ctrl{2} & \gate{T} & \qw & \ctrl{2} & \qw & \qw & \qw & \qw & \qw & \qw & \qw & \qw & \qw & \qw & \qw\\ 
					\nghost{ {q}_{3} :  } & \lstick{ {q}_{0} :  } & \gate{H} & \qw & \qw & \qw & \qw & \qw & \qw & \qw & \qw & \targ & \gate{T^\dagger} & \targ & \gate{T} & \targ & \gate{T^\dagger} & \targ & \qw & \gate{T} & \gate{H} & \qw & \qw & \qw & \qw & \qw & \qw & \qw & \qw & \qw & \qw & \qw & \qw\\ 
					\nghost{ {a}_{0} :  } & \lstick{ {a}_{0} :  } & \gate{H} & \gate{T} & \targ & \gate{T^\dagger} & \targ & \gate{T} & \targ & \gate{T^\dagger} & \gate{H} & \ctrl{-1} & \qw & \qw & \qw & \ctrl{-1} & \gate{T} & \qw & \targ & \gate{T^\dagger} & \qw & \targ & \gate{H} & \gate{T} & \targ & \gate{T^\dagger} & \targ & \gate{T} & \targ & \gate{T^\dagger} & \gate{H} & \qw & \qw\\ 
					\\ }}}
			\caption{Multi-controlled Toffoli gate with ancillary qubits}\label{fig:mcx_ancilla}
		\end{subfigure}
		\caption{Decompositions of the multi-controlled Toffoli gate}\label{fig:mcx}\vspace*{-3mm}
	\end{figure}
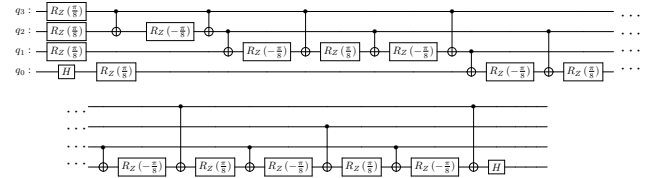
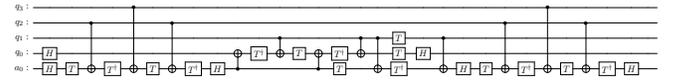
	
	Because of the coherence time and gate errors, it is important to optimize the number of elementary gate operations used in a quantum algorithm. It
	is not just a matter of execution time but of whether a meaningful result can be obtained at all from the execution of a
	quantum algorithm on a specific architecture. Many optimization schemes exist and optimization of quantum circuits is
	still an active area of research~\cite{hietalaVerifiedOptimizerQuantum2019,namAutomatedOptimizationLarge2018,itokoOptimizationQuantumCircuit2020}.
	
	The simplest of these optimization schemes is the \emph{single-qubit gate fusion}. Since any single qubit unitary
	represents a rotation of the Bloch sphere, any sequence of such gates also represents a rotation. Instead of performing
	each rotation separately, the complete rotation can be performed at once.
	
	Another kind of optimization involves using \emph{ancillary qubits}. Ancillary qubits are additional qubits apart from
	those required for the quantum algorithm. They can act as a sort of ``working memory'' to allow for a more compact
	representation of certain quantum gates or algorithms. Because the value of ancillary qubits is only important during
	the computation, they are not measured and do not factor into the final result.
	
	\begin{example}\label{ex:ancilla}
		Consider the quantum circuit in \autoref{fig:mcx_orig} which represents a multi-controlled Toffoli gate with three
		control qubits. This gate performs a Pauli $X$ gate on the last qubit if and only if all the previous qubits are in
		the $\ket{1}$ state. Otherwise it acts as the identity on all qubits. \autoref{fig:mcx_ancilla} shows an implementation of this gate with the addition of one ancillary qubit. This is
		now a quantum circuit acting on five qubits with a higher total gate count, using $33$ compared to the $31$ of the
		original circuit. This circuit uses less CNOT gates however, needing only $12$ CNOT operations compared to the $14$ of
		the previous circuit. This trade-off is desirable because of the higher gate error of CNOT gates.
	\end{example}
	
	Of course, optimizations are also employed in conjunction with compilation methods to optimize a compiled circuit while
	still obeying all restrictions enforced by the hardware. But sometimes optimizations are done before mapping. In fact,
	some available optimization methods are not designed to handle hardware-specific restrictions~\cite{kissingerReducingTcountZXcalculus2020}. Therefore,
	equivalence checking methods are not only relevant when verifying the result of a compilation from a high-level
	description but also when verifying the results of optimizations of an uncompiled circuit or even just verifying the
	equivalence of a high-level description with an uncompiled optimized version of the high-level description.
	
	Eventually, compilation and optimization yields a new circuit that might look quite different from the original high-level description.
	It is essential for the successful execution of a quantum computation to verify that the compiled circuit still implements the same functionality as the original one.
	To this end, methods to check the equivalence of quantum circuits are necessary.
	
	\section{Equivalence Checking}\label{cha:equ_checking}
	In order to discuss equivalence checking methods for quantum circuits we first need to precisely define the equivalence
	checking problem and aspects unique to equivalence checking in the quantum realm---namely permutations of the input
	and output layout of a quantum circuit, inaccuracies stemming from working with complex numbers, and ancillary qubits.

		In general, given two quantum circuits 
		\[G = g_0 \dots g_{m-1} \mbox{ and } G^\prime = g_0^\prime \dots g_{m^\prime - 1}^\prime\]
		with corresponding system matrices 
		\[U = U_{m-1}\cdots U_0 \mbox{ and } U^\prime = U'_{m'-1}\cdots U'_0,\]
		the \emph{equivalence checking problem for quantum circuits} asks whether 
		\[U = e^{i \theta} U^\prime \mbox{ or, equivalently, } U^\dagger U' = e^{i \theta}I,\]
		where $\theta \in (-\pi, \pi]$ denotes a physically unobservable global phase.

	So, in principle, checking the equivalence of two quantum circuits reduces to the construction and the comparison of the respective system matrices.
	While this is straightforward conceptually, it quickly becomes an increasingly difficult task due to the size of the involved matrices scaling exponentially with the number of qubits.
	Equivalence checking of quantum circuits has even been shown to be
	QMA-complete~\cite{janzingNonidentityCheckQMAcomplete2005}.
	
	Even this definition is lacking when talking about the results of compilation flows. Compilation and optimization can
	alter a circuit in such a way that two circuits can be considered equal even if they have different system matrices.
	
	Firstly, there are numerical inaccuracies.
	This is one of the biggest, yet hardly talked about, practical issues when actually conducting equivalence checking.
        Because quantum gates are described by matrices over~$\mathbbm{C}$, they are hard to accurately represent in memory. 
	Usually, these matrices are stored using floating point numbers which leads to imprecisions and rounding errors. 
	Therefore, comparing two matrices for exact equality becomes pointless in many practical cases. 
	Instead, the Hilbert-Schmidt inner product can be used to quantify the similarity between two matrices. 
	Let $\operatorname{tr}$ denote the trace of a matrix, i.e., the sum of its diagonal elements.
	Then, because $\operatorname{tr}(I) = 2^n$ for the identity transformation on $n$ qubits, one can check whether $\vert\operatorname{tr}(U^\dagger U^\prime)\vert \approx 2^n$ in order to conclude the equivalence of both circuits up to a given tolerance. 
	
	Secondly, compilation flows introduce SWAP operations into a circuit such that the resulting circuit conforms to the hardware
	topology. These techniques use a circuit's initial layout and output permutation as an additional degree of freedom for
	saving SWAP operations, as, e.g., illustrated in \autoref{ex:ghz_mapping}. Because of these SWAPs, two circuits might
	only be equivalent up to a reordering of the qubits. 
	Hence, in order to verify the equivalence of compilation flow results, any equivalence checking routine must be able to
	handle these kinds of permutations and accurately track which gate is performed on which qubit.
	
	Lastly, as discussed in~\autoref{sec:opt}, sometimes it is necessary to check the equivalence of two circuits that might
	not even operate on the same number of qubits due to the use of ancillary qubits. Ideally, an equivalence checking
	routine can also handle those cases. The difficulty comes from the fact that quantum circuits with
	differing numbers of qubits cannot represent the same unitary. Indeed, their unitaries do not even have the same
	dimensions. Ancillary qubits have a constant initial state, being either in the $\ket{0}$ or $\ket{1}$
	state.

        Given two quantum circuits         \vspace*{0.5mm}
        \[G = g_0 \dots g_{m-1} \mbox{ and } G^\prime = g_0^\prime \dots g_{m^\prime - 1}^\prime\]        \vspace*{0.5mm}
        with ancillaries qubits with corresponding system matrices         \vspace*{0.5mm}
        \[U = U_{m-1}\cdots U_0 \mbox{ and } U^\prime = U^\prime_{m'-1}\cdots U^\prime_0,\]        \vspace*{0.5mm}
        the \emph{equivalence checking problem for quantum circuits involving ancillary qubits} asks whether         \vspace*{0.5mm}
        \[U_{\text{anc}} = e^{i \theta} U^\prime_{\text{anc}} \mbox{ or, equivalently, } U_{\text{anc}}^\dagger U_{\text{anc}}^\prime = e^{i \theta}I,\]        \vspace*{0.5mm}
        where $\theta \in (-\pi, \pi]$ denotes a physically unobservable global phase and $U_{\text{anc}}$ is the
        unitary obtained from $U$ by fixing the ancillary qubits in $G$ to their constant state.

        \begin{example}\label{ex:ancilla}
          A CNOT gate can be trivially used as a Pauli $X$ gate by treating the control qubit as an ancillary with
  constant state $\ket{1}$. To conform with our
  construction that the ancilla is the last qubit, we consider the matrix of a CNOT where the target is on qubit $0$.

\[U = \left[ 
    \begin{smallmatrix}
      1&0&0&0\\
      0&0&0&1\\
      0&0&1&0\\
      0&1&0&0\\
    \end{smallmatrix}\right]   = \ket{00}\bra{00}+\ket{01}\bra{11}+\ket{10}\bra{10}+\ket{11}\bra{01}.\]

\noindent Fixing the last qubit gives:

  \begin{align*}
U_{\text{anc}} &= \bra{10}  \left[ 
    \begin{smallmatrix}
      1&0&0&0\\
      0&0&0&1\\
      0&0&1&0\\
      0&1&0&0\\
    \end{smallmatrix}\right]\ket{01}\ket{0}\bra{0} +
                                           \bra{11}  \left[ 
    \begin{smallmatrix}
      1&0&0&0\\
      0&0&0&1\\
      0&0&1&0\\
      0&1&0&0\\
    \end{smallmatrix}\right]\ket{01}\ket{1}\bra{0} \\
                                         &+\bra{10}  \left[ 
    \begin{smallmatrix}
      1&0&0&0\\
      0&0&0&1\\
      0&0&1&0\\
      0&1&0&0\\
    \end{smallmatrix}\right]\ket{11}\ket{0}\bra{1} +
                                           \bra{11}  \left[ 
    \begin{smallmatrix}
      1&0&0&0\\
      0&0&0&1\\
      0&0&1&0\\
      0&1&0&0\\
    \end{smallmatrix}\right]\ket{11}\ket{1}\bra{1} \\
    &= \ket{1}\bra{0} + \ket{0}\bra{1} =
      \begin{bmatrix}
        0&1\\1&0
      \end{bmatrix} = X.
  \end{align*}

\noindent  This shows that $U_\text{anc}$ of a CNOT gate with constant control of $\ket{1}$ is equivalent to a Pauli $X$ gate.
\end{example}
	
	\begin{figure}[t]
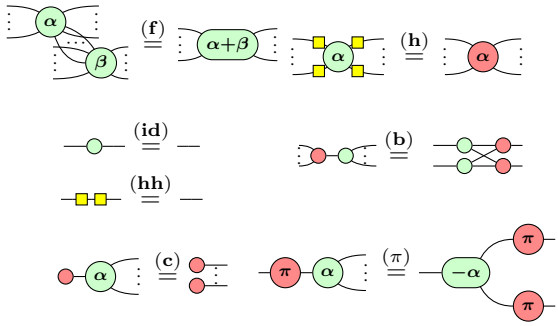

		\centering
		
		\begin{subfigure}[c]{0.49\linewidth}
			\centering
			\begin{equation*}
			\begin{ZX}[math baseline=wantedbaseline, row sep=0.3mm]
				\leftManyDots{} \zxZ[a=alpha]{\alpha}\rightManyDots{}  & \\
				\\
				&  \zxN{} \ar[r,3 dots] & \zxN{}\\
				\zxN[a=wantedbaseline]\\
				& \leftManyDots{} \zxZ[a=beta]{\beta} \rightManyDots{}\\
				\ar[from=alpha, to=beta, <']
				\ar[from=beta, to=alpha, <']
			\end{ZX} \overset{(\mathbf{f})}{=} \begin{ZX}[math baseline=wantedbaseline]
				\leftManyDots{} \zxZ[a=wantedbaseline]{\alpha + \beta} \rightManyDots{}
			\end{ZX}
		\end{equation*}
			
		\end{subfigure}
		\hspace{-1cm}
		\begin{subfigure}[c]{0.49\linewidth}
			\centering
			\begin{equation*}
			\begin{ZX}[math baseline=wantedbaseline, row sep=0.3mm]
				\zxN[a=upperL]{} &[\zxHCol] \zxN{}& & & [\zxHCol] \zxN[a=upperR]{}\\
				\zxN{}\ar[3 vdots]& [\zxHCol] \zxN{}& [\zxHCol]\zxZ[a=spider]{\alpha} &[\zxHCol] \zxN{} &[\zxHCol] \zxN[a=wantedbaseline]{}\ar[3 vdots]\\
				\zxN[a=lowerL]{} & & &  & \zxN[a=lowerR]{}
				\ar[from=spider, to=upperL, <', H]
				\ar[from=spider, to=upperR, <', H]
				\ar[from=spider, to=lowerL, <', H]
				\ar[from=spider, to=lowerR, <', H]
                              \end{ZX}\overset{(\mathbf{h})}{=}
			\begin{ZX}[row sep=0.3mm]
				\zxN[a=upperL]{} & & & & \zxN[a=upperR]{}\\
				\zxN{}\ar[3 vdots]& & \zxX[a=spider]{\alpha} & & \zxN[a=wantedbaseline]{}\ar[3 vdots]\\
				\zxN[a=lowerL]{} & & &  & \zxN[a=lowerR]{}
				\ar[from=spider, to=upperL, <']
				\ar[from=spider, to=upperR, <']
				\ar[from=spider, to=lowerL, <']
				\ar[from=spider, to=lowerR, <']
			\end{ZX}
	\end{equation*}	
	\end{subfigure}

		\begin{subfigure}[c]{0.49\linewidth}
			\centering
			\begin{equation*}
			\begin{ZX}[math baseline=wantedbaseline, row sep=0.3mm]
				\zxN{} \rar & \zxN{} \rar& \zxZ{} \rar& \zxN{} \rar &\zxN{}
			\end{ZX}\overset{(\mathbf{id})}{=}
			\begin{ZX}
				\zxN{} \rar& \zxN{} \rar& \zxN{} &
			\end{ZX}
		\end{equation*}
	\begin{equation*}
			\begin{ZX}
				\zxN{} \rar&\zxN{} \rar[H]& [\zxwCol]\zxN{} \rar[H]&[\zxwCol] \zxN{} \rar& \zxN{}
			\end{ZX}\overset{(\mathbf{hh})}{=}
			\begin{ZX}
				\zxN{} \rar& \zxN{} \rar& \zxN{} &
			\end{ZX}
	\end{equation*}
		\end{subfigure}
		\hspace{-1cm}
		\begin{subfigure}[c]{0.49\linewidth}
			\centering
			\begin{equation*}
			\begin{ZX}[math baseline=wantedbaseline, row sep=0.3mm]
				\zxN[a=upperL]{} & & & & \zxN[a=upperR]{}\\
				\zxN{}\ar[3 vdots]&  \zxX[a=leftS]{}\rar &[\zxwCol] \zxZ[a=rightS]{}& \zxN[a=wantedbaseline]{}\ar[3 vdots]\\
				\zxN[a=lowerL]{} & & &  & \zxN[a=lowerR]{}
				\ar[from=leftS, to=upperL, <']
				\ar[from=rightS, to=upperR, <']
				\ar[from=leftS, to=lowerL, <']
				\ar[from=rightS, to=lowerR, <']
			\end{ZX}\overset{(\mathbf{b})}{=}
			\begin{ZX}[math baseline=wantedbaseline, row sep=0.3mm]
				\zxN{} \rar & \zxZ[a=upperL]{} \ar[rr]& & \zxX[a=upperR]{} \rar & \zxN{} \\
				\zxN[a=wantedbaseline]\\
				\zxN{} \rar & \zxZ[a=lowerL]{} \ar[rr]& & \zxX[a=lowerR]{} \rar & \zxN{} \\
				\ar[from=upperL, to=lowerR]
				\ar[from=lowerL, to=upperR]
			\end{ZX}
	\end{equation*}
		\end{subfigure}

		\begin{subfigure}[c]{0.49\linewidth}
			\centering
			\begin{equation*}
			\begin{ZX}[row sep=0.3mm]
				\zxN[a=upperL]{} & & & & \zxN[a=upperR]{}\\
				& \zxX{} \rar & \zxZ[a=spider]{\alpha} & & \zxN{}\ar[3 vdots]\\
				\zxN[a=lowerL]{} & & &  & \zxN[a=lowerR]{}
				\ar[from=spider, to=upperR, <']
				\ar[from=spider, to=lowerR, <']
			\end{ZX}\overset{(\mathbf{c})}{=}
			\begin{ZX}[row sep=0.3mm]
				\zxX{} \ar[rr] & \zxN{}             & \zxN{} \\
				& \zxN{}\ar[3 vdots] &        \\
				\zxX{} \ar[rr] & \zxN{}             & \zxN{} \\
			\end{ZX}
		\end{equation*}
		\end{subfigure}
		\hspace{-1cm}
		\begin{subfigure}[c]{0.49\linewidth}
			\centering
			\begin{equation*}
			\begin{ZX}[row sep=0.3mm]
				\zxN[a=upperL]{} & & & & \zxN[a=upperR]{}\\
				\zxN{}\rar& \zxX{\pi} \rar & \zxZ[a=spider]{\alpha} & & \zxN{}\ar[3 vdots]\\
				\zxN[a=lowerL]{} & & &  & \zxN[a=lowerR]{}
				\ar[from=spider, to=upperR, <']
				\ar[from=spider, to=lowerR, <']
			\end{ZX}
			\overset{(\mathbf{\pi})}{=}
			\begin{ZX}[row sep=0.3mm]
				\zxN[a=upperL]{} &  &                         &  & \zxX[a=upperR]{\pi}\rar & \zxN{}\\
				\zxN{}\ar[rr]    &  & \zxZ[a=spider]{-\alpha} &  &       \\
				\zxN[a=lowerL]{} &  &                         &  & \zxX[a=lowerR]{\pi} \rar & \zxN{}
				\ar[from=spider, to=upperR, <']
				\ar[from=spider, to=lowerR, <']
			\end{ZX}
		\end{equation*}
		\end{subfigure}
		
		\caption{Axioms of the scalar-free ZX-calculus}\label{fig:zx_axioms}                \vspace*{-4mm}
              \end{figure}
              
	In order to avoid the emergence of a verification gap as for classical systems, automated software solutions for equivalence checking of quantum
	circuits have to be developed.
	To this end,
	various methods have been proposed~\cite{yamashitaFastEquivalencecheckingQuantum2010,berentSATEncodingQuantum2022,hongApproximateEquivalenceChecking2021,viamontesCheckingEquivalenceQuantum2007,cowtanGenericCompilationStrategy2020,amyLargescaleFunctionalVerification2019, burgholzerAdvancedEquivalenceChecking2021,wangXQDDbasedVerificationMethod2008}. 
	However, most of them either only work on small circuits, lack publicly available implementations or are based on paradigms established in classical computing that do not take the full picture of quantum computing into account. 
	Few methods exist that approach equivalence checking entirely from the perspective of quantum computing~\cite{burgholzerAdvancedEquivalenceChecking2021,cowtanGenericCompilationStrategy2020,amyLargescaleFunctionalVerification2019}.
	Even these existing approaches view the equivalence checking problem from completely different perspectives and a baseline indicating which paradigm is suited best for which use case is yet to be established.
	
	\section{ZX-Calculus}\label{cha:zx}

        The
	ZX-calculus~\cite{vandeweteringZXcalculusWorkingQuantum2020,coeckePicturingQuantumProcesses2018}
	is a graphical notation for quantum circuits equipped with a powerful set of rewrite rules that enable diagrammatic
	reasoning about quantum computing. It has been successfully applied to quantum circuit compilation and
	optimization~\cite{cowtanGenericCompilationStrategy2020,duncanGraphtheoreticSimplificationQuantum2019,kissingerReducingTcountZXcalculus2020}
	and to some extent also to equivalence checking of quantum
	circuits~\cite{kissingerReducingTcountZXcalculus2020}. The algorithm for equivalence checking using the ZX-calculus was
	only mentioned briefly as an alternative application of the optimization proposed in~\cite{kissingerReducingTcountZXcalculus2020}
        and has not been adapted to handle numerical inaccuracies, permutations of input and output layout of a compiled circuit, and ancillary qubits as mentioned in
	\autoref{cha:equ_checking}. In short, more work is needed to handle equivalence checking of compilation results with the
	ZX-calculus. In this section, we are going to discuss how this can be done as well as provide first results on the
	(in-)completeness of the ZX-calculus based equivalence checking algorithm.
	
	In order to do this, we are first going to introduce the necessary background on the ZX-calculus and the current
	state-of-the-art algorithm in equivalence checking with the ZX-calculus.

                        \vspace*{-2mm}
	\subsection{Basics of ZX-Calculus}

	\begin{figure}[t]
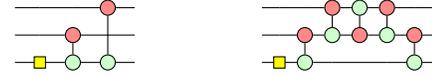

		\centering
		\begin{subfigure}[b]{.49\linewidth}
			\centering
			\begin{ZX}
				\zxN \rar   & [\zxwCol] \zxN \rar   & [\zxwCol] \zxN \rar             & [\zxwCol] \zxX{} \rar         & [\zxwCol] \zxN{} \\
				\zxN{} \rar & [\zxwCol] \zxN{} \rar & [\zxwCol] \zxX{} \ar[d] \ar[rr] & [\zxwCol] \zxN{}              & [\zxwCol] \zxN{} \\
				\zxN{} \rar & [\zxwCol] \zxH{} \rar & [\zxwCol] \zxZ{} \rar           & [\zxwCol] \zxZ{} \ar[uu] \rar & [\zxwCol] \zxN{} \\
			\end{ZX}
			
			\caption[ZX-Diagram of GHZ Circuit]{Uncompiled GHZ circuit}
			\label{fig:ghz_zx}    
		\end{subfigure}
		\hspace{-1cm}
		\begin{subfigure}[b]{.5\linewidth}
			\centering
			\begin{ZX}
				\zxN \rar   &  \zxN \rar   &  \zxN \rar            &  \zxX{} \rar        &  \zxZ{} \dar \rar & \zxX{} \ar[d]\rar &  \zxN{} \rar       & \zxN{} \\
				\zxN{} \rar &  \zxN{} \rar &  \zxX{} \ar[d] \ar[r] &  \zxZ{} \ar[u] \rar &  \zxX{}\rar       & \zxZ{} \rar       &  \zxX{} \rar       & \zxN{} \\
				\zxN{} \rar &  \zxH{} \rar &  \zxZ{} \rar          &  \zxN{}  \rar       &  \zxN{}\rar       & \zxN{} \rar       &  \zxZ{} \ar[u]\rar & \zxN{} \\
			\end{ZX}
			\caption[ZX-Diagram of GHZ Circuit]{Compiled GHZ circuit}
			\label{fig:ghz_mapped_zx}    
		\end{subfigure}
		
		\caption{ZX-diagrams of GHZ state preparation circuits}\label{fig:zx_diagrams}                \vspace*{-4mm}
              \end{figure}
              
	A ZX-diagram is made up of colored nodes (called \emph{spiders}) that
	are connected by wires (representing qubits, similar to quantum circuit notation). Each spider can either be green
	(Z-spider
	\begin{ZX}
		\zxZ{}
	\end{ZX}
	) or red (X-spider
	\begin{ZX}
		\zxX{}
	\end{ZX}
	) and is optionally attributed a scalar phase.
	
	ZX-diagrams can be composed just like quantum circuits. Horizontal composition or \emph{concatenation} (denoted $\circ$) is achieved by connecting the outputs of
	one diagram to the input of another. The bare wire
	``\begin{ZX}
		\zxN{}\ar[rr] & & \zxN{}
	\end{ZX}'' acts as the identity for concatenation. Vertical composition (denoted $\otimes$) is achieved by simply ``stacking'' two diagrams on top of each
	other. The \emph{empty diagram}
	\begin{ZX}
		\zxEmptyDiagram{}  
	\end{ZX} acts as the identity for vertical composition.
	Additionally, a ZX-diagram can carry a global phase that is annotated along the diagram. Since global phases are
	negligible in most cases, they are frequently omitted from ZX-diagrams and equations in the ZX-calculus usually hold up
	to a global phase. A spider with a phase of $\pm \pi$ is called a \emph{Pauli spider}. A spider with a phase
	$\alpha \in \{k~\frac{\pi}{2} \mid k \in \mathbbm{Z}\}$ is called a \emph{Clifford spider}. A Clifford spider that is not a Pauli spider is
	called a \emph{proper Clifford spider}. A ZX-diagram consisting entirely of Clifford (Pauli) spiders is called a
	\emph{Clifford (Pauli) ZX-diagram}.
	
	Any quantum circuit can be interpreted as a ZX-diagram. The reverse of
	this statement is not true, i.e., not every ZX-diagram can be interpreted as a quantum circuit because the ZX-diagram
	does not necessarily encode a unitary transformation. Every ZX-diagram does, however, have an interpretation as
	a linear map.                \vspace*{-2mm}

        \begin{align*}
        \llbracket \begin{ZX}[ampersand replacement=\&]
          \leftManyDots{} \zxZ[a=alpha]{\alpha}\rightManyDots{}
        \end{ZX} \rrbracket &=\ket{0 \dots 0}\bra{0 \dots 0} + e^{i \alpha}\ket{1 \dots 1}\bra{1 \dots 1}\\
        \llbracket\begin{ZX}[ampersand replacement=\&]
          \leftManyDots{} \zxX[a=alpha]{\alpha}\rightManyDots{}
        \end{ZX} \rrbracket &= \ket{+ \dots +}\bra{+ \dots +} + e^{i \alpha}\ket{- \dots -}\bra{- \dots -}          
        \end{align*}

        Here $\llbracket \cdot \rrbracket$ denotes the \emph{interpretation function} which maps a ZX-diagram to its corresponding
        linear map. Any linear map on qubits can then be built up from Z- and X-spiders by connecting and stacking diagrams.

	Spiders without inputs are called \emph{states}, whereas spiders with no outputs are called \emph{effects}. Interpreted
	as linear maps, states represent column vectors whereas effects represent row vectors. A ZX-diagram
	without inputs or outputs represents a number.

	The real power of ZX-diagrams becomes evident when adding rewrite rules to the language. The axioms of the scalar-free ZX-calculus
	are given in Fig.~\ref{fig:zx_axioms}. 

      		\begin{example}
		\label{ex:cnot_swap}
		To give a feel for how to work with ZX-diagrams, we are going to prove the well-known equivalence of a SWAP with 3 CNOT operations
		(as shown in Fig.~\ref{fig:ghz_mapped}). For this, we first need to prove the following rule, which is sometimes listed explicitly among the axioms for
		the ZX-calculus but can also be derived from the axioms as follows
		\begin{equation}
			\label{eq:antipode}
			\begin{ZX}[math baseline=wantedbaseline]
				\zxN{} \rar & \zxZ[a=z]{} & \zxX[a=x]{} \rar &\zxN[a=wantedbaseline]{} \\
				\ar[from=x, to=z, o']
				\ar[from=x, to=z, o.]
			\end{ZX} \overset{(\mathbf{f})}{=}
			\begin{ZX}[math baseline=wantedbaseline]
				\zxZ{}  \rar    & \zxZ[a=zl]{} & \zxX[a=xl]{}    & \zxX{} \lar \\
				\zxN{} \rar & \zxZ[a=z]{}  & \zxX[a=x]{} \rar &\zxN[a=wantedbaseline] \\
				\ar[from=z, to=xl]
				\ar[from=x, to=zl]
				\ar[from=xl, to=zl, o.]
				\ar[from=x, to=z, o']
			\end{ZX} \overset{(\mathbf{b})}{=} 
			\begin{ZX}[math baseline=wantedbaseline]
				\zxZ{} \ar[dr] & \zxN{} & \zxN{} &\zxX{} \ar[dl] \\
				\zxN{} \rar & \zxX[a=x]{} \rar  & \zxZ[a=z]{} \rar &\zxN[a=wantedbaseline] \\
			\end{ZX} \overset{(\mathbf{c})}{=} 
			\begin{ZX}[math baseline=wantedbaseline]
				& \zxZ{} \rar & \zxX{}           &                        \\
				\zxN{} \rar & \zxX[a=x]{} & \zxZ[a=z]{} \rar & \zxN[a=wantedbaseline] \\
			\end{ZX} =
			\begin{ZX}[math baseline=wantedbaseline]
				\zxN{} \rar & \zxX[a=x]{} & \zxZ[a=z]{} \rar &\zxN[a=wantedbaseline] \\
			\end{ZX}.
		\end{equation}
		
		With this we can proceed with
		\begin{equation}
			\label{eq:zx_cnot_swap}
			\begin{ZX}[math baseline=wantedbaseline]
				\zxN \rar            &  \zxX[a=c1x]{} \rar        &  \zxZ[a=c2z]{} \dar \rar & \zxX[a=c3x]{} \ar[d]\rar &  \zxN{} \\
				\zxN{} \ar[r]        &  \zxZ[a=c1z]{} \ar[u] \rar &  \zxX[a=c2x]{}\rar       & \zxZ[a=c3z]{} \rar       &  \zxN[a=wantedbaseline]{} \\
			\end{ZX} =
			\begin{ZX}[math baseline=wantedbaseline]
				\zxN \rar            &  \zxX[a=c1x]{}        &  \zxX[a=c2x]{} & \zxX[a=c3x]{} &   \zxN{} \lar\\
				\zxN{} \ar[r]        &  \zxZ[a=c1z]{} &  \zxZ[a=c2z]{}       & \zxZ[a=c3z]{}      & \zxN[a=wantedbaseline]{} \lar \\
				\ar[from=c1x, to=c1z]
				\ar[from=c2x, to=c2z]
				\ar[from=c3x, to=c3z]
				\ar[from=c1x, to=c2z, s]
				\ar[from=c1z, to=c2x, s]
				\ar[from=c2x, to=c3z, s]
				\ar[from=c2z, to=c3x, s]
			\end{ZX}\overset{(\mathbf{b})}{=}
			\begin{ZX}[math baseline=wantedbaseline]
				\zxN \rar            &  \zxX[a=c1x]{}        &  \zxZ[a=c2z]{} &   \zxN{} \lar\\
				\zxN{} \ar[r]        &  \zxZ[a=c1z]{} &  \zxX[a=c2x]{}       &  \zxN[a=wantedbaseline]{} \lar \\
				\ar[from=c1x, to=c1z]
				\ar[from=c2x, to=c2z]
				\ar[from=c1x, to=c2x, s]
				\ar[from=c1z, to=c2z, s]
			\end{ZX}\overset{(\mathbf{f})}{=} 
			\begin{ZX}[math baseline=wantedbaseline]
				\zxN[a=leftupper]           &  \zxZ[a=c1x]{}        &   \zxN[a=rightupper]{} \lar\\
				\zxN[a=leftlower]{}         &  \zxX[a=c1z]{}        &   \zxN[a=wantedbaseline]{} \lar \\
				\ar[from=c1x, to=c1z, -o] 
				\ar[from=c1x, to=c1z, o-]
				\ar[from=leftupper, to=c1z, s] 
				\ar[from=leftlower, to=c1x, s]
			\end{ZX}\overset{\eqref{eq:antipode}}{=}
			\begin{ZX}[math baseline=rightlower]
				\zxN[a=leftupper]{} & \zxNone|[a=c1x]{} & \zxN[a=rightupper]{} \\
				\zxN[a=leftlower]{} & \zxNone|[a=c1z]{} & \zxN[a=rightlower]{} \\
				\ar[from=leftupper,to=rightlower,s]
				\ar[from=rightupper,to=leftlower,s]
			\end{ZX}.
		\end{equation}
	\end{example}

        \vspace*{-2mm}
	\subsection{Equivalence Checking Graph-like ZX-diagrams}\label{sec:zx_equiv}

	\begin{figure*}
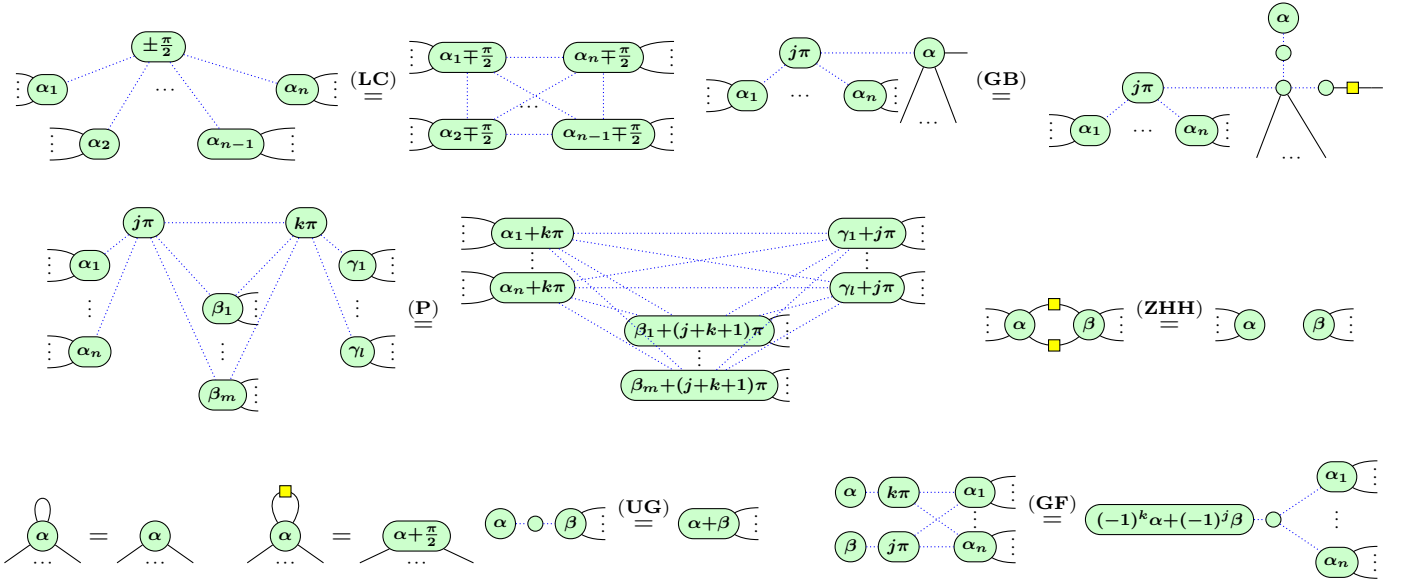

		\begin{subfigure}[c]{0.5\linewidth}
			\centering
				\begin{equation*}
				\label{eq:local_comp}
				\begin{ZX}
					\zxN{} & \zxN{}&\zxN{} & \zxZ[a=top]{\pm \frac{\pi}{2}} & \zxN{}&\zxN{} \\
					\leftManyDots{} \zxZ[a=a1]{\alpha_1} &\zxN{} \ar[rr, 3 dots]& \zxN{}& \zxN{}& \zxZ[a=an]{\alpha_n} \rightManyDots{} \\
					\\
					\zxN{} & \leftManyDots{} \zxZ[a=a2]{\alpha_2} & \zxN{}  & \zxZ[a=anm1]{\alpha_{n-1}} \rightManyDots{}
					\ar[from=top,to=a1,blue,densely dotted]
					\ar[from=top,to=a2,blue,densely dotted]
					\ar[from=top,to=anm1,blue,densely dotted]
					\ar[from=top,to=an,blue,densely dotted]
                                      \end{ZX} \overset{(\operatorname{\mathbf{LC}})}{=}   \begin{ZX}
					\leftManyDots{} \zxZ[a=a1]{\alpha_1 \mp \frac{\pi}{2}} &\zxN{} & \zxN{}& \zxN{}& \zxZ[a=an]{\alpha_n\mp
						\frac{\pi}{2}} \rightManyDots{} \\
					\\
					\\
					\zxN{} &  \zxN{} &\zxN{} \ar[rr, 3 dots]& \zxN{}& \zxN{}\\
					\leftManyDots{} \zxZ[a=a2]{\alpha_2\mp \frac{\pi}{2}} & \zxN{} & \zxN{} &  \zxN{} & \zxZ[a=anm1]{\alpha_{n-1}\mp \frac{\pi}{2}} \rightManyDots{}
					\ar[from=a1,to=a2,blue,densely dotted]
					\ar[from=a1,to=anm1,blue,densely dotted]
					\ar[from=a1,to=an,blue,densely dotted]
					\ar[from=a2,to=anm1,blue,densely dotted]
					\ar[from=a2,to=an,blue,densely dotted]
					\ar[from=anm1,to=an,blue,densely dotted]
				\end{ZX}
			\end{equation*}
		\end{subfigure}
		\begin{subfigure}[c]{0.5\linewidth}
		\begin{equation*}
			\label{eq:boundary}
			\begin{ZX}
				\zxN{}& \zxN{} & \zxZ[a=jp]{j\pi} & \zxN{} & \zxN{} & \zxN{} & \zxZ[a=a]{\alpha}\ar[ddl] \ar[ddr] \ar[rr]& \zxN{} & \zxN{}\\
				\leftManyDots{} \zxZ[a=a1]{\alpha_1} \ar[rr, 3 dots] & \zxN{} & \zxN{} & \zxZ[a=an]{\alpha_n} \rightManyDots{}\\
				\zxN{}& \zxN{} & \zxN{} & \zxN{} & \zxN{} & \zxN{} \ar[rr, 3 dots]& \zxN{} & \zxN{} & \zxN{}\\
				\ar[from=jp, to=a1, blue, densely dotted]
				\ar[from=jp, to=an, blue, densely dotted]
				\ar[from=jp, to=a, blue, densely dotted]
			\end{ZX} \overset{(\operatorname{\mathbf{GB}})}{=}
			\begin{ZX}
				\zxN{}                                     & \zxN{} & \zxN{}                                                     & \zxN{} & \zxN{} & \zxN{} & \zxN{} & \zxZ{\alpha}\ar[d,blue,densely dotted]                                               & [\zxwCol]\zxN{}            & [\zxwCol] \zxN{} & [\zxwCol] \zxN{}            & \zxN{} \\
				\zxN{}                           & \zxN{} & \zxN{}                                                     & \zxN{} & \zxN{} & \zxN{} & \zxN{} & \zxZ{}\ar[d,blue,densely dotted]                                               & [\zxwCol]\zxN{}            & [\zxwCol] \zxN{} & [\zxwCol] \zxN{}            & \zxN{} \\
				\zxN{}                                           & \zxN{} & \zxZ[a=v]{j\pi}\ar[dl, blue, densely dotted] \ar[dr, blue, densely dotted] & \zxN{} & \zxN{} & \zxN{} & \zxN{} & \zxZ[a=w]{}\ar[r, blue, densely dotted] \ar[ddr]    \ar[ddl] & [\zxwCol]\zxZ{} \ar[rr, H] & [\zxwCol] \zxN{} & [\zxwCol] \zxN{} \rar & \zxN{} \\
				\leftManyDots{} \zxZ[]{\alpha_1} \ar[rr, 3 dots] & \zxN{} & \zxZ[]{\alpha_n} \rightManyDots{}                                                                                                                                                                                                            \\
				\zxN{}& \zxN{} & \zxN{} & \zxN{} & \zxN{} & \zxN{} & \zxN{} \ar[rr, 3 dots]& \zxN{} & \zxN{} & \zxN{}\\
				\ar[from=v, to=w, blue, densely dotted]
			\end{ZX}
		\end{equation*}
\end{subfigure}

		\begin{subfigure}[c]{0.7\linewidth}
		\centering
			\begin{equation*}
			\label{eq:pivot}
			\begin{ZX}
				\zxN{} & \zxN{}                    & \zxZ[a=jp]{j \pi} & \zxN{} & \zxN{} & \zxN{} & \zxN{} & \zxN{} & \zxZ[a=kp]{k \pi} & \zxN{} \\
				\leftManyDots{} \zxZ[a=a1]{\alpha_1}\ar[dd, 3 vdots] & \zxN{} & \zxN{} & \zxN{} & \zxN{} & \zxN{} & \zxN{} & \zxN{}
				& \zxZ[a=c1]{\gamma_1} \ar[dd, 3 vdots] \rightManyDots{}\\
				\zxN{} & \zxN{} & \zxN{} & \zxN{} & \zxN{} & \zxZ[a=b1]{\beta_1} \rightManyDots{}\\
				\leftManyDots{} \zxZ[a=an]{\alpha_n} & \zxN{} & \zxN{} & \zxN{}  & \zxN{} & \zxN{} & \zxN{} & \zxN{} &
				\zxZ[a=cl]{\gamma_l}\rightManyDots{} \\
				\zxN{} & \zxN{} & \zxN{} & \zxN{} & \zxN{} &\zxZ[a=bm]{\beta_m} \rightManyDots{}\\
				\ar[from=jp, to=kp, blue, densely dotted]
				\ar[from=jp, to=a1, blue, densely dotted]
				\ar[from=jp, to=an, blue, densely dotted]
				\ar[from=jp, to=b1, blue, densely dotted]
				\ar[from=jp, to=bm, blue, densely dotted]
				\ar[from=kp, to=b1, blue, densely dotted]
				\ar[from=kp, to=bm, blue, densely dotted]
				\ar[from=kp, to=c1, blue, densely dotted]
				\ar[from=kp, to=cl, blue, densely dotted]
				\ar[from=b1, to=bm, 3 vdots]
			\end{ZX}  \overset{(\operatorname{\mathbf{P}})}{=}
			\begin{ZX}
				\leftManyDots{} \zxZ[a=a1]{\alpha_1+k\pi}\ar[dd, 3 vdots] & \zxN{} & \zxN{} & \zxN{} & \zxN{} & \zxN{} & \zxN{} & \zxN{}
				& \zxZ[a=c1]{\gamma_1+j\pi} \ar[dd, 3 vdots] \rightManyDots{}\\
				\\
				\leftManyDots{} \zxZ[a=an]{\alpha_n+k\pi} & \zxN{} & \zxN{} & \zxN{}  & \zxN{} & \zxN{} & \zxN{} & \zxN{} &
				\zxZ[a=cl]{\gamma_l+j\pi}\rightManyDots{} \\
				\zxN{} & \zxN{} & \zxN{} & \zxN{} & \zxN{} & \zxZ[a=b1]{\beta_1+(j+k+1)\pi} \rightManyDots{}\\
				\\
				\zxN{} & \zxN{} & \zxN{} & \zxN{} & \zxN{} &\zxZ[a=bm]{\beta_m+(j+k+1)\pi} \rightManyDots{}\\
				\ar[from=a1, to=c1, blue, densely dotted]
				\ar[from=a1, to=cl, blue, densely dotted]
				\ar[from=a1, to=b1, blue, densely dotted]
				\ar[from=a1, to=bm, blue, densely dotted]
				\ar[from=an, to=c1, blue, densely dotted]
				\ar[from=an, to=cl, blue, densely dotted]
				\ar[from=an, to=b1, blue, densely dotted]
				\ar[from=an, to=bm, blue, densely dotted]
				\ar[from=b1, to=c1, blue, densely dotted]
				\ar[from=b1, to=cl, blue, densely dotted]
				\ar[from=bm, to=c1, blue, densely dotted]
				\ar[from=bm, to=cl, blue, densely dotted]
				\ar[from=b1, to=bm, 3 vdots]
			\end{ZX}  
		\end{equation*}
	\end{subfigure}
		\begin{subfigure}[c]{0.3\linewidth}	
			\centering
	\begin{equation*}
		\label{eq:parallel}
		\begin{ZX}[math baseline=base]
			\leftManyDots{} \zxZ[a=base]{\alpha}\ar[rr,H,o'] \ar[rr,H,o.]  & [\zxwCol] \zxN{} & [\zxwCol] \zxZ{\beta} \rightManyDots{} \\
		\end{ZX} \overset{(\operatorname{\mathbf{ZHH}})}{=}
		\begin{ZX}[math baseline=base]
			\leftManyDots{} \zxZ[a=base]{\alpha} & [\zxwCol] \zxN{} & [\zxwCol] \zxZ{\beta} \rightManyDots{} \\
		\end{ZX}
	\end{equation*} 
 \end{subfigure}

		\begin{subfigure}[c]{0.3\linewidth}
	
	\begin{equation*}
		\begin{ZX}[math baseline=base]
			\zxN{} & \zxN{} & \zxZ[a=base]{\alpha} \zxLoop{} \ar[dll] \ar[drr] & \zxN{} & \zxN{} \\
			\zxN{} & \zxN{} \ar[rr, 3 dots]& \zxN{}                                   & \zxN{} & \zxN{} \\ 
		\end{ZX} =   \begin{ZX}[math baseline=base]
			\zxN{} & \zxN{} & \zxZ[a=base]{\alpha}  \ar[dll] \ar[drr] & \zxN{} & \zxN{} \\
			\zxN{} & \zxN{} \ar[rr, 3 dots]& \zxN{}                                   & \zxN{} & \zxN{} \\ 
		\end{ZX}
	\qquad
	\begin{ZX}[math baseline=base]
		\zxN{} & \zxN{} & \zxZ[a=base]{\alpha} \zxLoop[90][30][H]{} \ar[dll] \ar[drr] & \zxN{} & \zxN{} \\
		\zxN{} & \zxN{} \ar[rr, 3 dots]& \zxN{}                                   & \zxN{} & \zxN{} \\ 
	\end{ZX} =
	\begin{ZX}[math baseline=base]
		\zxN{} & \zxN{} & \zxZ[a=base]{\alpha+\frac{\pi}{2}} \ar[dll] \ar[drr] & \zxN{} & \zxN{} \\
		\zxN{} & \zxN{} \ar[rr, 3 dots]& \zxN{}                                   & \zxN{} & \zxN{} \\ 
	\end{ZX}
\end{equation*}
\end{subfigure}
		\begin{subfigure}[c]{.3\linewidth}
				\begin{equation*}
					\label{eq:unary_gadget}
					\begin{ZX}
						\zxZ{\alpha} \ar[r, blue, densely dotted] & \zxZ{} \ar[r, blue, densely dotted] & \zxZ{\beta} \rightManyDots{}
					\end{ZX} \overset{(\operatorname{\mathbf{UG}})}{=}
					\begin{ZX}
						\zxZ{\alpha + \beta} \rightManyDots{}
					\end{ZX}
				\end{equation*}
	\end{subfigure}
\begin{subfigure}[c]{0.4\linewidth}
	\begin{equation*}
		\label{eq:gadget_fusion}
		\begin{ZX}[math baseline=base]
			\zxZ[a=a]{\alpha} \ar[r,blue,densely dotted] & \zxZ[a=ga]{k \pi} & \zxN{} & \zxN{} & \zxZ[a=a1]{\alpha_1} \rightManyDots{}\\
			\zxN[a=base]{}\\
			\zxZ[a=b]{\beta} \ar[r,blue,densely dotted] & \zxZ[a=gb]{j \pi} & \zxN{} & \zxN{} & \zxZ[a=an]{\alpha_n} \rightManyDots{}\\
			\ar[from=ga, to=a1, blue, densely dotted]
			\ar[from=ga, to=an, blue, densely dotted]
			\ar[from=gb, to=a1, blue, densely dotted]
			\ar[from=gb, to=an, blue, densely dotted]
			\ar[from=a1, to=an, 3 vdots]
		\end{ZX} \overset{(\operatorname{\mathbf{GF}})}{=}
		\begin{ZX}[math baseline=ga]
			\zxN{}                                       & \zxN{} & \zxN{} & \zxN{} & \zxZ[a=a1]{\alpha_1} \rightManyDots{} \\
			\zxZ{(-1)^k \alpha+ (-1)^j \beta}\ar[r,blue,densely dotted] & \zxZ[a=ga]{}                                                     \\
			\zxN{}                                       & \zxN{} & \zxN{} & \zxN{} & \zxZ[a=an]{\alpha_n} \rightManyDots{} 
			\ar[from=ga, to=a1, blue, densely dotted]
			\ar[from=ga, to=an, blue, densely dotted]
			\ar[from=a1, to=an, 3 vdots]    
		\end{ZX}
	\end{equation*} 
\end{subfigure}
\caption{Rewrite system for graph-like ZX-diagrams}\label{fig:graph_like_rewriting}
\end{figure*}

	Equivalence checking with the ZX-calculus can be done in one of two ways, by either rewriting the diagram of both circuits into
	one another (as in Ex.~\ref{ex:cnot_swap}) or by
	inverting one diagram, composing the diagrams, and simplifying as much as possible. This is sometimes called an
	\emph{equivalence checking miter}. If the composed diagram simplifies to
	a diagram composed only of bare wires, it is either the identity or contains swaps, i.e., resembles a permutation.

	\begin{example}
		\label{ex:zx_equiv}
		Consider again the circuits $G$ from Fig.~\ref{fig:ghz} and $G^\prime$ from Fig.~\ref{fig:ghz_mapped}. Their
		respective ZX-diagrams are shown in Fig.~\ref{fig:ghz_zx} and Fig.~\ref{fig:ghz_mapped_zx}. Since all phases in all
		spiders are 0, the inverse of each diagram is obtained by just reversing the diagram. Using the rewrite rules of the
		ZX-calculus to prove the identity of the circuits proceeds as follows:

		\begin{ZX}[math baseline=wantedbaseline, ampersand replacement=\&]
			\zxN \rar   \&  \zxX{} \rar         \&  \zxN \rar            \&  \zxN \rar   \&  \zxN \rar   \&  \zxN \rar            \&  \zxX{} \rar        \&  \zxZ{} \dar \rar \& \zxX{} \ar[d]\rar \&  \zxN{} \rar       \& \zxN{}                 \\
			\zxN{} \rar \&  \zxN{} \rar         \&  \zxX{} \ar[d] \ar[r] \&  \zxN{} \rar \&  \zxN{} \rar \&  \zxX{} \ar[d] \ar[r] \&  \zxZ{} \ar[u] \rar \&  \zxX{}\rar       \& \zxZ{} \rar       \&  \zxX{} \rar       \& \zxN{a=wantedbaseline} \\
			\zxN{} \rar \&  \zxZ{} \ar[uu] \rar \&  \zxZ{} \rar          \&  \zxH{} \rar \&  \zxH{} \rar \&  \zxZ{} \rar          \&  \zxN{}  \rar       \&  \zxN{}\rar       \& \zxN{} \rar       \&  \zxZ{} \ar[u]\rar \& \zxN{}                 \\
		\end{ZX} $\overset{\text{\eqref{eq:zx_cnot_swap}}}{=} $
		\begin{ZX}[math baseline=wantedbaseline, ampersand replacement=\&]
			\zxN \rar   \&  \zxX{} \rar         \&  \zxN \rar            \&  \zxN \rar   \&  \zxN \rar   \&  \zxN \rar            \&  \zxN[a=s1ul]{} \&  \zxN{}      \&  \zxN[a=s1ur]{} \rar  \&  \zxN{}                   \rar\&\zxN\\
			\zxN{} \rar \&  \zxN{} \rar         \&  \zxX{} \ar[d] \ar[r] \&  \zxN{} \rar \&  \zxN{} \rar \&  \zxX{} \ar[d] \ar[r] \&  \zxN[a=s1ll]{} \&  \zxN{}      \&  \zxN[a=s1lr]{}  \rar \&  \zxX[a=wantedbaseline]{} \rar\&\zxN\\
			\zxN{} \rar \&  \zxZ{} \ar[uu] \rar \&  \zxZ{} \rar          \&  \zxH{} \rar \&  \zxH{} \rar \&  \zxZ{} \rar          \&  \zxN{} \rar    \&  \zxN{} \rar \&  \zxN{} \rar          \&  \zxZ{} \ar[u]            \rar\&\zxN\\
			\ar[from=s1ul, to=s1lr,s]
			\ar[from=s1ll, to=s1ur,s]
		\end{ZX} $=$ 
		\begin{ZX}[math baseline=s1lr, ampersand replacement=\&]
			\zxN \rar   \& \zxX{} \rar         \& \zxN \rar            \& \zxN \rar   \& \zxN \rar   \& \zxN \rar            \& \zxX[a=s1ul]{}      \& \zxN{}      \& \zxN[a=s1ur]{} \\
			\zxN{} \rar \& \zxN{} \rar         \& \zxX{} \ar[d] \ar[r] \& \zxN{} \rar \& \zxN{} \rar \& \zxX{} \ar[d] \ar[r] \& \zxN[a=s1ll]{}      \& \zxN{}      \& \zxN[a=s1lr]{} \\
			\zxN{} \rar \& \zxZ{} \ar[uu] \rar \& \zxZ{} \rar          \& \zxH{} \rar \& \zxH{} \rar \& \zxZ{} \rar          \& \zxZ{} \ar[uu] \rar \& \zxN{} \rar \& \zxN{}         \\
			\ar[from=s1ul, to=s1lr,s]
			\ar[from=s1ll, to=s1ur,s]
		\end{ZX} $\overset{(\textbf{hh})}{=}$
		\begin{ZX}[math baseline=s1lr, ampersand replacement=\&]
			\zxN \rar   \& \zxX{} \rar         \& \zxN \rar            \& \zxN \rar            \& \zxX[a=s1ul]{}      \& \zxN{}      \& \zxN[a=s1ur]{}  \\
			\zxN{} \rar \& \zxN{} \rar         \& \zxX{} \ar[d] \ar[r] \& \zxX{} \ar[d] \ar[r] \& \zxN[a=s1ll]{}      \& \zxN{}      \& \zxN[a=s1lr]{}  \\
			\zxN{} \rar \& \zxZ{} \ar[uu] \rar \& \zxZ{} \rar          \& \zxZ{} \rar          \& \zxZ{} \ar[uu] \rar \& \zxN{} \rar \& \zxN{}  \\
			\ar[from=s1ul, to=s1lr,s]
			\ar[from=s1ll, to=s1ur,s]
		\end{ZX} $\overset{(\textbf{f})}{=}$		
		\begin{ZX}[math baseline=s1lr, ampersand replacement=\&]
			\zxN \rar   \& \zxX{} \rar         \& \zxN \rar          \& \zxX[a=s1ul]{}      \& \zxN{}      \& \zxN[a=s1ur]{}  \\
			\zxN{} \rar \& \zxN{} \rar         \& \zxX[a=x]{} \ar[r] \& \zxN[a=s1ll]        \& \zxN{}      \& \zxN[a=s1lr]{} \\
			\zxN{} \rar   \& \zxZ{} \ar[uu] \rar \& \zxZ[a=z]{} \rar   \& \zxZ{} \ar[uu] \rar \& \zxN{} \rar \& \zxN{}  \\
			\ar[from=x,to=z,o-]
			\ar[from=x,to=z,-o]
			\ar[from=s1ul, to=s1lr,s]
			\ar[from=s1ll, to=s1ur,s]
		\end{ZX} $\overset{\textbf{\eqref{eq:antipode}}}{=}$ 
		\begin{ZX}[math baseline=s1lr, ampersand replacement=\&]
			\zxN \rar   \& \zxX{} \rar         \& \zxN \rar          \& \zxX{}      \rar    \& \zxN[a=s1ul]{} \& \zxN{}      \& \zxN[a=s1ur]{} \\
			\zxN{} \rar \& \zxN{} \rar         \& \zxX[a=x]{} \ar[r] \& \zxN \rar           \& \zxN[a=s1ll]{} \& \zxN{}      \& \zxN[a=s1lr]{} \\
			\zxN{} \rar \& \zxZ{} \ar[uu] \rar \& \zxZ[a=z]{} \rar   \& \zxZ{} \ar[uu] \rar \& \zxN{} \rar    \& \zxN{} \rar \& \zxN{}   \\
			\ar[from=s1ul, to=s1lr,s]
			\ar[from=s1ll, to=s1ur,s]
		\end{ZX} $\overset{(\textbf{id})}{=}$
		\begin{ZX}[math baseline=s1lr, ampersand replacement=\&]
			\zxN \rar   \& \zxX{} \rar         \& \zxX{}      \rar    \& \zxN[a=s1ul]{} \& \zxN{}      \& \zxN[a=s1ur]{} \\
			\zxN{} \rar \& \zxNone|{} \rar         \& \zxN \rar           \& \zxN[a=s1ll]{} \& \zxN{}      \& \zxN[a=s1lr]{} \\
			\zxN{} \rar \& \zxZ{} \ar[uu] \rar \& \zxZ{} \ar[uu] \rar \& \zxN{} \rar    \& \zxN{} \rar \& \zxN{}         \\
			\ar[from=s1ul, to=s1lr,s]
			\ar[from=s1ll, to=s1ur,s]
		\end{ZX} $\overset{(\textbf{f})}{=}$
		\begin{ZX}[math baseline=s1lr, ampersand replacement=\&]
			\zxN \rar   \& \zxX[a=x]{} \rar \& \zxN[a=s1ul]{} \& \zxN{}      \& \zxN[a=s1ur]{} \\
			\zxN{} \ar[rr] \& \zxNone|{}      \& \zxN[a=s1ll]{} \& \zxN{}      \& \zxN[a=s1lr]{} \\
			\zxN{} \rar \& \zxZ[a=z]{} \rar \& \zxN{} \rar    \& \zxN{} \rar \& \zxN{}         \\
			\ar[from=x,to=z,o-]
			\ar[from=x,to=z,-o]
			\ar[from=s1ul, to=s1lr,s]
			\ar[from=s1ll, to=s1ur,s]
		\end{ZX} $\overset{\textbf{\eqref{eq:antipode}}}{=}$
		\begin{ZX}[math baseline=s1lr, ampersand replacement=\&]
			\zxN \rar   \& \zxX[a=x]{} \rar \& \zxN[a=s1ul]{} \& \zxN{}      \& \zxN[a=s1ur]{} \\
			\zxN{} \ar[rr] \& \zxNone|{}  \& \zxN[a=s1ll]{} \& \zxN{}      \& \zxN[a=s1lr]{} \\
			\zxN{} \rar \& \zxZ[a=z]{} \rar \& \zxN{} \rar    \& \zxN{} \rar \& \zxN{}     \\
			\ar[from=s1ul, to=s1lr,s]
			\ar[from=s1ll, to=s1ur,s]
		\end{ZX} $\overset{(\textbf{id})}{=}$
		\begin{ZX}[math baseline=s1lr, ampersand replacement=\&]
			\zxN[a=s1ul]{} \& \zxNone|{}      \& \zxN[a=s1ur]{} \\
			\zxN[a=s1ll]{} \& \zxNone|{}      \& \zxN[a=s1lr]{} \\
			\zxN{} \ar[rr]    \& \zxNone|{}  \& \zxN{}\\
			\ar[from=s1ul, to=s1lr,s]
			\ar[from=s1ll, to=s1ur,s]
		\end{ZX}
		
		The diagram contains a SWAP which permutes qubit $Q_1$ and $Q_2$. Since this is what we expect from the output
		permutation shown in \autoref{fig:ghz_mapped} it can be concluded that the circuits are equivalent.
	\end{example}
	
	This example shows that the ZX-calculus cannot only show the equivalence of circuits but that it can also provide a
	proof certificate in the form of the order of rewrite rules that are applied to derive the identity. 

        The set of equations in \autoref{fig:zx_axioms} is not well suited for automated equivalence checking. In
        \autoref{eq:antipode} we needed to use the spider fusion rule in both directions. This is undesirable in
        automated rewriting where \emph{terminating} rewriting systems are
        preferable. In~\cite{duncanGraphtheoreticSimplificationQuantum2019}, the authors introduce an alternative
        structure for ZX-diagrams coupled with additional rewrite rules.
	
        A ZX-diagram is \emph{graph-like} when:
        \begin{enumerate}
        \item All spiders are Z-spiders.
        \item Z-spiders are only connected via Hadamard edges.
        \item There are no parallel Hadamard edges or self-loops.
        \item Every input or output is connected to a Z-spider and every Z-spider is connected to at most one input or output.
        \end{enumerate}
        In graph-like ZX-diagrams, a spider connected to an input or output is called a \emph{boundary} spider. Otherwise, it is
        called an \emph{interior} spider.

	Most importantly, every ZX-diagram is equal to a graph-like
	ZX-diagram~\cite{duncanGraphtheoreticSimplificationQuantum2019}. Every ZX-diagram can be rewritten to its equivalent
	graph-like form using the basic rules given in \autoref{fig:zx_axioms}. Instead of rigorously defining this rewriting
	procedure we give an intuition with the following example.
	
	\begin{example}
		The ZX-diagram of the GHZ state preparation circuit from \autoref{fig:ghz_zx} can easily be transformed to its
		equivalent graph-like form by applying rules $(\mathbf{id})$ and $(\mathbf{h})$.
                                \vspace*{1mm}
		
		\resizebox{0.9\linewidth}{!}{
			\begin{ZX}[ampersand replacement=\&]
				\zxN \rar   \&  \zxN \rar   \&  \zxN \rar            \&  \zxX{} \rar        \&  \zxZ{} \ar[dd] \rar \& \zxX{} \ar[dd]\rar \&
				\zxN{} \rar       \& \zxN{} \\
				\\
				\zxN{} \rar \&  \zxN{} \rar \&  \zxX{} \ar[dd] \ar[r] \&  \zxZ{} \ar[uu] \rar \&  \zxX{}\rar       \& \zxZ{} \rar       \&
				\zxX{} \rar       \& \zxN{} \\
				\\
				\zxN{} \rar \&  \zxH{} \rar \&  \zxZ{} \rar          \&  \zxN{}  \rar       \&  \zxN{}\rar       \& \zxN{} \rar       \&  \zxZ{} \ar[uu]\rar \& \zxN{} \\
			\end{ZX} =
			\begin{ZX}[ampersand replacement=\&]
				\zxN| \rar         \& \zxH{} \rar              \& [\zxwCol]\zxN{} \rar                \& [\zxwCol]\zxZ{} \ar[r, H]          \&[\zxwCol] \zxZ{}
				\ar[dd,H] \ar[r,H] \& [\zxwCol]\zxZ{} \ar[dd,H]\ar[r, H] \&[\zxwCol] \zxN{} \rar                \& [\zxwCol]\zxN{} \\
				\\
				\zxN|{} \rar       \& \zxH{} \ar[r]          \& [\zxwCol]\zxZ{} \ar[dd, H] \ar[r, H] \& [\zxwCol]\zxZ{} \ar[uu, H] \ar[r,H] \&[\zxwCol] \zxZ{}\ar[r,H]
				\& [\zxwCol]\zxZ{}    \ar[r,H]           \&[\zxwCol] \zxZ{} \ar[r,H]                \& [\zxwCol]\zxN{} \\
				\\
				\zxN|{} \rar       \& \zxH{} \rar              \& [\zxwCol]\zxZ{} \rar                \& [\zxwCol]\zxN{}  \rar              \&[\zxwCol] \zxN{}\rar
				\& [\zxwCol]\zxN{} \rar              \& [\zxwCol]\zxZ{} \ar[uu, H]\rar       \&[\zxwCol] \zxN{} \\
			\end{ZX} =
			\begin{ZX}[ampersand replacement=\&]
				\zxN| \rar         \& \zxH{} \rar              \& [\zxwCol]\zxN{} \rar                \& [\zxwCol]\zxZ{} \ar[r, H]          \&[\zxwCol] \zxZ{}
				\ar[dd,H] \ar[r,H] \& [\zxwCol]\zxZ{} \ar[dd,H]\ar[r, H] \&[\zxwCol] \zxN{} \rar                \& [\zxwCol]\zxN{} \\
				\\
				\zxN|{} \rar       \& \zxH{} \ar[r]          \& [\zxwCol]\zxZ{} \ar[dd, H] \ar[r, H] \& [\zxwCol]\zxZ{} \ar[uu, H] \ar[r,H] \&[\zxwCol] \zxZ{}\ar[r,H]
				\& [\zxwCol]\zxZ{}    \ar[r,H]           \&[\zxwCol] \zxZ{} \ar[r,H]                \& [\zxwCol]\zxN{} \\
				\\
				\zxN|{} \rar       \& \zxH{} \rar              \& [\zxwCol]\zxZ{} \rar                \& [\zxwCol]\zxH{}  \rar              \&[\zxwCol] \zxZ{}\rar
				\& [\zxwCol]\zxH{} \rar              \& [\zxwCol]\zxZ{} \ar[uu, H]\rar       \&[\zxwCol] \zxN{} \\
			\end{ZX}=
			\begin{ZX}[ampersand replacement=\&]
				\zxN| \ar[r, blue, densely dotted]         \& \zxN{} \ar[r, blue, densely dotted]              \& [\zxwCol]\zxN{} \ar[r, blue, densely dotted]                \& [\zxwCol]\zxZ{} \ar[r, blue, densely dotted]          \&[\zxwCol] \zxZ{}
				\ar[dd,blue,densely dotted] \ar[r,blue,densely dotted] \& [\zxwCol]\zxZ{} \ar[dd,blue,densely dotted]\ar[r, blue, densely dotted] \&[\zxwCol] \zxN{} \ar[r, blue, densely dotted]                \& \zxN{} \\
				\\
				\zxN|{} \ar[r, blue, densely dotted]       \& \zxN{} \ar[r,blue, densely dotted]          \& [\zxwCol]\zxZ{} \ar[dd, blue, densely dotted] \ar[r,
				blue, densely dotted] \& [\zxwCol]\zxZ{} \ar[uu, blue, densely dotted] \ar[r, blue, densely dotted] \&[\zxwCol] \zxZ{}\ar[r, blue, densely dotted]
				\& [\zxwCol]\zxZ{}    \ar[r, blue, densely dotted]           \&[\zxwCol] \zxZ{} \ar[r, blue, densely dotted]                \& \zxN{} \\
				\\
				\zxN|{} \ar[r, blue, densely dotted]       \& \zxN{} \ar[r, blue, densely dotted]              \& [\zxwCol]\zxZ{} \ar[r, blue, densely dotted]                \&
				[\zxwCol]\zxN{}  \ar[r, blue, densely dotted]              \&[\zxwCol] \zxZ{}\ar[r, blue, densely dotted]
				\& [\zxwCol]\zxN{} \ar[r, blue, densely dotted]              \& [\zxwCol]\zxZ{} \ar[uu, blue, densely dotted]\rar       \& \zxN{} \\
		\end{ZX}}
	\end{example}
	Graph-like diagrams allow for the formulation of a normalizing rewrite system. The rules of this system are given in \autoref{fig:graph_like_rewriting}. Adding rules $(\mathbf{f})$, $(\mathbf{id})$ and $(\mathbf{hh})$ (applied from left to right) to this system allows for the definition of a simplification algorithm that reduces every ZX-diagram into a \emph{reduced gadget form}~\cite{kissingerReducingTcountZXcalculus2020}.

          This automated rewriting of diagrams into reduced gadget form allows for the definition of an equivalence checking algorithm. Given two quantum circuits $G$ and $G^\prime$ we can check them for equivalence by taking their respective representations as ZX-diagrams $D$ and $D^\prime$, combining them to $D^\dagger D^\prime$ and simplifying the combined diagram to reduced gadget form. If the reduced gadget form is the identity diagram---the ZX-diagram consisting only of bare wires---then $G$ and $G^\prime$ are equivalent. Otherwise, nothing can be concluded about the relation of $G$ and $G^\dagger$ because there are generally multiple reduced gadget forms for a ZX-diagram.

                          \vspace*{-1mm}
\section{Equivalence Checking\\Compilation Flows Using the ZX-Calculus}\label{sec:zx_comp}

The original ZX-calculus equivalence checking algorithm proposed in \cite{vandeweteringZXcalculusWorkingQuantum2020} has been introduced as a byproduct of the optimization algorithm proposed in that work.
It has, therefore, not been expanded to handle the more technical aspects of quantum circuit equivalence checking necessary to check the results of compilation flows.
In the following, we will remedy this by showing how inaccuracies, permutations, and ancillae can be handled in ZX-calculus equivalence checking.

                \vspace*{-2mm}
        \subsection{Handling Inaccuracies}\label{sec:zx_inaccuracies}
	
	The equivalence checking routine based on the ZX-calculus is an exact
	method. Hence, when considering two quantum circuits $G$ and $G^\prime$,
	where $G^\prime$ is equivalent to $G$ up to some small error,
	the ZX-calculus is unable to conclude equivalence. But can anything be concluded about the reduced gadget form
	of $D^\dagger D^\prime$ where $\llbracket D \rrbracket \approx \llbracket D^\prime \rrbracket$?
	
\begin{figure}[t]
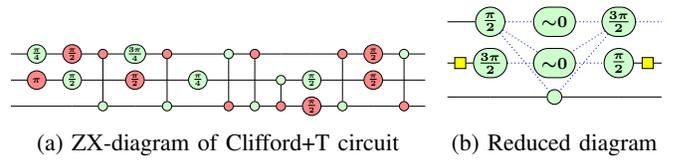

		\begin{subfigure}[b]{0.64\linewidth}
			\centering
                        \resizebox{\linewidth}{!}{
			\begin{ZX}[ampersand replacement=\&]
				\zxN{}\rar \& [\zxWCol] \zxFracZ{\pi}{4}\rar \& [\zxWCol] \zxFracX{\pi}{2}\rar \& [\zxWCol] \zxX{}\ar[dd]\rar \& [\zxWCol] \zxFracZ{3\pi}{4}\rar \& [\zxWCol] \zxX{}\rar \ar[dd] \& [\zxWCol] \zxN{}\rar           \& [\zxWCol] \zxZ{}\ar[dd]\rar \& [\zxWCol] \zxX{}\ar[dd]\rar \& [\zxWCol] \zxN{}\rar       \& [\zxWCol] \zxN{}\rar           \& [\zxWCol] \zxX{}\ar[dd]\rar \& [\zxWCol] \zxFracX{\pi}{2}\rar \& [\zxWCol] \zxZ{}\ar[dd]\rar \& [\zxWCol] \zxN{} \\
				\zxN{}\rar \& [\zxWCol] \zxX{\pi}    \rar        \& [\zxWCol] \zxFracZ{\pi}{2}\rar \& [\zxWCol] \zxN{}\rar        \& [\zxWCol] \zxFracX{\pi}{2}\rar  \& [\zxWCol] \zxN{}\rar         \& [\zxWCol] \zxFracZ{\pi}{4}\rar \& [\zxWCol] \zxN{}\rar        \& [\zxWCol] \zxN{}\rar        \& [\zxWCol] \zxZ{}\ar[d]\rar \& [\zxWCol] \zxFracZ{\pi}{2}\rar \& [\zxWCol] \zxN{}\rar        \& [\zxWCol] \zxFracX{\pi}{2}\rar \& [\zxWCol] \zxN{}\rar        \& [\zxWCol] \zxN{} \\
				\zxN{}\rar \& [\zxWCol] \zxN{}\rar           \& [\zxWCol] \zxN{}\rar           \& [\zxWCol] \zxZ{}    \rar    \& [\zxWCol] \zxN{}\rar            \& [\zxWCol] \zxZ{} \rar        \& [\zxWCol] \zxN{}\rar           \& [\zxWCol] \zxX{}  \rar      \& [\zxWCol] \zxZ{}\rar        \& [\zxWCol] \zxX{}\rar       \& [\zxWCol] \zxFracX{\pi}{2}\rar \& [\zxWCol] \zxZ{}\rar        \& [\zxWCol] \zxN{}\rar           \& [\zxWCol] \zxX{}\rar            \& [\zxWCol] \zxN{}   \end{ZX}
                            }
			\caption{ZX-diagram of Clifford+T circuit}
			\label{fig:zx_clifford_t}    
		\end{subfigure}
		\begin{subfigure}[b]{0.35\linewidth}
                  \centering
                  \resizebox{\linewidth}{!}{
			\begin{ZX}[ampersand replacement=\&]
				\zxN{}\rar     \& [\zxWCol] \zxFracZ{\pi}{2}  \ar[r,blue,densely dotted]\ar[dr,blue,densely dotted]\ar[ddr,blue,densely dotted] \& [\zxWCol] \zxZ{\sim 0}\ar[r,blue,densely dotted] \& [\zxWCol] \zxFracZ{3\pi}{2}\ar[r,blue,densely dotted]\ar[dl,blue,densely dotted]\ar[ddl,blue,densely dotted] \& [\zxWCol] \zxN{} \\
				\zxN{}\ar[r,H] \& [\zxWCol] \zxFracZ{3\pi}{2} \ar[r,blue,densely dotted]                                                        \& [\zxWCol] \zxZ{\sim 0}\ar[r,blue,densely dotted] \& [\zxWCol] \zxFracZ{\pi}{2}\ar[r,H]                                                                           \& [\zxWCol] \zxN{} \\
				\zxN{}\rar     \& [\zxWCol] \zxN{}\ar[r]                                                                                        \& [\zxWCol] \zxZ{}  \ar[r]                         \& [\zxWCol] \zxN{}\ar[r]                                                                                     \& [\zxWCol] \zxN{} \\
                              \end{ZX}
                            }
                            \caption{Reduced diagram}\label{fig:zx_clifford_t_err}    
		\end{subfigure}
		\caption{Equivalence checking with the ZX-calculus in the presence of few small errors}\label{fig:zx_err}\vspace*{-5mm}
	\end{figure}

	To give an intuition, consider the Clifford+T circuit given in \autoref{fig:zx_clifford_t}. Introducing an error of
	$10^{-15}$ in the phase of two spiders and checking the equivalence of the original and the erroneous circuit yields the
	ZX-diagram shown in \autoref{fig:zx_clifford_t_err}. The phases indicated with $\sim \alpha$ means that the phase is
	$\alpha \pm 10^{-15}$. It is not at all obvious that this diagram is close to the
	identity. However, if we were to round the phases and simplified
	further, we would indeed be able to derive the identity. 
	
	One strategy is to interleave simplification to normal form with detection and rounding of phases close to $k \frac{\pi}{2}$ for some $k \in \mathbb{Z}$. This allows equivalence checking of quantum circuits that differ by small numerical inaccuracies in some continuous parameters. The corresponding algorithm is obviously not correct
	in a formal sense, i.e., it can attest two non-equivalent circuits to be equivalent, but that is the whole point. The threshold for rounding $\epsilon$ can be used to tune the degree to which errors are allowed. However, it does not give any
	indication about the absolute error.
	
	To clarify this point, consider a ZX-diagram $M$ after simplification and the corresponding
	ZX-diagram $M^\prime$ obtained after rounding and simplifying again. The tolerance $\epsilon$ can not be used to assess
	$\operatorname{tr}(\llbracket D^\dagger \rrbracket \llbracket D^\prime \rrbracket)$---the Hilbert-Schmidt inner
	product discussed in \autoref{cha:equ_checking}. 
	
	A question one might ask is why the rounding doesn't already occur on the diagrams $D$ and
	$D^\prime$. The reason is that even phases that are not nice fractions of $\pi$ (or very small fractions) might cancel during simplification due
	to the rules $\operatorname{\mathbf{UG}}$ and $\operatorname{\mathbf{GF}}$. Thus rounding before simplifying would increase
	the total error made during the equivalence check. 
		
	This way of handling inaccuracies is still lacking. As discussed above it is hard to gauge the tolerance required in
	order to ensure the absolute error allowed is within some bound. Given the ZX-diagram $M$ after full simplification, how can we determine whether $\vert\operatorname{tr}(\llbracket M \rrbracket)\vert \approx 2^n$?
	The diagrammatic trace is defined as follows:
        		\begin{equation*}
			\operatorname{tr} \Bigg({
				\tikzset{
					my box/.style={inner sep=4pt, draw, thick, fill=white,anchor=center}
				}
				\begin{ZX}[
					execute at end picture={
						\node[
						my box,
						node on layer=box,
						fit=(f1)(f2), 
						label={[node on layer=box]center:D},
						] {};
					}
					]
					\zxN{} \ar[dd, 3 vdots]\rar & [\zxWCol] \zxNoneDouble+[alias=f1]{} \rar & [\zxWCol] \zxN{} \ar[dd, 3 vdots] \\
					\zxN{}                         & [\zxWCol] \zxNoneDouble+{}                & [\zxWCol] \zxN                        \\
					\zxN{}  \rar                   & [\zxWCol]\zxNoneDouble+[alias=f2]{} \rar  & [\zxWCol] \zxN{}                      \\
				\end{ZX}
			}\Bigg) = 
			{
				\tikzset{
					my box/.style={inner sep=4pt, draw, thick, fill=white,anchor=center}
				}
				\begin{ZX}[
					execute at end picture={
						\node[
						my box,
						node on layer=box,
						fit=(f1)(f2), 
						label={[node on layer=box]center:D},
						] {};
					}
					]
					\zxN{}     \ar[IO,C,ddddddd,wc] \rar                  & [\zxWCol] \zxN{}         \rar                 & [\zxWCol] \zxN{} \ar[IO,C-,ddddddd,wc]            \\
					\zxN{}                    \ar[rr, 3 vdots]  & [\zxWCol] \zxN{}                          & [\zxWCol] \zxN{}                  \\
					\zxN{}     \ar[IO,C,dd,wc] \rar                  & [\zxWCol] \zxN{}         \rar                 & [\zxWCol] \zxN{}
					\ar[IO,C-,dd,wc]            \\
					\\
					\zxN{} \ar[dd, 3 vdots]\rar & [\zxWCol] \zxNoneDouble+[alias=f1]{} \rar & [\zxWCol] \zxN{} \ar[dd, 3 vdots] \\
					\zxN{}                      & [\zxWCol] \zxN{}                          & [\zxWCol] \zxN{}                  \\
					\zxN{}                     & [\zxWCol] \zxNoneDouble+{}                & [\zxWCol] \zxN                    \\
					\zxN{}  \rar               & [\zxWCol]\zxNoneDouble+[alias=f2]{} \rar  & [\zxWCol] \zxN{}                  \\
				\end{ZX}
			} 
		\end{equation*}
        Unfortunately, this definition is hardly
	helpful if we want to actually compute the trace. In order to compute the trace, further simplifications have to be made
	after the inputs and outputs have been connected. This doesn't necessarily enable the ZX-diagram to be simplified to a
	point where calculations are practical. A possible solution to this problem is to leave the ZX-calculus framework
	entirely. ZX-diagrams are, in essence, tensor
	networks~\cite{coeckePicturingQuantumProcesses2018,biamonteTensorNetworksNutshell2017}. Therefore methods from the
	tensor network domain can be used to compute the trace of a ZX-diagram.

        \vspace*{-2mm}
	\subsection{Handling Permutations}\label{sec:zx_perm}
	
	Handling SWAPs in ZX-diagrams is a trivial matter. Since SWAPs are nothing but edges connecting
	spiders acting on different qubits, they do not add much complexity to a ZX-diagram.
	To correct permutations of the initial layout, it has to be ensured that wires are connected accordingly when
	constructing $D^\dagger \circ D^\prime$. But since SWAPs incur such little overhead in ZX-diagrams, the
	initial layout can also just be encoded into the original diagrams themselves before performing the equivalence
	check. The wires can then be connected in the usual fashion, i.e.\ by connecting the $i$-th output wire of $D^\dagger$
	with the $i$-th input wire of $D^\prime$.
	
	Output permutations can be handled in a similar fashion as with decision diagrams, by comparing the permutation of wires
	after fully simplifying $D^\dagger \circ D^\prime$ with the expected permutation. Once again the permutation can also
	just be handled by encoding the SWAPs directly into the diagrams.
	
	As discussed in \autoref{sec:quant-circ-comp}, permutations of input and output layouts are performed in order to save
	CNOT gates that need to be executed on the quantum hardware. When converting a compiled circuit to a ZX-diagram,
	\autoref{eq:zx_cnot_swap} can be used to reconstruct compiled SWAP gates if they are not optimized away. Since a SWAP in
	the ZX-calculus is only a crossing of the wires, this reconstruction can greatly improve the performance of the
	equivalence check, decreasing runtime by up to two orders of magnitude.

        \begin{example}\label{ex:swap_zx}
          The output permutation of the qubits in \autoref{fig:ghz_mapped_zx} can be directly encoded back into the circuit via a SWAP at the end of the circuit.
          Additionally, the $3$ CNOTs can be converted back into a SWAP, yielding the following ZX-diagram:

          \begin{center}           
          \begin{ZX}
            \zxN \rar   & \zxN \rar   & \zxN \rar            & \zxN[a=leftupper1]{} & \zxNone|[a=c1x]{} & \zxN[a=rightupper1]{} \rar & \zxN{} \rar       & \zxN{} \rar & \zxN[a=leftupper]{} & \zxNone|[a=c1x]{} & \zxN[a=rightupper]{} \\
            \zxN{} \rar & \zxN{} \rar & \zxX{} \ar[d] \ar[r] & \zxN[a=leftlower1]{} & \zxNone|[a=c1z]{} & \zxN[a=rightlower1]{} \rar & \zxX{} \rar       & \zxN{} \rar & \zxN[a=leftlower]{} & \zxNone|[a=c1z]{} & \zxN[a=rightlower]{} \\
            \zxN{} \rar & \zxH{} \rar & \zxZ{} \rar          & \zxN{}  \rar         & \zxN{}\rar        & \zxN{} \rar           & \zxZ{} \ar[u]\rar & \zxN{} \rar & \zxN{} \rar         & \zxN{} \rar       & \zxN{} 
            \ar[from=leftupper,to=rightlower,s]
            \ar[from=rightupper,to=leftlower,s]
            \ar[from=leftupper1,to=rightlower1,s]
            \ar[from=rightupper1,to=leftlower1,s]
          \end{ZX}
        \end{center}
        
          This ZX-diagram is equivalent to the one in \autoref{fig:ghz_zx} up to an untangling of the wires.
        \end{example}

                        \vspace*{-3mm}
	\subsection{Handling Ancillaries}\label{sec:zx_ancilla}
	
	In the ZX-calculus the equivalence checking problem using ancillaries is reducible to the ancilla-free case in a
	straightforward fashion. Remember that ancillaries are qubits that have a constant initial state and end in the same
	state. This is easily translated into the diagrammatic language of the
	ZX-calculus, by replacing each input and output belonging to an ancilla qubit by the respective state and effect, which
	are just $X$-spiders with a phase of either $0$ or $\pi$.
	
	\begin{example}\label{ex:zx_ancilla}
          In \autoref{ex:ancilla} it was shown how fixing the control qubit of a CNOT gate to the $\ket{1}$ state, transforms it into a Pauli $X$ gate. 
          This can also be shown in the ZX-calculus with only
		basic applications of the axioms. We start by applying the $\ket{1}$ state and effect to the
		ancillary line and proceed to simplify.
                                \vspace*{-3mm}

		\begin{equation*}        
			\begin{ZX}[math baseline=base, ampersand replacement=\&]
				\zxN{} \rar \& [\zxWCol] \zxZ{}\ar[dd] \rar{} \& [\zxWCol] \zxN[a=base]{}\\
				\zxN{}\\
				\zxN{} \rar \& [\zxWCol] \zxX{} \rar{} \& [\zxWCol] \zxN{}\\
			\end{ZX} \rightarrow   \begin{ZX}[math baseline=base, ampersand replacement=\&]
				\zxX{\pi} \rar \& [\zxWCol] \zxZ{}\ar[dd] \rar{} \& [\zxWCol] \zxX[a=base]{\pi}\\
				\zxN{}\\
				\zxN{} \rar \& [\zxWCol] \zxX{} \rar{} \& [\zxWCol] \zxN{}\\
			\end{ZX} \overset{(\pi)}{=} \begin{ZX}[math baseline=base, ampersand replacement=\&]
				\zxN{} \& [\zxWCol] \zxN{}  \& [\zxWCol] \zxX{\pi}\rar \& [\zxWCol]\zxX{\pi}\\
				\zxN[a=base]{} \&[\zxWCol] \zxX{\pi}\dar\\
				\zxN{} \rar \& [\zxWCol] \zxX{} \rar{} \& [\zxWCol] \zxN{}\\
			\end{ZX} \overset{(\textbf{f})}{=}
			\begin{ZX}[math baseline=base, ampersand replacement=\&]
                          \zxN{}\& [\zxWCol] \zxX{\pi}\rar \& [\zxWCol]\zxX[a=base]{\pi} \\
                          \zxN{} \rar \& [\zxWCol] \zxX{\pi} \rar{} \& [\zxWCol] \zxN{}           \\
			\end{ZX}
                      \end{equation*}
                                      \vspace*{-1mm}
		
                      Since we ignore scalars, the right-hand side indeed implements a Pauli $X$ gate (
		\begin{ZX}
			\zxN{}\rar & \zxX{\pi} \rar & \zxN{}
		\end{ZX}
		).
	\end{example}
	
                \vspace*{-1mm}
	\section{Completeness}\label{sec:zx_completeness}
	
	A natural question
	to ask is whether the ZX-calculus is powerful enough to derive the identity for any pair of functionally equivalent
	circuits. The good news is that the ruleset provided in this paper is complete for circuits solely composed of Clifford
	gates~\cite{backensZXcalculusCompleteStabilizer2013}. The bad news is that, in order to achieve completeness for
	universal quantum computing, the ruleset has to be extended with a rule involving complicated iterated
	trigonometric functions~\cite{vilmartNearoptimalAxiomatisationZXcalculus2018}, which makes it difficult to apply in
	automated reasoning.

	\begin{figure}[t]
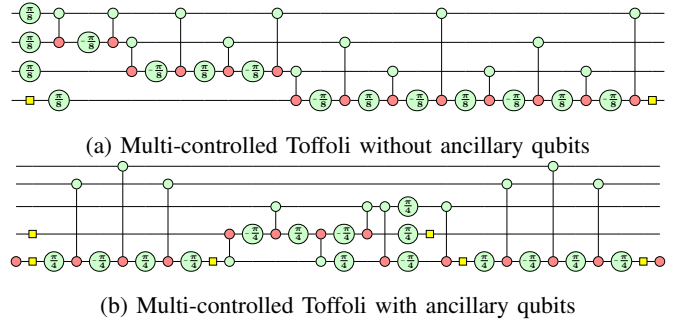

		\centering
		\begin{subfigure}[b]{\linewidth}
			\centering
			\resizebox{\linewidth}{!}{
			\begin{ZX}[ampersand replacement=\&]
				\zxN{}\rar \& \zxFracZ{\pi}{8}\rar \& \zxZ{}\rar\dar       \& \zxN{}\rar             \& \zxZ{}\rar\dar \& \zxN{}\rar     \& \zxN{}\rar             \& \zxZ{}\rar\ar[dd] \& \zxN{}\rar           \& \zxN{}\rar     \& \zxN{}\rar             \& \zxZ{}\rar\ar[dd] \& \zxN{}\rar     \& \zxN{}\rar             \& \zxN{}\rar        \& \zxN{}\rar           \& \zxN{}\rar     \& \zxN{}\rar             \& \zxZ{}\rar\ar[ddd] \& \zxN{}\rar           \& \zxN{}\rar     \& \zxN{}\rar             \& \zxN{}\rar        \& \zxN{}\rar           \& \zxN{}\rar     \& \zxN{}\rar             \& \zxZ{}\rar\ar[ddd] \& \zxN{}\rar \& \zxN{} \\ 
				\zxN{}\rar \& \zxFracZ{\pi}{8}\rar \& \zxX{}\rar           \& \zxFracZ-{\pi}{8}\rar \& \zxX{}\rar     \& \zxZ{}\rar\dar \& \zxN{}\rar             \& \zxN{}\rar        \& \zxN{}\rar           \& \zxZ{}\rar\dar \& \zxN{}\rar             \& \zxN{}\rar        \& \zxN{}\rar     \& \zxN{}\rar             \& \zxZ{}\rar\ar[dd] \& \zxN{}\rar           \& \zxN{}\rar     \& \zxN{}\rar             \& \zxN{}\rar         \& \zxN{}\rar           \& \zxN{}\rar     \& \zxN{}\rar             \& \zxZ{}\rar\ar[dd] \& \zxN{}\rar           \& \zxN{}\rar     \& \zxN{}\rar             \& \zxN{}\rar         \& \zxN{}\rar \& \zxN{} \\
				\zxN{}\rar \& \zxFracZ{\pi}{8}\rar \& \zxN{}\rar           \& \zxN{}\rar             \& \zxN{}\rar     \& \zxX{}\rar     \& \zxFracZ-{\pi}{8}\rar \& \zxX{}\rar        \& \zxFracZ{\pi}{8}\rar \& \zxX{}\rar     \& \zxFracZ-{\pi}{8}\rar \& \zxX{}\rar        \& \zxZ{}\rar\dar \& \zxN{}\rar             \& \zxN{}\rar        \& \zxN{}\rar           \& \zxZ{}\rar\dar \& \zxN{}\rar             \& \zxN{}\rar         \& \zxN{}\rar           \& \zxZ{}\rar\dar \& \zxN{}\rar             \& \zxN{}\rar        \& \zxN{}\rar           \& \zxZ{}\rar\dar \& \zxN{}\rar             \& \zxN{}\rar         \& \zxN{}\rar \& \zxN{} \\
				\zxN{}\rar \& \zxH{}\rar           \& \zxFracZ{\pi}{8}\rar \& \zxN{}\rar             \& \zxN{}\rar     \& \zxN{}\rar     \& \zxN{}\rar             \& \zxN{}\rar        \& \zxN{}\rar           \& \zxN{}\rar     \& \zxN{}\rar             \& \zxN{}\rar        \& \zxX{}\rar     \& \zxFracZ-{\pi}{8}\rar \& \zxX{}\rar        \& \zxFracZ{\pi}{8}\rar \& \zxX{}\rar     \& \zxFracZ-{\pi}{8}\rar \& \zxX{}\rar         \& \zxFracZ{\pi}{8}\rar \& \zxX{}\rar     \& \zxFracZ-{\pi}{8}\rar \& \zxX{}\rar        \& \zxFracZ{\pi}{8}\rar \& \zxX{}\rar     \& \zxFracZ-{\pi}{8}\rar \& \zxX{}\rar         \& \zxH{}\rar \& \zxN{} \\
			\end{ZX}
		}
			\caption{Multi-controlled Toffoli without ancillary qubits}\label{fig:zx_mcx_no_ancilla}
		\end{subfigure}
		
		\begin{subfigure}[b]{\linewidth}
			\centering
			\resizebox{\linewidth}{!}{
			\begin{ZX}[ampersand replacement=\&]
				\zxN{}\rar \& \zxN{}\rar \& \zxN{}\rar           \& \zxN{}\rar         \& \zxN{}\rar            \& \zxZ{}\rar\ar[dddd] \& \zxN{}\rar           \& \zxN{}\rar         \& \zxN{}\rar            \& \zxN{}\rar \& \zxN{}\rar       \& \zxN{}\rar            \& \zxN{}\rar     \& \zxN{}\rar           \& \zxN{}\rar     \& \zxN{}\rar            \& \zxN{}\rar     \& \zxN{}\rar         \& \zxN{}\rar            \& \zxN{}\rar \& \zxN{}\rar        \& \zxN{}\rar \& \zxN{}\rar           \& \zxN{}\rar         \& \zxN{}\rar            \& \zxZ{}\rar\ar[dddd] \& \zxN{}\rar           \& \zxN{}\rar         \& \zxN{}\rar            \& \zxN{}\rar \& \zxN{}\\
				\zxN{}\rar \& \zxN{}\rar \& \zxN{}\rar           \& \zxZ{}\rar\ar[ddd] \& \zxN{}\rar            \& \zxN{}\rar          \& \zxN{}\rar           \& \zxZ{}\rar\ar[ddd] \& \zxN{}\rar            \& \zxN{}\rar \& \zxN{}\rar       \& \zxN{}\rar            \& \zxN{}\rar     \& \zxN{}\rar           \& \zxN{}\rar     \& \zxN{}\rar            \& \zxN{}\rar     \& \zxN{}\rar         \& \zxN{}\rar            \& \zxN{}\rar \& \zxN{}\rar        \& \zxN{}\rar \& \zxN{}\rar           \& \zxZ{}\rar\ar[ddd] \& \zxN{}\rar            \& \zxN{}\rar          \& \zxN{}\rar           \& \zxZ{}\rar\ar[ddd] \& \zxN{}\rar            \& \zxN{}\rar \& \zxN{}\\
				\zxN{}\rar \& \zxN{}\rar \& \zxN{}\rar           \& \zxN{}\rar         \& \zxN{}\rar            \& \zxN{}\rar          \& \zxN{}\rar           \& \zxN{}\rar         \& \zxN{}\rar            \& \zxN{}\rar \& \zxN{}\rar       \& \zxN{}\rar            \& \zxZ{}\rar\dar \& \zxN{}\rar           \& \zxN{}\rar     \& \zxN{}\rar            \& \zxZ{}\rar\dar \& \zxZ{}\rar\dar[dd] \& \zxFracZ{\pi}{4}\rar  \& \zxN{}\rar \& \zxZ{}\rar\ar[dd] \& \zxN{}\rar \& \zxN{}\rar           \& \zxN{}\rar         \& \zxN{}\rar            \& \zxN{}\rar          \& \zxN{}\rar           \& \zxN{}\rar         \& \zxN{}\rar            \& \zxN{}\rar \& \zxN{}\\
				\zxN{}\rar \& \zxH{}\rar \& \zxN{}\rar           \& \zxN{}\rar         \& \zxN{}\rar            \& \zxN{}\rar          \& \zxN{}\rar           \& \zxN{}\rar         \& \zxN{}\rar            \& \zxN{}\rar \& \zxX{}\rar\ar[d] \& \zxFracZ-{\pi}{4}\rar \& \zxX{}\rar     \& \zxFracZ{\pi}{4}\rar \& \zxX{}\rar\dar \& \zxFracZ-{\pi}{4}\rar \& \zxX{}\rar     \& \zxN{}\rar         \& \zxFracZ{\pi}{4}\rar  \& \zxH{}\rar \& \zxN{}\rar        \& \zxN{}\rar \& \zxN{}\rar           \& \zxN{}\rar         \& \zxN{}\rar            \& \zxN{}\rar          \& \zxN{}\rar           \& \zxN{}\rar         \& \zxN{}\rar            \& \zxN{}\rar \& \zxN{}\\
				\zxX{}\rar \& \zxH{}\rar \& \zxFracZ{\pi}{4}\rar \& \zxX{}\rar         \& \zxFracZ-{\pi}{4}\rar \& \zxX{}\rar          \& \zxFracZ{\pi}{4}\rar \& \zxX{}\rar         \& \zxFracZ-{\pi}{4}\rar \& \zxH{}\rar \& \zxZ{}\rar       \& \zxN{}\rar            \& \zxN{}\rar     \& \zxN{}\rar           \& \zxZ{}\rar     \& \zxFracZ{\pi}{4}\rar   \& \zxN{}\rar     \& \zxX{}\rar         \& \zxFracZ-{\pi}{4}\rar \& \zxN{}\rar \& \zxX{}\rar        \& \zxH{}\rar \& \zxFracZ{\pi}{4}\rar \& \zxX{}\rar         \& \zxFracZ-{\pi}{4}\rar \& \zxX{}\rar          \& \zxFracZ{\pi}{4}\rar \& \zxX{}\rar         \& \zxFracZ-{\pi}{4}\rar \& \zxH{}\rar \& \zxX{}\\
			\end{ZX}
		}
			\caption{Multi-controlled Toffoli with ancillary qubits}\label{fig:zx_mcx_ancilla}
		\end{subfigure}
		\caption{ZX-diagrams of the multi-controlled Toffoli gate}\label{fig:zx_mcx}
	\end{figure}

	\begin{figure}[t]
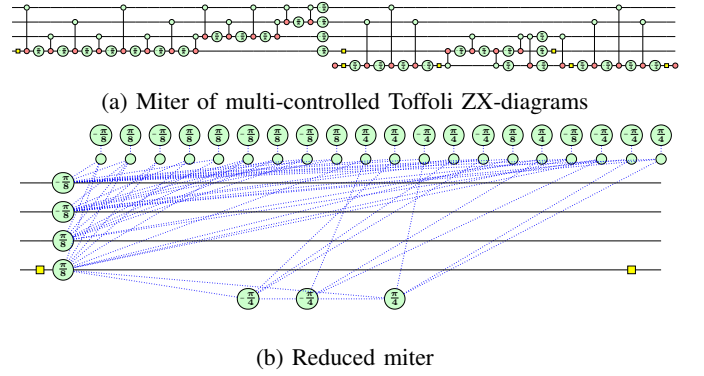

		\centering
		\begin{subfigure}[b]{\linewidth}    
			\resizebox{\linewidth}{!}{%
				\centering
				\begin{ZX}[ampersand replacement=\&]
					\zxN{}\rar \& \zxN{}\rar \& \zxZ{}\rar\ar[ddd] \& \zxN{}\rar           \& \zxN{}\rar     \& \zxN{}\rar            \& \zxN{}\rar        \& \zxN{}\rar           \& \zxN{}\rar     \& \zxN{}\rar            \& \zxZ{}\rar\ar[ddd] \& \zxN{}\rar           \& \zxN{}\rar     \& \zxN{}\rar            \& \zxN{}\rar        \& \zxN{}\rar           \& \zxN{}\rar     \& \zxZ{}\rar\ar[dd] \& \zxN{}\rar           \& \zxN{}\rar     \& \zxN{}\rar            \& \zxZ{}\rar\ar[dd] \& \zxN{}\rar           \& \zxN{}\rar     \& \zxZ{}\rar\dar \& \zxN{}\rar           \& \zxZ{}\rar\dar        \& \zxFracZ-{\pi}{8}\rar \&       \zxN{}\rar \& \zxN{}\rar \& \zxN{}\rar           \& \zxN{}\rar         \& \zxN{}\rar            \& \zxZ{}\rar\ar[dddd] \& \zxN{}\rar           \& \zxN{}\rar         \& \zxN{}\rar            \& \zxN{}\rar \& \zxN{}\rar       \& \zxN{}\rar            \& \zxN{}\rar     \& \zxN{}\rar           \& \zxN{}\rar     \& \zxN{}\rar            \& \zxN{}\rar     \& \zxN{}\rar         \& \zxN{}\rar            \& \zxN{}\rar \& \zxN{}\rar        \& \zxN{}\rar \& \zxN{}\rar           \& \zxN{}\rar         \& \zxN{}\rar            \& \zxZ{}\rar\ar[dddd] \& \zxN{}\rar           \& \zxN{}\rar         \& \zxN{}\rar            \& \zxN{}\rar \& \zxN{}  \\ 
					\zxN{}\rar \& \zxN{}\rar \& \zxN{}\rar         \& \zxN{}\rar           \& \zxN{}\rar     \& \zxN{}\rar            \& \zxZ{}\rar\ar[dd] \& \zxN{}\rar           \& \zxN{}\rar     \& \zxN{}\rar            \& \zxN{}\rar         \& \zxN{}\rar           \& \zxN{}\rar     \& \zxN{}\rar            \& \zxZ{}\rar\ar[dd] \& \zxN{}\rar           \& \zxN{}\rar     \& \zxN{}\rar        \& \zxN{}\rar           \& \zxZ{}\rar\dar \& \zxN{}\rar            \& \zxN{}\rar        \& \zxN{}\rar           \& \zxZ{}\rar\dar \& \zxX{}\rar     \& \zxFracZ{\pi}{8}\rar \& \zxX{}\rar            \& \zxFracZ-{\pi}{8}\rar \&       \zxN{}\rar \& \zxN{}\rar \& \zxN{}\rar           \& \zxZ{}\rar\ar[ddd] \& \zxN{}\rar            \& \zxN{}\rar          \& \zxN{}\rar           \& \zxZ{}\rar\ar[ddd] \& \zxN{}\rar            \& \zxN{}\rar \& \zxN{}\rar       \& \zxN{}\rar            \& \zxN{}\rar     \& \zxN{}\rar           \& \zxN{}\rar     \& \zxN{}\rar            \& \zxN{}\rar     \& \zxN{}\rar         \& \zxN{}\rar            \& \zxN{}\rar \& \zxN{}\rar        \& \zxN{}\rar \& \zxN{}\rar           \& \zxZ{}\rar\ar[ddd] \& \zxN{}\rar            \& \zxN{}\rar          \& \zxN{}\rar           \& \zxZ{}\rar\ar[ddd] \& \zxN{}\rar            \& \zxN{}\rar \& \zxN{}  \\
					\zxN{}\rar \& \zxN{}\rar \& \zxN{}\rar         \& \zxN{}\rar           \& \zxZ{}\rar\dar \& \zxN{}\rar            \& \zxN{}\rar        \& \zxN{}\rar           \& \zxZ{}\rar\dar \& \zxN{}\rar            \& \zxN{}\rar         \& \zxN{}\rar           \& \zxZ{}\rar\dar \& \zxN{}\rar            \& \zxN{}\rar        \& \zxN{}\rar           \& \zxZ{}\rar\dar \& \zxX{}\rar        \& \zxFracZ{\pi}{8}\rar \& \zxX{}\rar     \& \zxFracZ-{\pi}{8}\rar \& \zxX{}\rar        \& \zxFracZ{\pi}{8}\rar \& \zxX{}\rar     \& \zxN{}\rar     \& \zxN{}\rar           \& \zxN{}\rar            \& \zxFracZ-{\pi}{8}\rar \&       \zxN{}\rar \& \zxN{}\rar \& \zxN{}\rar           \& \zxN{}\rar         \& \zxN{}\rar            \& \zxN{}\rar          \& \zxN{}\rar           \& \zxN{}\rar         \& \zxN{}\rar            \& \zxN{}\rar \& \zxN{}\rar       \& \zxN{}\rar            \& \zxZ{}\rar\dar \& \zxN{}\rar           \& \zxN{}\rar     \& \zxN{}\rar            \& \zxZ{}\rar\dar \& \zxZ{}\rar\dar[dd] \& \zxFracZ{\pi}{4}\rar  \& \zxN{}\rar \& \zxZ{}\rar\ar[dd] \& \zxN{}\rar \& \zxN{}\rar           \& \zxN{}\rar         \& \zxN{}\rar            \& \zxN{}\rar          \& \zxN{}\rar           \& \zxN{}\rar         \& \zxN{}\rar            \& \zxN{}\rar \& \zxN{}  \\
					\zxN{}\rar \& \zxH{}\rar \& \zxX{}\rar         \& \zxFracZ{\pi}{8}\rar \& \zxX{}\rar     \& \zxFracZ-{\pi}{8}\rar \& \zxX{}\rar        \& \zxFracZ{\pi}{8}\rar \& \zxX{}\rar     \& \zxFracZ-{\pi}{8}\rar \& \zxX{}\rar         \& \zxFracZ{\pi}{8}\rar \& \zxX{}\rar     \& \zxFracZ-{\pi}{8}\rar \& \zxX{}\rar        \& \zxFracZ{\pi}{8}\rar \& \zxX{}\rar     \& \zxN{}\rar        \& \zxN{}\rar           \& \zxN{}\rar     \& \zxN{}\rar            \& \zxN{}\rar        \& \zxN{}\rar           \& \zxN{}\rar     \& \zxN{}\rar     \& \zxN{}\rar           \& \zxN{}\rar\&\zxFracZ-{\pi}{8}\rar \& \zxN{}\rar            \& \zxH{}\rar \& \zxN{}\rar           \& \zxN{}\rar         \& \zxN{}\rar            \& \zxN{}\rar          \& \zxN{}\rar           \& \zxN{}\rar         \& \zxN{}\rar            \& \zxN{}\rar \& \zxX{}\rar\ar[d] \& \zxFracZ-{\pi}{4}\rar \& \zxX{}\rar     \& \zxFracZ{\pi}{4}\rar \& \zxX{}\rar\dar \& \zxFracZ-{\pi}{4}\rar \& \zxX{}\rar     \& \zxN{}\rar         \& \zxFracZ{\pi}{4}\rar  \& \zxH{}\rar \& \zxN{}\rar        \& \zxN{}\rar \& \zxN{}\rar           \& \zxN{}\rar         \& \zxN{}\rar            \& \zxN{}\rar          \& \zxN{}\rar           \& \zxN{}\rar         \& \zxN{}\rar            \& \zxN{}\rar \& \zxN{}              \\
					\zxN{} \& \zxN{} \& \zxN{}         \& \zxN{}           \& \zxN{}     \& \zxN{}            \& \zxN{}        \& \zxN{}           \& \zxN{}     \& \zxN{}            \& \zxN{}         \& \zxN{}           \& \zxN{}     \& \zxN{}            \& \zxN{}        \& \zxN{}           \& \zxN{}     \& \zxN{}        \& \zxN{}           \& \zxN{}     \& \zxN{}            \& \zxN{}        \& \zxN{}           \& \zxN{}     \& \zxN{}     \& \zxN{}           \& \zxN{}            \& \zxN{}            \&       \zxX{}\rar \& \zxH{}\rar \& \zxFracZ{\pi}{4}\rar \& \zxX{}\rar         \& \zxFracZ-{\pi}{4}\rar \& \zxX{}\rar          \& \zxFracZ{\pi}{4}\rar \& \zxX{}\rar         \& \zxFracZ-{\pi}{4}\rar \& \zxH{}\rar \& \zxZ{}\rar       \& \zxN{}\rar            \& \zxN{}\rar     \& \zxN{}\rar           \& \zxZ{}\rar     \& \zxFracZ{\pi}{4}\rar   \& \zxN{}\rar     \& \zxX{}\rar         \& \zxFracZ-{\pi}{4}\rar \& \zxN{}\rar \& \zxX{}\rar        \& \zxH{}\rar \& \zxFracZ{\pi}{4}\rar \& \zxX{}\rar         \& \zxFracZ-{\pi}{4}\rar \& \zxX{}\rar          \& \zxFracZ{\pi}{4}\rar \& \zxX{}\rar         \& \zxFracZ-{\pi}{4}\rar \& \zxH{}\rar \& \zxX{} \\
			\end{ZX}}
                      \caption{Miter of multi-controlled Toffoli ZX-diagrams}\label{fig:zx_mcx_miter}
		\end{subfigure}

                \vspace*{1mm}		
		\begin{subfigure}[b]{\linewidth}
				\resizebox{\linewidth}{!}{%
			\begin{ZX}[ampersand replacement = \&]
				\zxN{}     \& \zxN{}     \& \zxN{}                 \& \zxN{}       \& \zxFracZ[]-{\pi}{8}\ar[d,blue,densely dotted] \& \zxFracZ[]{\pi}{8}\ar[d,blue,densely dotted] \& \zxFracZ[]-{\pi}{8}\ar[d,blue,densely dotted] \& \zxFracZ[]{\pi}{8}\ar[d,blue,densely dotted] \& \zxFracZ[]{\pi}{8}\ar[d,blue,densely dotted] \& \zxFracZ[]-{\pi}{8}\ar[d,blue,densely dotted] \& \zxFracZ[]{\pi}{8}\ar[d,blue,densely dotted] \& \zxFracZ[]-{\pi}{8}\ar[d,blue,densely dotted] \& \zxFracZ[]{\pi}{8}\ar[d,blue,densely dotted] \& \zxFracZ[]{\pi}{4}\ar[d,blue,densely dotted] \& \zxFracZ[]{\pi}{4}\ar[d,blue,densely dotted] \& \zxFracZ[]-{\pi}{4}\ar[d,blue,densely dotted] \& \zxFracZ[]{\pi}{4}\ar[d,blue,densely dotted] \& \zxFracZ[]-{\pi}{4}\ar[d,blue,densely dotted] \& \zxFracZ[]{\pi}{8}\ar[d,blue,densely dotted] \& \zxFracZ[]{\pi}{4}\ar[d,blue,densely dotted] \& \zxFracZ[]-{\pi}{8}\ar[d,blue,densely dotted] \& \zxFracZ[]-{\pi}{4}\ar[d,blue,densely dotted] \& \zxFracZ[]-{\pi}{4}\ar[d,blue,densely dotted] \& \zxFracZ[]{\pi}{4}\ar[d,blue,densely dotted] \\
				\zxN{}     \& \zxN{}     \&\zxN{}     \& \zxN{}     \& \zxZ[a=40]{}           \& \zxZ[a=37]{} \& \zxZ[a=34]{}                                  \& \zxZ[a=31]{}                                 \& \zxZ[a=21]{}                                  \& \zxZ[a=18]{}                                 \& \zxZ[a=15]{}                                 \& \zxZ[a=12]{}                                  \& \zxZ[a=9]{}                                  \& \zxZ[a=6]{}                                   \& \zxZ[a=57]{}                                 \& \zxZ[a=93]{}                                 \& \zxZ[a=60]{}                                 \& \zxZ[a=63]{}                                  \& \zxZ[a=67]{}                                 \& \zxZ[a=73]{}                                  \& \zxZ[a=77]{}                                 \& \zxZ[a=79]{}                                 \& \zxZ[a=87]{}                                  \& \zxZ[a=90]{}                                                                                                                                 \\
				\zxN{}\rar \& \zxN{}\rar \& \zxFracZ[a=4]-{\pi}{8} \&              \&                                               \&                                              \&                                               \&                                              \&                                              \&                                               \&                                              \&                                               \&                                              \&                                              \&                                              \&                                               \&                                              \&                                               \&                                              \&                                              \&                                               \&                                               \&                                               \& \zxN[a=a]{}                                  \\
				\zxN{}\rar \& \zxN{}\rar \& \zxFracZ[a=5]-{\pi}{8} \&              \&                                               \&                                              \&                                               \&                                              \&                                              \&                                               \&                                              \&                                               \&                                              \&                                              \&                                              \&                                               \&                                              \&                                               \&                                              \&                                              \&                                               \&                                               \&                                               \& \zxN[a=b]{}                                  \\
				\zxN{}\rar \& \zxN{}\rar \& \zxFracZ[a=11]{\pi}{8} \&              \&                                               \&                                              \&                                               \&                                              \&                                              \&                                               \&                                              \&                                               \&                                              \&                                              \&                                              \&                                               \&                                              \&                                               \&                                              \&                                              \&                                               \&                                               \&                                               \& \zxN[a=c]{}                                  \\
				\zxN{}\rar \& \zxH{}\rar \& \zxFracZ[a=97]{\pi}{8} \&              \&                                               \&                                              \&                                               \&                                              \&                                              \&                                               \&                                              \&                                               \&                                              \&                                              \&                                              \&                                               \&                                              \&                                               \&                                              \&                                              \&                                               \&                                               \& \zxH[a=blah]{}\rar                            \& \zxN{}                                       \\
				\&            \&                        \&              \&                                               \&                                              \&                                               \&                                              \&                                              \& \zxFracZ[a=65]-{\pi}{4}                       \&                                              \& \zxFracZ[a=69]-{\pi}{4}                       \&                                              \&                                              \& \zxFracZ[a=83]{\pi}{4}                       \&                                               \&                                              \&                                               \&                                              \&                                              \&                                               \&                                                                                                                                              \\
				\ar[from=4,to=90,blue,densely dotted]
                                \ar[from=4,to=63,blue,densely dotted]
				\ar[from=4,to=93,blue,densely dotted]
				\ar[from=4,to=57,blue,densely dotted]
				\ar[from=4,to=9,blue,densely dotted]
				\ar[from=4,to=12,blue,densely dotted]
				\ar[from=4,to=77,blue,densely dotted]
				\ar[from=4,to=21,blue,densely dotted]
				\ar[from=4,to=67,blue,densely dotted]
				\ar[from=4,to=37,blue,densely dotted]
				\ar[from=4,to=40,blue,densely dotted]
                                \ar[from=5,to=63,blue,densely dotted]
				\ar[from=5,to=87,blue,densely dotted]
				\ar[from=5,to=90,blue,densely dotted]
				\ar[from=5,to=60,blue,densely dotted]
				\ar[from=5,to=9,blue,densely dotted]
				\ar[from=5,to=12,blue,densely dotted]
				\ar[from=5,to=77,blue,densely dotted]
				\ar[from=5,to=15,blue,densely dotted]
				\ar[from=5,to=37,blue,densely dotted]
				\ar[from=5,to=18,blue,densely dotted]
				\ar[from=5,to=31,blue,densely dotted]
                                \ar[from=6,to=69,blue,densely dotted]
                                \ar[from=6,to=97,blue,densely dotted]
                                \ar[from=11,to=12,blue,densely dotted]
				\ar[from=11,to=79,blue,densely dotted]
				\ar[from=11,to=18,blue,densely dotted]
				\ar[from=11,to=15,blue,densely dotted]
				\ar[from=11,to=73,blue,densely dotted]
				\ar[from=11,to=34,blue,densely dotted]
				\ar[from=11,to=37,blue,densely dotted]
				\ar[from=11,to=21,blue,densely dotted]
				\ar[from=11,to=40,blue,densely dotted]
				\ar[from=18,to=97,blue,densely dotted]
				\ar[from=31,to=97,blue,densely dotted]
				\ar[from=34,to=97,blue,densely dotted]
                                \ar[from=37,to=97,blue,densely dotted]
                                \ar[from=40,to=97,blue,densely dotted]
                                \ar[from=57,to=65,blue,densely dotted]
                                \ar[from=60,to=65,blue,densely dotted]
				\ar[from=63,to=65,blue,densely dotted]
                                \ar[from=65,to=69,blue,densely dotted]
				\ar[from=65,to=97,blue,densely dotted]
                                \ar[from=67,to=97,blue,densely dotted]
				\ar[from=69,to=73,blue,densely dotted]
				\ar[from=69,to=79,blue,densely dotted]
				\ar[from=69,to=83,blue,densely dotted]
                                \ar[from=77,to=97,blue,densely dotted]
                                \ar[from=79,to=97,blue,densely dotted]
				\ar[from=83,to=87,blue,densely dotted]
				\ar[from=83,to=90,blue,densely dotted]
				\ar[from=83,to=93,blue,densely dotted]
				\ar[from=83,to=97,blue,densely dotted]
				\ar[from=4,to=a]
				\ar[from=5,to=b]
				\ar[from=11,to=c]
				\ar[from=97,to=blah]
			\end{ZX}
		}
                \caption{Reduced miter}\label{fig:zx_mcx_miter_reduced}
		\end{subfigure}
		
		\caption{Counterexample to completeness}
                \vspace*{-5mm}
	\end{figure}
	The question of completeness arises naturally in the context of rewriting. Given two circuits $G$ and $G^\prime$, can we prove their (non-)equivalence using the ZX-calculus rewriting strategy?
	
	As the ZX-calculus is complete for Clifford ZX-diagrams, it is not surprising that automated equivalence checking with the ZX-calculus is also
	complete for Clifford ZX-diagrams.
	
	\begin{theorem}\label{thm:cliff_comp}
		Given two quantum Circuits $G$ and $G^\prime$ consisting only of Clifford gates with corresponding ZX-diagrams $D$ and $D^\prime$ the only reduced gadget form of $D^\dagger D^\prime$ is the identity diagram.
	\end{theorem}\vspace*{-2mm}
	
	\begin{proof}
          The rules $\operatorname{\mathbf{LC}}$ and $\operatorname{\mathbf{P}}$ remove every interior Clifford spider from a ZX-diagram. Since all
          spiders in $D^\dagger D^\prime$ are Clifford, there are no
		more interior spiders left after simplifying. After simplification $D^\dagger D^\prime$ must therefore be of the form
                \vspace*{-1mm}
                
                \begin{center}
			{  \tikzset{
					/zx/user overlay nodes/.style={
						zxH/.append style={dashed,inner sep=2mm}
                                              }}
                                            \resizebox{.5\linewidth}{!}{
				\begin{ZX}[ampersand replacement=\&]
					\zxN{} \rar                                                                             \& [\zxwCol]\zxH{} \rar \& [\zxwCol] \zxFracZ{j_0 \pi}{2} \ar[rrrr, blue, densely dotted]\ar[d, blue, densely dotted]\ar[dr, blue, densely dotted]     \& [\zxwCol] \zxN{} \& [\zxwCol] \zxN{}                  \& [\zxwCol] \zxN{}       \& [\zxwCol]
					\zxFracZ{k_0 \pi}{2} \rar\ar[dl, blue, densely dotted]\ar[d, blue, densely dotted]      \& [\zxwCol] \zxH{}\rar \& [\zxwCol]\zxN{} \\
					\zxN{}                                                                                  \& \zxN{}               \& \zxN{} \ar[r,3 dots]                                                                                                        \& \zxN{}           \& \zxN{}           \ar[dd,3 vdots] \& \zxN{}                 \& \zxN{}\ar[l,3 dots]
					\& \zxN{}                                 \\
					\\
					\zxN{}                                                                                  \& \zxN{}               \& \zxN{}\ar[r,3 dots]                                                                                                         \& \zxN{}           \& \zxN{}                            \& \zxN{}                 \& \zxN{}\ar[l,3 dots]
					\& \zxN{}                                 \\
					\zxN{} \rar                                                                             \& [\zxwCol]\zxH{} \rar \& [\zxwCol] \zxFracZ{j_{n-1} \pi}{2} \ar[rrrr, blue, densely dotted]\ar[u, blue, densely dotted]\ar[ur, blue, densely dotted] \& [\zxwCol] \zxN{} \& [\zxwCol]                         \& \zxN{}[\zxwCol] \zxN{} \& [\zxwCol]
					\zxFracZ{k_{n-1} \pi}{2} \rar \ar[ul, blue, densely dotted]\ar[u, blue, densely dotted] \& [\zxwCol] \zxH{}\rar \& [\zxwCol]\zxN{}                                                                                                                                                                                                                
				\end{ZX}},
                            }
                            \end{center}                \vspace*{-1mm}

		i.e.\ a Clifford ZX-diagram with only boundary spiders, where all boundary spiders (that are in fact connected) are connected via Hadamard edges and
		where there must be exactly one Hadamard box on each line. In fact this structure is independent of the fact that
		$\llbracket D^\dagger D^\prime \rrbracket = I_n$---every Clifford \mbox{ZX-Diagram} has this reduced gadget form. W.l.o.g.\ assume that the Hadamard boxes are all on the left-hand side and call this
		diagram $D_{\text{simp}}$
		
                We need to show $j_i=k_i=0$ for all $0 \leq i < n$ and that there are no
		connections between spiders on different lines. To break this problem into simpler sub-problems we are going to use the
		following helpful trick:
                \resizebox{\linewidth}{!}{%
		\begin{ZX}[ampersand replacement=\&]
                  \zxZ{} \rar                                                                             \& [\zxwCol]\zxH{} \rar \& [\zxwCol] \zxFracZ{j_0 \pi}{2} \ar[rrrr, blue, densely dotted]\ar[d, blue, densely dotted]\ar[dr, blue, densely dotted]     \& [\zxwCol] \zxN{} \& [\zxwCol] \zxN{}                  \& [\zxwCol] \zxN{}       \& [\zxwCol]
			\zxFracZ{k_0 \pi}{2} \rar\ar[dl, blue, densely dotted]\ar[d, blue, densely dotted]      \& [\zxwCol] \zxN{}\rar \& [\zxwCol]\zxX{} \\
			\zxN{}                                                                                  \& \zxN{}               \& \zxN{} \ar[r,3 dots]                                                                                                        \& \zxN{}           \& \zxN{}           \ar[dd,3 vdots] \& \zxN{}                 \& \zxN{}\ar[l,3 dots]
			\& \zxN{}                                 \\
			\\
			\zxN{}                                                                                  \& \zxN{}               \& \zxN{}\ar[r,3 dots]                                                                                                         \& \zxN{}           \& \zxN{}                            \& \zxN{}                 \& \zxN{}\ar[l,3 dots]
			\& \zxN{}                                 \\
			\zxN{} \rar                                                                             \& [\zxwCol]\zxH{} \rar \& [\zxwCol] \zxFracZ{j_{n-1} \pi}{2} \ar[rrrr, blue, densely dotted]\ar[u, blue, densely dotted]\ar[ur, blue, densely dotted] \& [\zxwCol] \zxN{} \& [\zxwCol]                         \& \zxN{}[\zxwCol] \zxN{} \& [\zxwCol]
			\zxFracZ{k_{n-1} \pi}{2} \rar \ar[ul, blue, densely dotted]\ar[u, blue, densely dotted] \& [\zxwCol] \zxN{}\rar \& [\zxwCol]\zxN{}                   
		\end{ZX} $\overset{\textbf{(h)}}{=}$    \begin{ZX}[ampersand replacement=\&]
			\zxN{}                                                                              \& [\zxwCol]\zxX{} \rar \& [\zxwCol] \zxFracZ{j_0 \pi}{2} \ar[rrrr, blue, densely dotted]\ar[d, blue, densely dotted]\ar[dr, blue, densely dotted]     \& [\zxwCol] \zxN{} \& [\zxwCol] \zxN{}                  \& [\zxwCol] \zxN{}       \& [\zxwCol]
			\zxFracZ{k_0 \pi}{2} \rar\ar[dl, blue, densely dotted]\ar[d, blue, densely dotted]      \& [\zxwCol] \zxN{}\rar \& [\zxwCol]\zxX{} \\
			\zxN{}                                                                                  \& \zxN{}               \& \zxN{} \ar[r,3 dots]                                                                                                        \& \zxN{}           \& \zxN{}           \ar[dd,3 vdots] \& \zxN{}                 \& \zxN{}\ar[l,3 dots]
			\& \zxN{}                                 \\
			\\
			\zxN{}                                                                                  \& \zxN{}               \& \zxN{}\ar[r,3 dots]                                                                                                         \& \zxN{}           \& \zxN{}                            \& \zxN{}                 \& \zxN{}\ar[l,3 dots]
			\& \zxN{}                                 \\
			\zxN{} \rar                                                                             \& [\zxwCol]\zxH{} \rar \& [\zxwCol] \zxFracZ{j_{n-1} \pi}{2} \ar[rrrr, blue, densely dotted]\ar[u, blue, densely dotted]\ar[ur, blue, densely dotted] \& [\zxwCol] \zxN{} \& [\zxwCol]                         \& \zxN{}[\zxwCol] \zxN{} \& [\zxwCol]
			\zxFracZ{k_{n-1} \pi}{2} \rar \ar[ul, blue, densely dotted]\ar[u, blue, densely dotted] \& [\zxwCol] \zxN{}\rar \& [\zxwCol]\zxN{}                                                                   
                      \end{ZX} $\overset{\textbf{(c)}}{=}$
                    }
                \resizebox{\linewidth}{!}{%
                    \begin{ZX}[ampersand replacement=\&]
			\zxN{}                                                                                  \& [\zxwCol]\zxN{}      \& [\zxwCol] \zxN                                                                                                              \& [\zxwCol] \zxX{}\ar[rr,blue,densely dotted]          \& [\zxwCol] \zxN{}                 \& [\zxwCol] \zxX{}                 \& [\zxwCol]
			\zxN                                                                                    \& [\zxwCol] \zxN{} \& [\zxwCol]\zxN{} \\
			\zxN{}                                                                                  \& \zxN{}               \& \zxX{}\ar[d,blue,densely dotted] \ar[r,3 dots]                                                                              \& \zxX{}\ar[d,blue,densely dotted] \& \zxN{}           \ar[dd,3 vdots] \& \zxX{}\ar[d,blue,densely dotted] \& \zxX{}\ar[d,blue,densely dotted]\ar[l,3 dots]
			\& \zxN{}                                 \\
			\zxN{}                                                                                  \& \zxN{}               \& \zxN{}                                                                                                                      \& \zxN{}                           \& \zxN{}                           \& \zxN{}                           \& \zxN{}
			\& \zxN{}                                 \\
			\zxN{}                                                                                  \& \zxN{}               \& \zxN{}\ar[r,3 dots]                                                                                                         \& \zxN{}                           \& \zxN{}                           \& \zxN{}                           \& \zxN{}\ar[l,3 dots]
			\& \zxN{}                                 \\
			\zxN{} \rar                                                                             \& [\zxwCol]\zxH{} \rar \& [\zxwCol] \zxFracZ{j_{n-1} \pi}{2} \ar[rrrr, blue, densely dotted]\ar[u, blue, densely dotted]\ar[ur, blue, densely dotted] \& [\zxwCol] \zxN{}                 \& [\zxwCol]                        \& \zxN{}[\zxwCol] \zxN{}           \& [\zxwCol]
			\zxFracZ{k_{n-1} \pi}{2} \rar \ar[ul, blue, densely dotted]\ar[u, blue, densely dotted] \& [\zxwCol] \zxN{}\rar \& [\zxwCol]\zxN{}                                                                   
		\end{ZX}$\overset{\textbf{(h)}}{=}$
		\begin{ZX}[ampersand replacement=\&]
			\zxN{}                                                                                  \& [\zxwCol]\zxN{}      \& [\zxwCol] \zxN                                                                                                              \& [\zxwCol] \zxZ{}\ar[rr]          \& [\zxwCol] \zxN{}                 \& [\zxwCol] \zxX{}                 \& [\zxwCol]
			\zxN                                                                                    \& [\zxwCol] \zxN{} \& [\zxwCol]\zxN{} \\
			\zxN{}                                                                                  \& \zxN{}               \& \zxZ{}\dar \ar[r,3 dots]                                                                              \& \zxZ{}\dar \& \zxN{}           \ar[dd,3 vdots] \& \zxZ{}\dar \& \zxZ{}\dar\ar[l,3 dots]
			\& \zxN{}                                 \\
			\zxN{}                                                                                  \& \zxN{}               \& \zxN{}                                                                                                                      \& \zxN{}                           \& \zxN{}                           \& \zxN{}                           \& \zxN{}
			\& \zxN{}                                 \\
			\zxN{}                                                                                  \& \zxN{}               \& \zxN{}\ar[r,3 dots]                                                                                                         \& \zxN{}                           \& \zxN{}                           \& \zxN{}                           \& \zxN{}\ar[l,3 dots]
			\& \zxN{}                                 \\
			\zxN{} \rar                                                                             \& [\zxwCol]\zxH{} \rar \& [\zxwCol] \zxFracZ{j_{n-1} \pi}{2} \ar[rrrr, blue, densely dotted]\ar[u, blue, densely dotted]\ar[ur, blue, densely dotted] \& [\zxwCol] \zxN{}                 \& [\zxwCol]                        \& \zxN{}[\zxwCol] \zxN{}           \& [\zxwCol]
			\zxFracZ{k_{n-1} \pi}{2} \rar \ar[ul, blue, densely dotted]\ar[u, blue, densely dotted] \& [\zxwCol] \zxN{}\rar \& [\zxwCol]\zxN{}                                                                   
                      \end{ZX} $\overset{\textbf{(f)}}{=}$
                    }
                    \resizebox{.5\linewidth}{!}{%
		\begin{ZX}[ampersand replacement=\&]
			\zxN{}                                                                                  \& [\zxwCol]\zxN{}      \& [\zxwCol] \zxN                                                                                                              \& [\zxwCol] \zxZ{}\ar[rr]          \& [\zxwCol] \zxN{}                 \& [\zxwCol] \zxX{}                 \& [\zxwCol]
			\zxN                                                                                    \& [\zxwCol] \zxN{} \& [\zxwCol]\zxN{} \\
			\zxN{}                                                                                  \& \zxN{}               \& \zxN{}                                                                               \& \zxN{} \& \zxN{}           \ar[dd,3 vdots] \& \zxN{} \& \zxN{}
			\& \zxN{}                                 \\
			\zxN{}                                                                                  \& \zxN{}               \& \zxN{}                                                                                                                      \& \zxN{}                           \& \zxN{}                           \& \zxN{}                           \& \zxN{}
			\& \zxN{}                                 \\
			\zxN{}                                                                                  \& \zxN{}               \& \zxN{}\ar[r,3 dots]                                                                                                         \& \zxN{}                           \& \zxN{}                           \& \zxN{}                           \& \zxN{}\ar[l,3 dots]
			\& \zxN{}                                 \\
			\zxN{} \rar                                                                             \& [\zxwCol]\zxH{} \rar \& [\zxwCol] \zxFracZ{j_{n-1} \pi}{2} \ar[rrrr, blue, densely dotted]\ar[u, blue, densely dotted]\ar[ur, blue, densely dotted] \& [\zxwCol] \zxN{}                 \& [\zxwCol]                        \& \zxN{}[\zxwCol] \zxN{}           \& [\zxwCol]
			\zxFracZ{k_{n-1} \pi}{2} \rar \ar[ul, blue, densely dotted]\ar[u, blue, densely dotted] \& [\zxwCol] \zxN{}\rar \& [\zxwCol]\zxN{}                                                                   
                      \end{ZX}
                      }
		
		The spider fusion in the last equality is due to the fact that all spiders in the second row have a phase of $0$ and
		all the spiders they are connected to are $Z$-spiders (all spiders in the diagram are $Z$-spiders). With
                this trick, we
		can effectively eliminate a line from the diagram because the diagram manipulations only effect the connections from
		the first row and not the connections between other rows. The rest of the diagram still has to act as the identity on
		the rest of the qubits. This is due to the fact that if $\llbracket D_{\text{simp}} \rrbracket = \llbracket
		\begin{ZX}
			\zxN{}\rar&[\zxwCol]\zxN{}\rar{}&[\zxwCol]\zxN{}\\
			\zxN{}\rar&[\zxwCol]\zxN{}\ar[dd,3 vdots]\rar{}&\zxN{}\\
			\\
			\zxN{}\rar&[\zxwCol]\zxN{}\rar{}&[\zxwCol]\zxN{}\\
		\end{ZX}
		\rrbracket$ then

                \vspace*{-2mm}
                \begin{equation*}
			\begin{ZX}
				\zxZ{} \rar                                                                             & [\zxwCol]\zxH{} \rar & [\zxwCol] \zxFracZ{j_0 \pi}{2} \ar[rrrr, blue, densely dotted]\ar[d, blue, densely dotted]\ar[dr, blue, densely dotted]     & [\zxwCol] \zxN{} & [\zxwCol] \zxN{}                  & [\zxwCol] \zxN{}       & [\zxwCol]
				\zxFracZ{k_0 \pi}{2} \rar\ar[dl, blue, densely dotted]\ar[d, blue, densely dotted]      & [\zxwCol] \zxN{}\rar & [\zxwCol]\zxX{} \\
				\zxN{}                                                                                  & \zxN{}               & \zxN{} \ar[r,3 dots]                                                                                                        & \zxN{}           & \zxN{}           \ar[dd,3 vdots] & \zxN{}                 & \zxN{}\ar[l,3 dots]
				& \zxN{}                                 \\
				\\
				\zxN{}                                                                                  & \zxN{}               & \zxN{}\ar[r,3 dots]                                                                                                         & \zxN{}           & \zxN{}                            & \zxN{}                 & \zxN{}\ar[l,3 dots]
				& \zxN{}                                 \\
				\zxN{} \rar                                                                             & [\zxwCol]\zxH{} \rar & [\zxwCol] \zxFracZ{j_{n-1} \pi}{2} \ar[rrrr, blue, densely dotted]\ar[u, blue, densely dotted]\ar[ur, blue, densely dotted] & [\zxwCol] \zxN{} & [\zxwCol]                         & \zxN{}[\zxwCol] \zxN{} & [\zxwCol]
				\zxFracZ{k_{n-1} \pi}{2} \rar \ar[ul, blue, densely dotted]\ar[u, blue, densely dotted] & [\zxwCol] \zxN{}\rar & [\zxwCol]\zxN{}                                                                                                                                                                                                                
			\end{ZX} =   \begin{ZX}
				\zxZ{}\rar&\zxN{}\rar{}&\zxX{}\\
				\zxN{}\rar&\zxN{}\ar[dd,3 vdots]\rar{}&\zxN{}\\
				\\
				\zxN{}\rar&\zxN{}\rar{}&\zxN{}\\
			\end{ZX}.
                      \end{equation*}\vspace*{-2mm}
                      
		If two ZX-diagrams have the same interpretation they can be replaced with each other in every context. This is a simple
		consequence of the soundness of the ZX-calculus.
		
		With the introduced trick, we can effectively reduce $D_\text{simp}$ until only one line remains. In this case the
		remaining line
		\begin{ZX}
			\zxN{} \rar                                                                             & [\zxwCol]\zxH{} \rar & [\zxwCol] \zxFracZ{j_t \pi}{2} \ar[rr, blue, densely dotted]     & [\zxwCol] \zxN{}      & [\zxwCol]
			\zxFracZ{k_t \pi}{2}    \rar{}   & [\zxwCol] \zxN{} 
		\end{ZX} still has to represent the identity. We prove that $j_t=k_t=0$ by concrete calculation of the matrix of the
		diagram.
                \vspace*{-3mm}

		\begin{align*}
                  & \llbracket  \begin{ZX}[ampersand replacement=\&]
				\zxN{} \rar                                                                             \& [\zxwCol]\zxH{} \rar \& [\zxwCol] \zxFracZ{j_t \pi}{2} \ar[rr, blue, densely dotted]     \& [\zxwCol] \zxN{}      \& [\zxwCol]
				\zxFracZ{k_t \pi}{2}    \rar{}   \& [\zxwCol] \zxN{} 
			\end{ZX}\rrbracket \overset{\textbf{(h)}}{=}\llbracket  \begin{ZX}[ampersand replacement=\&]
				\zxN{} \rar                                                                             \& [\zxwCol]\zxN{} \rar \& [\zxwCol] \zxFracX{j_t \pi}{2} \ar[rr]     \& [\zxwCol] \zxN{}      \& [\zxwCol]
				\zxFracZ{k_t \pi}{2}    \rar{}   \& [\zxwCol] \zxN{} 
                              \end{ZX}\rrbracket =\\
                  &(\ket{+}\bra{+} + e^{ij_t\frac{\pi}{2}}\ket{-}\bra{-}) (\ket{0}\bra{0} +
			e^{ik_t\frac{\pi}{2}}\ket{1}\bra{1}) = \\ 
                  & \frac{1}{2}
                   \begin{bmatrix}
				1+e^{i j_t \frac{\pi}{2}} & (1- e^{i j_t\frac{\pi}{2}})e^{i k_t\frac{\pi}{2}}\\
				1-e^{i j_t \frac{\pi}{2}}& (1 + e^{i j_t\frac{\pi}{2}})e^{i k_t\frac{\pi}{2}}
			\end{bmatrix}
		\end{align*}                
		
		Since this matrix has to equal $\begin{bmatrix} 1&0\\0&1\end{bmatrix}$ the constraints force $e^{i j_t \frac{\pi}{2}} =
		e^{i k_t\frac{\pi}{2}} = 1$ which can only be true if $j_t = k_t = 0$. Thus we can conclude that all spiders in
		$D_\text{simp}$ have a phase of $0$.
		
		To show that no spiders belonging to different lines can be connected, we use our trick again. But this time we reduce
		down to two lines.

                \vspace*{-2mm}
		\begin{equation*}
			\begin{ZX}
				\zxN{} \rar                                                                             & [\zxwCol]\zxH{} \rar & [\zxwCol] \zxZ{} \ar[rr, blue, densely dotted]\ar[d, blue, densely dotted]\ar[drr, blue, densely dotted]     &  [\zxwCol] \zxN{}       & [\zxwCol]
				\zxZ{} \rar\ar[d, blue, densely dotted]      & [\zxwCol] \zxN{}\rar & [\zxwCol]\zxN{} \\
				\zxN{} \rar                                                                             & [\zxwCol]\zxH{} \rar & [\zxwCol] \zxZ{} \ar[rr, blue, densely dotted]\ar[u, blue, densely dotted]\ar[urr, blue, densely dotted] &  [\zxwCol] \zxN{} & [\zxwCol]
				\zxZ{} \rar & [\zxwCol] \zxN{}\rar & [\zxwCol]\zxN{}                                                                                                                                                                                                                
			\end{ZX}
		\end{equation*}\vspace*{-2mm}
		
		The Hadamard edges between spiders on different lines may or may not exist. We are going to see that for the purpose of
		this proof we do not need to make a case distinction on all possible combinations of connections. Using a similar
		strategy as for the line removal trick we obtain
		
		\begin{align*}
		&	\begin{ZX}[ampersand replacement=\&]
				\zxZ{} \rar                                                                             \& [\zxwCol]\zxH{} \rar \& [\zxwCol] \zxZ{} \ar[rr, blue, densely dotted]\ar[d, blue, densely dotted]\ar[drr, blue, densely dotted]     \&  [\zxwCol] \zxN{}       \& [\zxwCol]
				\zxZ{} \rar\ar[d, blue, densely dotted]      \& [\zxwCol] \zxN{}\rar \& [\zxwCol]\zxN{} \\
				\zxN{} \rar                                                                             \& [\zxwCol]\zxH{} \rar \& [\zxwCol] \zxZ{} \ar[rr, blue, densely dotted]\ar[u, blue, densely dotted]\ar[urr, blue, densely dotted] \&  [\zxwCol] \zxN{} \& [\zxwCol]
				\zxZ{} \rar \& [\zxwCol] \zxN{}\rar \& [\zxwCol]\zxX{}          
			\end{ZX} \overset{\textbf{(h)}}{=}  \begin{ZX}[ampersand replacement=\&]
				\zxN{}                                                                              \& [\zxwCol]\zxX{} \rar \& [\zxwCol] \zxZ{} \ar[rr, blue, densely dotted]\ar[d, blue, densely dotted]\ar[drr, blue, densely dotted]     \&  [\zxwCol] \zxN{}       \& [\zxwCol]
				\zxZ{} \rar\ar[d, blue, densely dotted]      \& [\zxwCol] \zxN{}\rar \& [\zxwCol]\zxN{} \\
				\zxN{} \rar                                                                             \& [\zxwCol]\zxH{} \rar \& [\zxwCol] \zxZ{} \ar[rr, blue, densely dotted]\ar[u, blue, densely dotted]\ar[urr, blue, densely dotted] \&  [\zxwCol] \zxN{} \& [\zxwCol]
				\zxZ{} \rar \& [\zxwCol] \zxN{}\rar \& [\zxwCol]\zxX{}          
			\end{ZX} \overset{\textbf{(c)}}{=}  \begin{ZX}[ampersand replacement=\&]
				\zxX{}\rar                                            \& [\zxwCol]\zxZ{}      \& [\zxwCol] \zxN{}                                                                           \& [\zxwCol] \zxX{}\ar[r,blue,dotted] \& [\zxwCol]
				\zxZ{} \rar\ar[d, blue, densely dotted] \& [\zxwCol] \zxN{}\rar \& [\zxwCol]\zxN{} \\
				\zxN{}                                                \& \zxN{}               \& \zxX{}                                                                                     \& \zxN{}               \& \zxX{}
				\& \zxN{}               \& \zxN{}          \\
				\zxN{} \rar                                           \& [\zxwCol]\zxH{} \rar \& [\zxwCol] \zxZ{}\ar[u, blue, densely dotted]\ar[uurr, blue, densely dotted] \& [\zxwCol] \zxN{}                   \& [\zxwCol]
				\zxN{}                                                \& [\zxwCol] \zxN{}     \& [\zxwCol]\zxN{}          
			\end{ZX} \overset{\textbf{(h)}}{=} \\
			&\begin{ZX}[ampersand replacement=\&]
				\zxX{}\rar                                            \& [\zxwCol]\zxZ{}      \& [\zxwCol] \zxN{}                                                                           \& [\zxwCol] \zxZ{}\ar[r] \& [\zxwCol]
				\zxZ{} \rar\ar[d] \& [\zxwCol] \zxN{}\rar \& [\zxwCol]\zxN{} \\
				\zxN{}                                                \& \zxN{}               \& \zxZ{}                                                                                     \& \zxN{}               \& \zxZ{}
				\& \zxN{}               \& \zxN{}          \\
				\zxN{} \rar                                           \& [\zxwCol]\zxH{} \rar \& [\zxwCol] \zxZ{}\ar[u]\ar[uurr, blue, densely dotted] \& [\zxwCol] \zxN{}                   \& [\zxwCol]
				\zxN{}                                                \& [\zxwCol] \zxN{}     \& [\zxwCol]\zxN{}          
			\end{ZX} \overset{\textbf{(f)}}{=} 
			\begin{ZX}[ampersand replacement=\&]
				\zxX{}\rar                                            \& [\zxwCol]\zxZ{}      \& [\zxwCol] \zxN{}                                                                           \& [\zxwCol] \zxN{} \& [\zxwCol]
				\zxZ{} \rar \& [\zxwCol] \zxN{}\rar \& [\zxwCol]\zxN{} \\
				\zxN{}                                                \& \zxN{}               \& \zxN{}                                                                                     \& \zxN{}               \& \zxN{}
				\& \zxN{}               \& \zxN{}          \\
				\zxN{} \rar                                           \& [\zxwCol]\zxH{} \rar \& [\zxwCol] \zxZ{}\ar[uurr, blue, densely dotted] \& [\zxwCol] \zxN{}                   \& [\zxwCol]
				\zxN{}                                                \& [\zxwCol] \zxN{}     \& [\zxwCol]\zxN{}          
			\end{ZX} \overset{\textbf{(h)}}{=} 
			\begin{ZX}[ampersand replacement=\&]
				\zxX{}\rar                                            \& [\zxwCol]\zxZ{}      \& [\zxwCol] \zxN{}                                                                           \& [\zxwCol] \zxN{} \& [\zxwCol]
				\zxZ{} \rar \& [\zxwCol] \zxN{}\rar \& [\zxwCol]\zxN{} \\
				\zxN{}                                                \& \zxN{}               \& \zxN{}                                                                                     \& \zxN{}               \& \zxN{}
				\& \zxN{}               \& \zxN{}          \\
				\zxN{} \rar                                           \& [\zxwCol]\zxN{} \rar \& [\zxwCol] \zxX{}\ar[uurr] \& [\zxwCol] \zxN{}                   \& [\zxwCol]
				\zxN{}                                                \& [\zxwCol] \zxN{}     \& [\zxwCol]\zxN{}          
			\end{ZX} 
		\end{align*}\vspace*{1mm}
		
		Since this diagram has to be the identity on each line and we input
		\begin{ZX}
			\zxZ{} \rar & \zxN{}
		\end{ZX} in the first line and \begin{ZX}
			\zxN{} \rar & \zxX{}
		\end{ZX} on the second line, they also have to be the output on their respective lines. But this can only be the case if
		the diagonal connection between the first and second line does not exist. Therefore there is also no connection in the
		original diagram, i.e.\ it has to look like
                
		\begin{equation*}
			\begin{ZX}
				\zxN{} \rar                                                                             & [\zxwCol]\zxH{} \rar & [\zxwCol] \zxZ{} \ar[rr, blue, densely dotted]\ar[d, blue, densely dotted]\ar[drr, blue, densely dotted]     &  [\zxwCol] \zxN{}       & [\zxwCol]
				\zxZ{} \rar\ar[d, blue, densely dotted]      & [\zxwCol] \zxN{}\rar & [\zxwCol]\zxN{} \\
				\zxN{} \rar                                                                             & [\zxwCol]\zxH{} \rar & [\zxwCol] \zxZ{} \ar[rr, blue, densely dotted]\ar[u, blue, densely dotted] &  [\zxwCol] \zxN{} & [\zxwCol]
				\zxZ{} \rar & [\zxwCol] \zxN{}\rar & [\zxwCol]\zxN{}                                                                                                                                                                                                                
			\end{ZX}.
		\end{equation*}
		
		Similar reasoning can be applied to conclude that the remaining inter-line connections cannot exist if the diagram is
		equal to the identity diagram.
		
		We have proven that all phases in $D_\text{simp}$ are $0$ and that there are no connections between spiders on different
		lines. Therefore $D_\text{simp}$ looks like

		\vspace*{1mm}
		\begin{equation*}
			\begin{ZX}
				\zxN{} \rar & [\zxWCol] \zxH{}\rar & [\zxWCol] \zxZ{} \ar[rr, blue, densely dotted] & [\zxWCol] \zxN{}                & [\zxWCol] \zxZ{} \ar[r] & [\zxWCol] \zxN{}                        \\
				\zxN{} \rar & [\zxWCol]\zxH{} \rar & [\zxWCol] \zxZ{} \ar[rr, blue, densely dotted] & [\zxWCol] \zxN{}\ar[dd,3 vdots] & [\zxWCol] \zxZ{} \ar[r] & [\zxWCol] \zxN{}  \\
				\\
				\zxN{} \rar & [\zxWCol] \zxH{}\rar & [\zxWCol] \zxZ{} \ar[rr, blue, densely dotted] & [\zxWCol] \zxN{}                & [\zxWCol] \zxZ{} \ar[r] & [\zxWCol] \zxN{}                        \\
			\end{ZX}.
		\end{equation*}

                        Rules $(\mathbf{hh})$, $(\mathbf{f})$ and $(\mathbf{id})$ remove identity spiders, fuse spiders and cancel
		adjacent Hadamard boxes as much as possible. Thus the diagram is further simplified. Then, 
		
		\begin{equation*}  
			\begin{ZX}
				\zxN{} \rar & [\zxWCol] \zxH{}\rar & [\zxWCol] \zxZ{} \rar & [\zxWCol] \zxH{}\rar                & [\zxWCol] \zxZ{} \ar[r] & [\zxWCol] \zxN{} \rar & [\zxWCol] \zxN{} \\
				\zxN{} \rar & [\zxWCol]\zxH{} \rar & [\zxWCol] \zxZ{} \rar & [\zxWCol] \zxH{}\rar\ar[dd,3 vdots] & [\zxWCol] \zxZ{} \ar[r] & [\zxWCol] \zxN{}\rar  & [\zxWCol] \zxN{} \\
				\\
				\zxN{} \rar & [\zxWCol] \zxH{}\rar & [\zxWCol] \zxZ{} \rar & [\zxWCol] \zxH{}\rar                & [\zxWCol] \zxZ{} \ar[r] & [\zxWCol] \zxN{}\rar  & [\zxWCol] \zxN{} \\
			\end{ZX}
			\overset{\textbf{(id)}}{=}
			\begin{ZX}
				\zxN{} \rar & [\zxWCol] \zxH{}\rar & [\zxWCol] \zxN{} \rar & [\zxWCol] \zxH{}\rar                & [\zxWCol] \zxN{} \ar[r] & [\zxWCol] \zxN{} \rar & [\zxWCol] \zxN{} \\
				\zxN{} \rar & [\zxWCol]\zxH{} \rar & [\zxWCol] \zxN{} \rar & [\zxWCol] \zxH{}\rar\ar[dd,3 vdots] & [\zxWCol] \zxN{} \ar[r] & [\zxWCol] \zxN{}\rar  & [\zxWCol] \zxN{} \\
				\\
				\zxN{} \rar & [\zxWCol] \zxH{}\rar & [\zxWCol] \zxN{} \rar & [\zxWCol] \zxH{}\rar                & [\zxWCol] \zxN{} \ar[r] & [\zxWCol] \zxN{}\rar  & [\zxWCol] \zxN{} \\
			\end{ZX}  \overset{\textbf{(hh)}}{=}
			\begin{ZX}
				\zxN{} \rar & [\zxWCol] \zxN{}\rar & [\zxWCol] \zxN{} \rar & [\zxWCol] \zxN{}\rar                & [\zxWCol] \zxN{} \ar[r] & [\zxWCol] \zxN{} \rar & [\zxWCol] \zxN{} \\[\zxHRow]
				\zxN{} \rar & [\zxWCol]\zxN{} \rar & [\zxWCol] \zxN{} \rar & [\zxWCol] \zxN{}\rar\ar[dd,3 vdots] & [\zxWCol] \zxN{} \ar[r] & [\zxWCol] \zxN{}\rar  & [\zxWCol] \zxN{} \\[\zxHRow]
				\\[\zxHRow]
				\zxN{} \rar & [\zxWCol] \zxN{}\rar & [\zxWCol] \zxN{} \rar & [\zxWCol] \zxN{}\rar                & [\zxWCol] \zxN{} \ar[r] & [\zxWCol] \zxN{}\rar  & [\zxWCol] \zxN{} \\[\zxHRow]
			\end{ZX}
		\end{equation*}\vspace*{1mm}

                Due to symmetry the proof still works if some of the Hadamard boxes are at the outputs.
	\end{proof}

	\autoref{thm:cliff_comp} establishes a baseline for what equivalences can be proven via automated reasoning with the
	ZX-calculus. Next, we want to look at completeness from a different perspective, by showing that rewriting to reduced gadget form is
	not sufficient for proving the equivalence of arbitrary equivalent circuits. In particular, we are going to show that this
	algorithm is not even sufficient for proving equivalence of reversible circuits. 
	
	\vspace*{1mm}
	\begin{theorem}\label{thm:zx_incomplete}
		There exist two equivalent quantum Circuits $G$ and $G^\prime$---using ancillary qubits---with corresponding
                ZX-diagrams $D$ and $D^\prime$ where $D^\dagger D^\prime$ possesses a reduced gadget form that is not the identity diagram.
	\end{theorem}
	
	\begin{proof}\label{proof:zx_incomplete}
		Unfortunately the proof of this theorem is not achieved through cunning manipulation of diagrams, kets and bras but by
		brute-force calculation.
		Consider the ZX-diagrams in \autoref{fig:zx_mcx} which are the ZX-diagrams of the circuits in
		\autoref{fig:mcx} where the ancillary qubit's input and output has been set to $\ket{0} =
		\begin{ZX}
			\zxX{} \rar & \zxN{}
		\end{ZX}
		$. Taking the adjoint of the diagram in \autoref{fig:zx_mcx_no_ancilla}, and concatenating the two
		diagrams, yields the diagram in \autoref{fig:zx_mcx_miter}. A reduced gadget form of this diagrams is shown in \autoref{fig:zx_mcx_miter_reduced}. No further simplifications can be made but this diagram is clearly not the identity. It can be checked by (tedious) computation of
		the corresponding matrices that the ZX-diagram in \autoref{fig:zx_mcx_miter} does actually implement the identity
		transformation.
	\end{proof}

	It is not surprising at all that equivalence checking via simplification to reduced gadget form is not complete in general. Because the equivalence checking
	problem is \mbox{QMA-complete} and since NP is a subset of QMA, it would be entirely unexpected that
	the ZX-calculus based equivalence checking algorithm solves the equivalence checking problem, given that a reduced gadget form can be derived in polynomial
	time with respect to the number of spiders of the original diagram~\cite{kissingerReducingTcountZXcalculus2020}. But the proof by counterexample shows that it cannot even show the
	equivalence of two circuits (involving ancillaries) even when they are fairly simple.
	\vspace*{3mm}
	\section{Case Study}\label{cha:experiments}
	The basic equivalence checking routine based on the \mbox{ZX-calculus} is publicly available via the Python library
	pyzx~\cite{kissingerPyZXLargeScale2019}. Since pyzx does not support layout permutations, inaccuracies, or ancillary
	qubits, and because Python is inherently slower than a compiled programming language, the \mbox{ZX-calculus} based
	equivalence checking algorithm has been re-implemented in C++ and integrated into the publicly available QCEC tool~(\mbox{\url{https://github.com/cda-tum/qcec}}) which is part of the Munich Quantum Toolkit (MQT, formerly known as JKQ~\cite{willeJKQJKUTools2020}). This re-implementation has additional features that allow for handling of the mentioned problems.

        To properly evaluate the resulting implementation of the ZX-calculus based equivalence checking algorithm, two state-of-the-art equivalence checking tools have been considered as a comparison:
        First, the proposed ZX-calculus equivalence checker is compared against an approach based on path-sums~\cite{amyLargescaleFunctionalVerification2019} on a large set of random Clifford circuits. This is done in order to assess how the proposed checker---which we proved to be complete for Clifford circuits---performs in relation to another Clifford-complete method.
 
        Second, an extensive comparison is performed against the complete equivalence checking approach based on decision diagrams proposed in \cite{burgholzerAdvancedEquivalenceChecking2021} to see how the incomplete \mbox{ZX-calculus} checker compares on a broad range of quantum circuits.

        The path-sum equivalence checker is publicly available via the \emph{Feynver} tool which is part of the Feynman toolset.
        The decision diagram based equivalence checker is also publicly available via \emph{QCEC}. For the remainder of this section \enquote{QCEC} explicitly refers to the decision diagram based equivalence checker proposed in~\cite{burgholzerAdvancedEquivalenceChecking2021} and implemented in QCEC.

        Before the experimental setup and the results are discussed, we will briefly introduce the basics of the two other equivalence checking methods considered for the evaluation.
	
	\subsection{Equivalence Checking Using Decision Diagrams}
        \emph{Decision
          Diagrams}~\cite{viamontesGatelevelSimulationQuantum2003,wangXQDDbasedVerificationMethod2008,niemannQMDDsEfficientQuantum2016,millerQMDDDecisionDiagram2006,zulehnerHowEfficientlyHandle2019}
        are a data structure used for efficiently representing complex matrices. Using \emph{redundancies} in the
        representation of a matrix, decision diagrams can often represent an exponentially large matrix using only
        polynomial resources. This makes them great candidates for use in equivalence checking of quantum circuits, as a
        quantum circuit is just another way of writing a unitary matrix.

        Recall that two quantum circuits $G=g_0\cdots g_m$ and $G^\prime=g^\prime_0 \cdots g^\prime_{m^\prime}$ are
        equivalent if $G^\dagger G^\prime =  g_m^\dagger \cdots g_0^\dagger  g^\prime_0 \cdots g^\prime_{m^\prime} =
        I$. Similar to equivalence checking using ZX-diagrams, decision diagrams can be used to efficiently carry out
        the matrix multiplication of $G^\dagger$ and $G^\prime$. The idea is to start constructing the functionality of the combined circuit from the
        ``middle'' and alternating between applications of $G^\dagger$ and $G^\prime$, such that the
        decision diagram being constructed remains as close to the identity as
        possible~\cite{burgholzerAdvancedEquivalenceChecking2021}. This is desirable because the $n$-qubit identity matrix only
        requires linear space when represented as a decision diagram instead of $2^n \times 2^n$ complex numbers for the entire matrix.

        	\begin{figure}[t]
          \centering

	\begin{subfigure}[b]{\linewidth}
		\centering
		\begin{subfigure}[b]{.49\linewidth}
                  \includegraphics[width=\linewidth]{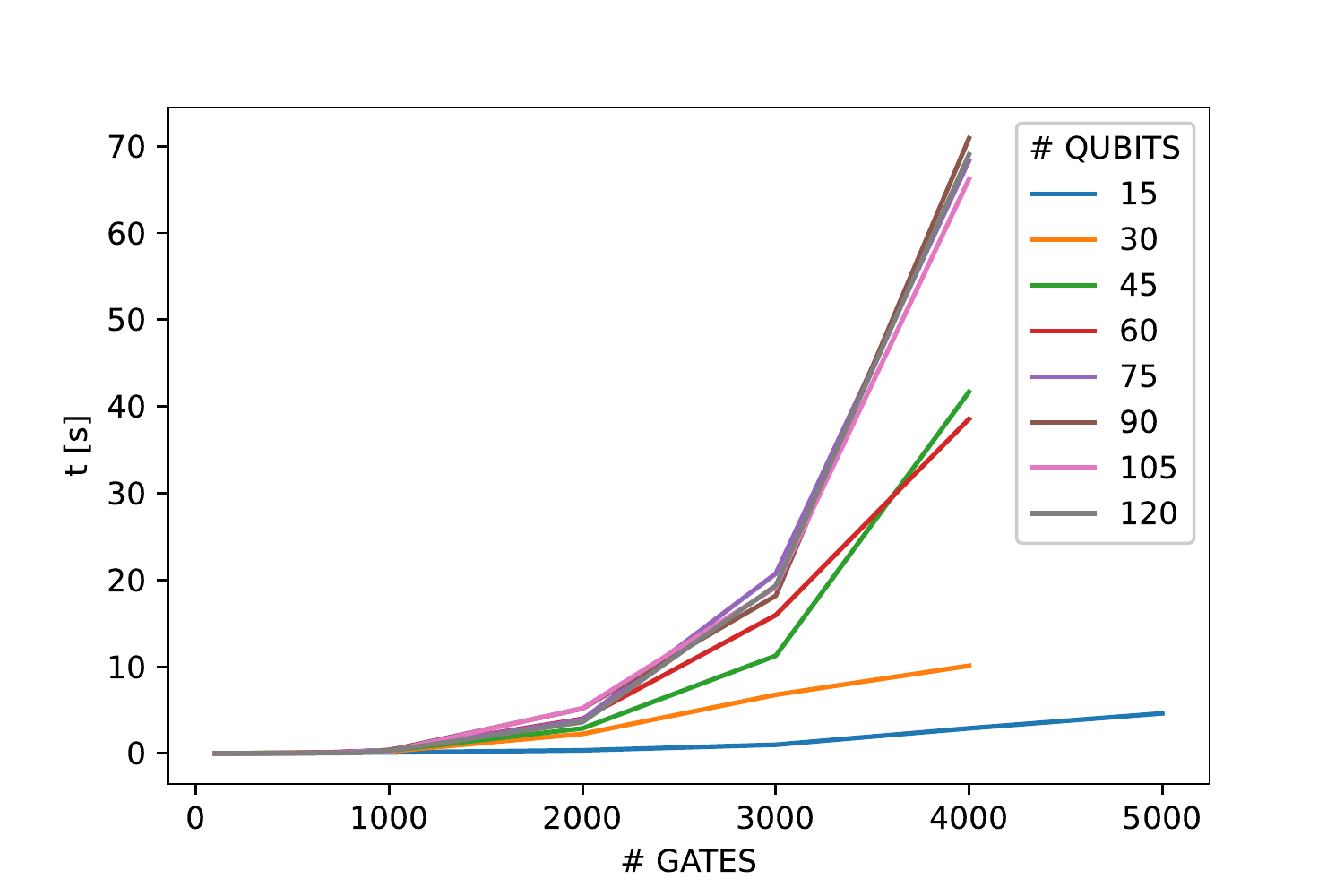}
                  \captionsetup{labelformat=empty}
			\caption{ZX-calculus checker}
		\end{subfigure}
		\begin{subfigure}[b]{.49\linewidth}
                  \includegraphics[width=\linewidth]{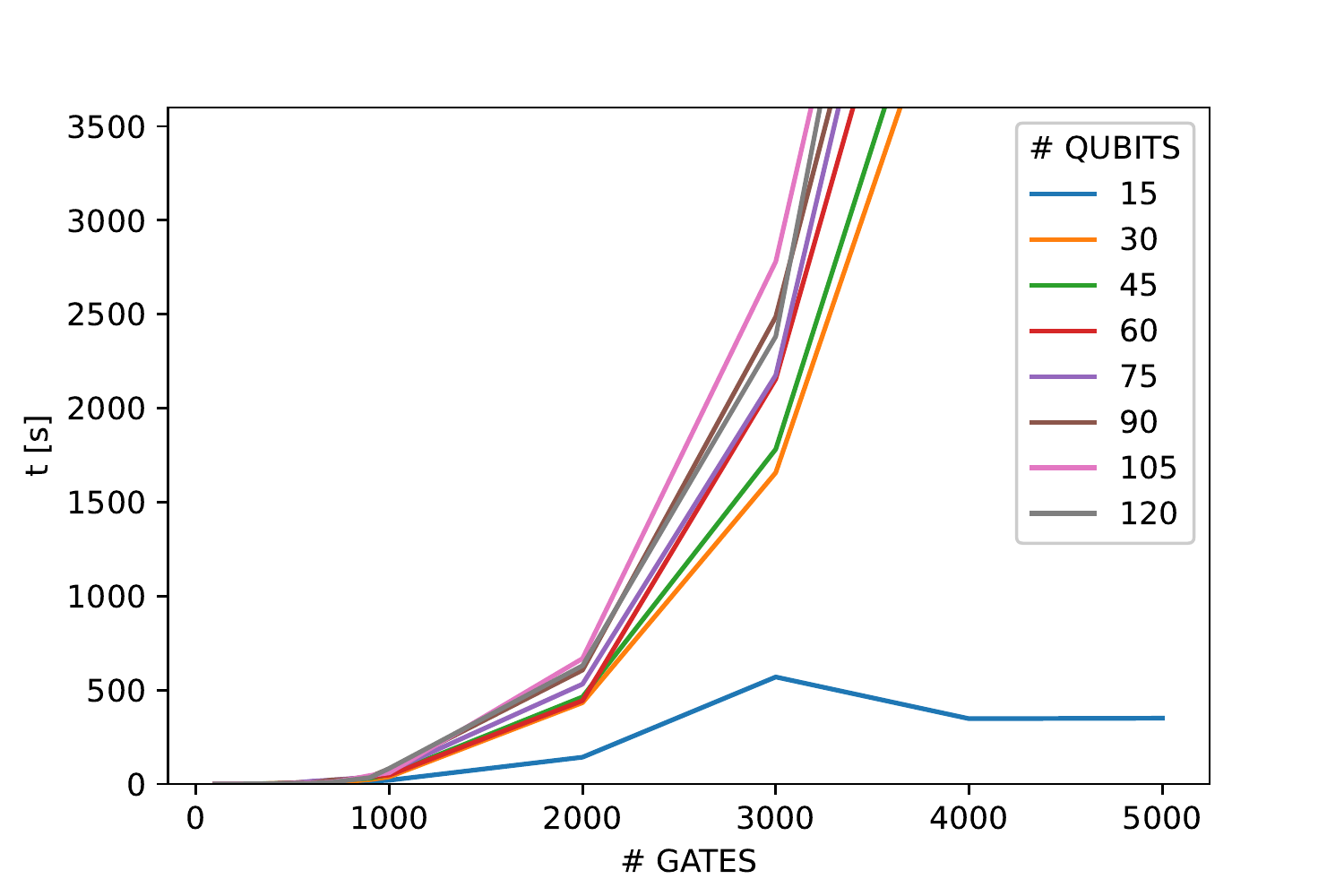}
                  \captionsetup{labelformat=empty}
			\caption{Feynver}
		\end{subfigure}\setcounter{subfigure}{0}
		\caption{Runtimes for differing gate count}\label{fig:random_clifford_gates}
              \end{subfigure}
              
	\begin{subfigure}[b]{\linewidth}
		\centering
		\begin{subfigure}[b]{.49\linewidth}
                  \includegraphics[width=\linewidth]{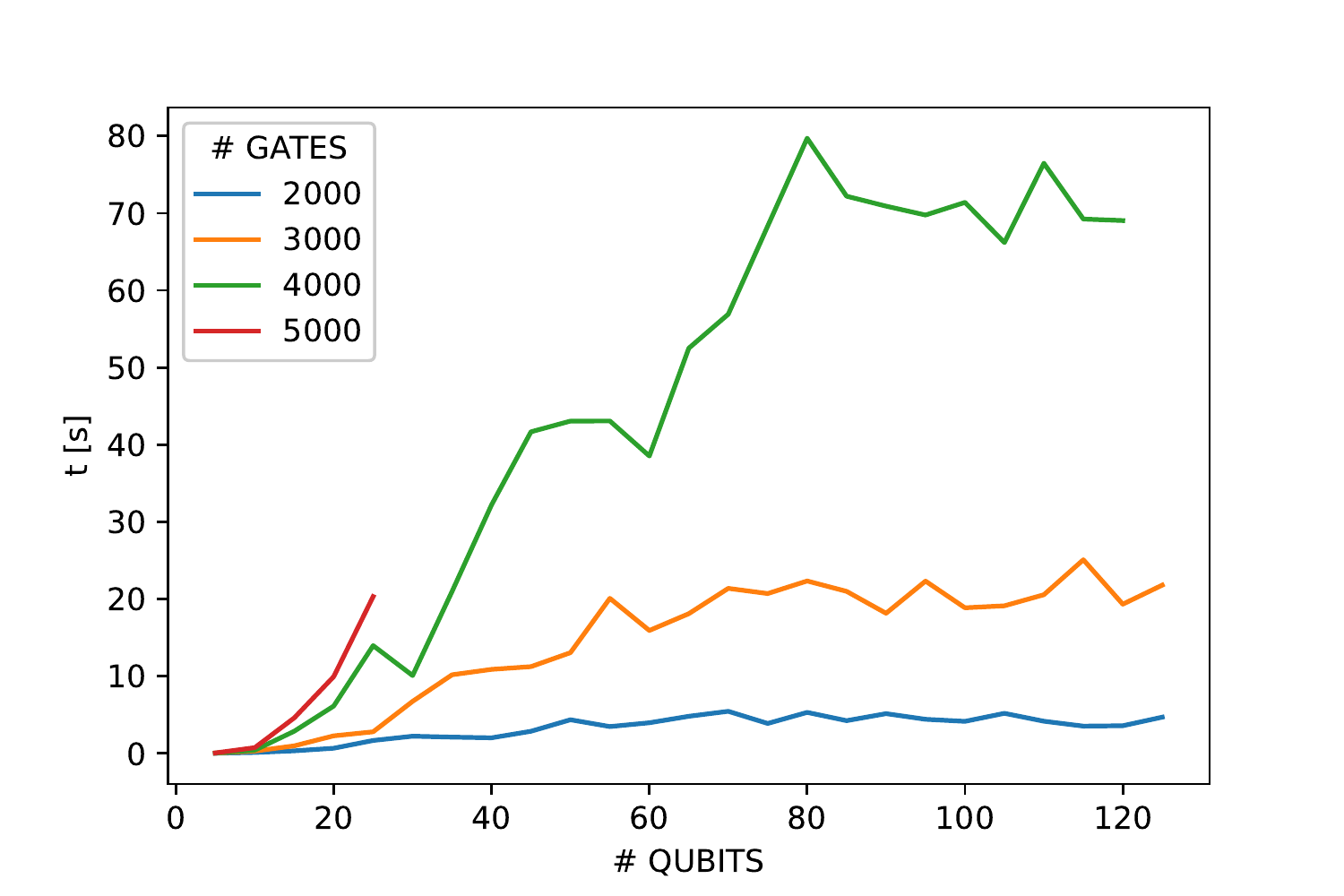}
                  \captionsetup{labelformat=empty}
			\caption{ZX-calculus checker}
		\end{subfigure}
		\begin{subfigure}[b]{.49\linewidth}
                  \includegraphics[width=\linewidth]{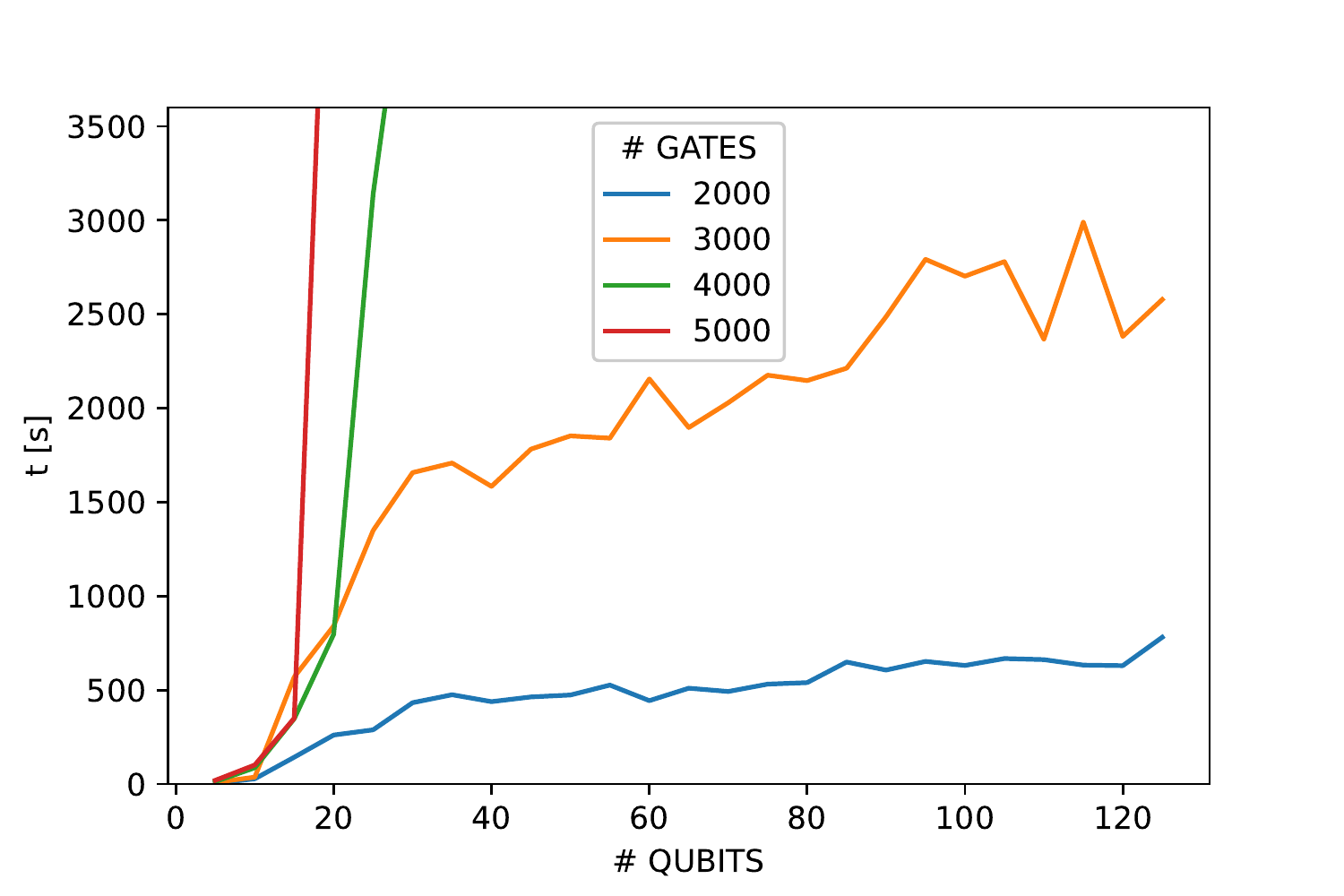}
                  \captionsetup{labelformat=empty}
			\caption{Feynver}
		\end{subfigure}\setcounter{subfigure}{1}
		\caption{Runtimes for differing qubit count}\label{fig:random_clifford_qubits}
              \end{subfigure}                
          \caption{Equivalence checking random Clifford benchmarks}
          \label{fig:results_clifford}
        \end{figure}
        
        \subsection{Equivalence Checking Using Path-Sums}
        Path-sums~\cite{amyLargescaleFunctionalVerification2019} are an abstract representation of quantum circuits in the form of multivariate polynomials over Boolean variables.
        Similar to Feynman path integrals, the idea of \mbox{path-sums} is to encode the action of a unitary as a sum over all possible input-output basis states of a quantum computation.
        This symbolic representation of a quantum circuit allows for handling Clifford + $R_Z(\frac{\pi}{2k})~k \in \mathbb{Z}$, circuits on a high abstraction level.
        
        The symbolic treatment of quantum circuits as path-sums allows for the formulation of a set of rewrite rules which---similar to ZX-calculus rewriting---can be used to successively reduce a path-sum into a normal form in polynomial time.
        While this approach is not complete in general, it is complete for Clifford circuits~\cite{amyLargescaleFunctionalVerification2019}.
        Thus path-sum rewriting serves as an alternative complete approach to proving equivalence of Clifford circuits.
        
	\subsection{Experimental Setup}\label{sec:experimental-setup}
        
	While there is no explicit configuration for the ZX-calculus and path-sum equivalence checker, QCEC has different methods with their respective parameters based on~\cite{burgholzerAdvancedEquivalenceChecking2021,burgholzerRandomStimuliGeneration2021, burgholzerVerifyingResultsIBM2020}. For
	the evaluations involving decision diagrams, we compare the ZX-calculus based equivalence checking routine with the combined approach as presented
	in~\cite{burgholzerAdvancedEquivalenceChecking2021}. For QCEC, we run the equivalence checking routine  in parallel with a
	sequence of 16 simulation runs. If the simulations manage to prove \mbox{non-equivalence} of the circuits, the equivalence
	checking routine is terminated early.

	In order to compare the methods, various benchmarks have been considered. 	
	All benchmarks are provided in the form of QASM~\cite{crossOpenQASMBroaderDeeper2021} files, which serves as a common
	language for the ZX-calculus tool and QCEC. All circuits have been compiled using \emph{qiskit-terra} 0.18.3, either with the optimization
	level $O1$ or $O2$ depending on the benchmark set. Before checking the Clifford Circuits with Feynver, the circuits had to be translated to a format supported by Feynver, but since only Clifford circuits were considered, this was a trivial matter.

        To compare the scaling of the ZX-checker and Feynver, a large set of random Clifford circuits with a varying number of qubits and gate counts has been generated and verified using both tools.
        Each of these circuits has been optimized with $O2$.
        The resulting runtimes can be seen in \autoref{fig:results_clifford}.
	
	The remainder of the comparison was done against QCEC.
        QCEC has been previously
	evaluated on a benchmark set of reversible circuits (from~\cite{willeRevLibOnlineResource2008}) which are mapped
	to suitable quantum architectures.  We also use these in our evaluation as well as a selection of common quantum
	circuits that are available as part of the \emph{MQT Bench} benchmark set~\cite{quetschlichMQTBenchBenchmarking2022a}.
        For each benchmark, we consider three configurations. 
	First, two circuits that are indeed equivalent are used as input. 
	Then, two instances are created where errors are injected into one of the circuits---one with a random gate removed and
	one where the control and target of one CNOT gate have been swapped.

                	\begin{table}[t]
		\centering
		\caption{Slightly optimized reversible circuits }\label{tab:rev1}
		\resizebox{\linewidth}{!}{%
			\begin{tabular}{l r r r @{\hskip 0.2in} r r @{\hskip 0.4in} r r @{\hskip 0.5in} r r}
				\toprule
				\multicolumn{4}{c}{Benchmark}    &  \multicolumn{2}{c@{\hskip 0.2in}}{Equivalent} &  \multicolumn{2}{c@{\hskip 0.5in}}{1 Gate
					Missing} & \multicolumn{2}{c}{Flipped CNOT}  \\ \hline                                                                    
				Name & $n$ & $|G|$ & $|G^\prime|$   & $t_{\text{zx}}$[\si{s}]   & $t_{\text{qcec}}$[\si{s}] & $t_{\text{zx}}$[\si{s}]& $t_{\text{qcec}}$[\si{s}] & $t_{\text{zx}}$[\si{s}] & $t_{\text{qcec}}$[\si{s}] 
				\begin{filecontents*}{reversible_o0_o1.csv}
Benchmark,nqubits,ngates,ngatesprime,zxequ,qcecequ,zxgate,qcecgate,zxcnot,qceccnot
9symml-195,20,15701,13662,8.04,7.37,2.04,0.12,8.25,0.11
dist-223,20,22778,19458,11.82,11.88,24.65,0.16,9.13,0.17
clip-206,53,23782,20872,12.53,20.49,2.18,0.29,8.92,0.31
hwb7-61,20,14680,12557,13.45,1.56,0.56,0.11,13.68,0.10
life-238,20,12114,10441,13.55,3.78,1.57,0.10,9.92,0.09
hwb7-60,20,14096,12167,19.24,0.83,1.14,0.10,51.74,0.11
alu2-199,53,21474,19007,19.76,28.05,2.04,0.26,19.16,0.29
sym9-193,20,15944,13327,21.70,6.17,1.89,0.11,24.64,0.11
example2-231,53,22017,19411,23.63,18.36,5.12,0.27,25.64,0.25
hwb7-59,20,18435,15616,40.81,2.53,0.75,0.18,20.55,0.15
sym9-148,20,17061,14044,75.53,0.60,6.59,0.11,34.26,0.11
urf2-277,20,33348,30940,76.59,3.36,2.03,0.26,77.75,0.26
add6-196,53,30296,25460,82.48,14.94,3.13,0.38,95.46,0.34
urf2-153,20,55243,47155,214.21,210.77,6.54,0.35,162.52,0.39
hwb8-117,20,23596,20697,259.78,1.54,43.95,0.22,27.28,0.17
hwb8-116,20,23353,20996,274.82,2.68,18.58,0.15,272.41,0.16
urf2-161,20,100189,92597,403.99,5.08,5.48,0.69,25.83,1.32
hwb8-114,20,45079,38733,453.13,38.64,16.01,0.31,40.13,0.36
hwb8-118,20,52692,45784,521.59,63.61,85.34,0.43,617.56,0.39
urf2-154,20,52432,44615,587.01,145.44,7.97,0.54,16.57,0.49
hwb8-115,20,45247,39172,734.52,30.47,9.78,0.28,467.74,0.28
hwb8-113,20,52905,46230,1211.62,68.62,22.03,0.46,1476.73,0.37
urf5-159,20,65250,57099,1325.16,45.17,25.78,0.39,659.13,0.46
plus63mod4096-163,53,94520,82195,$>$3600,87.94,48.61,1.28,$>$3600,0.94
plus63mod8192-164,53,125612,111720,$>$3600,412.22,$>$3600,1.22,78.41,1.60
urf3-279,20,172651,157462,$>$3600,416.96,158.13,1.38,$>$3600,1.50
urf1-150,20,156575,134216,$>$3600,$>$3600,51.37,1.06,3084.44,1.34
urf6-281,20,98462,94097,$>$3600,$>$3600,604.79,1.30,$>$3600,1.28
\end{filecontents*}
\csvreader[head to column names]{reversible_o0_o1.csv}{}{%
					\\\Benchmark & \nqubits & \ngates & \ngatesprime & \zxequ & \qcecequ & \zxgate & \qcecgate & \zxcnot & \qceccnot
				}%
		\end{tabular}}\vspace*{-2mm}
	\end{table}

	The benchmark set of reversible circuits is compiled to the to the $65$-qubit IBM Manhattan architecture using
	optimization level $O0$ (no optimizations), $O1$ (slight optimizations) and $O2$ (advanced optimizations). The circuits
	compiled to $O1$ and $O2$ were checked against the unoptimized compiled circuit.
	
	For the quantum circuits, we distinguish two use cases: The first is concerned with verifying the compilation result of a high-level circuit.
	To this end, the circuits are compiled to the $65$-qubit IBM Manhattan architecture with a gate-set comprised of arbitrary single qubit rotations and the CNOT gate. 
	The second use case is about verifying the equivalence of two different implementations of the same functionality---an
	original circuit and an optimized version ($O2$). 
	
	In the following, we summarize the results of our evaluations by means of a representative subset of benchmarks.
	The results for the reversible benchmarks are shown in \autoref{tab:rev1} and \autoref{tab:rev2}. The results for the
	quantum benchmarks are shown in \autoref{tab:benchmarks}.
	
	For further analysis of the influence of the size of the circuits on the runtimes of the equivalence checking routines,
	we consider a set of random quantum circuits comprised of CNOT, Hadamard and T gates with specific
	numbers of qubits and gates. Every gate in this set of benchmarks has a $20\%$ chance of being a Hadamard gate and a
	$20\%$ chance of being a T gate. For this benchmark set only equivalent instances have been considered. Every circuit
	has been checked twice, once against a slightly optimized version ($O1$) and once against a highly optimized version
	($O2$).The resulting runtimes can be seen in \autoref{fig:random_gates} and \autoref{fig:random_qubits}.
	
	All computations were conducted on a \SI{4.2}{\giga\hertz} Intel i7-7700K machine running Ubuntu 18.04 and
	\SI{32}{\gibi\byte} main memory. Each benchmark was run with a hard timeout of \SI{1}{\hour} for each method.
	
	\vspace*{-3mm}
	\subsection{Discussion}\label{sec:discussion}
		
	\begin{table}[t]
		\centering
		\caption{Highly optimized reversible circuits}\label{tab:rev2}
		\resizebox{\linewidth}{!}{%
			\begin{tabular}{l r r r @{\hskip 0.2in} r r @{\hskip 0.4in} r r @{\hskip 0.5in} r r}
				\toprule
				\multicolumn{4}{c}{Benchmark}    &  \multicolumn{2}{c@{\hskip 0.2in}}{Equivalent} &  \multicolumn{2}{c@{\hskip 0.5in}}{1 Gate
					Missing} & \multicolumn{2}{c}{Flipped CNOT}  \\ \hline                                                                    
				Name & $n$ & $|G|$ & $|G^\prime|$   & $t_{\text{zx}}$[\si{s}]   & $t_{\text{qcec}}$[\si{s}] & $t_{\text{zx}}$[\si{s}]& $t_{\text{qcec}}$[\si{s}] & $t_{\text{zx}}$[\si{s}] & $t_{\text{qcec}}$[\si{s}] 
				\begin{filecontents*}{reversible_o0_o2.csv}
Benchmark,nqubits,ngates,ngatesprime,zxequ,qcecequ,zxgate,qcecgate,zxcnot,qceccnot
9symml-195,20,15701,13159,42.82,5.41,1.96,0.13,2.39,0.13
dist-223,20,22778,18721,16.44,11.96,3.45,0.16,2.49,0.17
clip-206,53,23782,21407,11.14,15.65,2.00,0.43,1.37,0.39
hwb7-61,20,14680,12741,34.99,0.52,0.47,0.18,0.46,0.17
life-238,20,12114,9869,8.73,2.80,1.12,0.10,0.69,0.10
alu2-199,53,21474,19080,26.23,4.27,7.00,0.31,6.23,0.36
sym9-193,20,15944,13908,7.76,11.21,1.41,0.14,1.07,0.13
example2-231,53,22017,19207,38.71,51.12,3.13,0.31,3.40,0.31
hwb7-59,20,18435,16027,15.68,1.24,0.46,0.15,0.50,0.16
sym9-148,20,17061,14002,284.87,0.96,1.07,0.11,1.08,0.11
urf2-277,20,33348,29711,68.28,16.10,2.95,0.32,2.68,0.31
add6-196,53,30296,26201,154.88,30.21,10.77,0.45,11.97,0.43
urf2-153,20,55243,46103,296.04,42.08,3.12,0.48,3.18,0.43
hwb8-117,20,23596,20413,123.04,17.21,2.56,0.20,3.03,0.20
hwb8-116,20,23353,20519,323.72,13.97,11.78,0.17,10.28,0.18
urf2-161,20,100189,91018,405.14,122.92,3.11,1.15,3.16,1.33
hwb8-114,20,45079,38398,1263.37,9.11,10.15,0.33,10.63,0.35
hwb8-118,20,52692,44901,712.38,13.65,8.40,0.68,5.93,0.41
urf2-154,20,52432,44306,1150.70,20.02,5.53,0.57,5.54,0.52
hwb8-115,20,45247,38205,269.99,13.15,6.26,0.40,8.26,0.32
hwb8-113,20,52905,45833,1036.07,11.59,15.25,0.41,16.68,0.48
urf5-159,20,65250,56089,2369.23,18.12,21.70,0.52,30.52,0.47
plus63mod4096-163,53,94520,83835,$>$3600,1249.90,78.66,1.27,73.99,1.12
plus63mod8192-164,53,125612,111258,$>$3600,$>$3600,313.11,1.49,300.36,1.61
urf3-279,20,172651,153474,$>$3600,$>$3600,40.35,1.40,40.53,1.75
urf1-150,20,156575,132898,$>$3600,$>$3600,103.27,1.17,105.97,1.16
urf6-281,20,98462,88750,$>$3600,$>$3600,198.70,1.36,187.00,1.32
\end{filecontents*}
\csvreader[head to column names]{reversible_o0_o2.csv}{}{%
					\\\Benchmark & \nqubits & \ngates & \ngatesprime & \zxequ & \qcecequ & \zxgate & \qcecgate & \zxcnot & \qceccnot
				}%
		\end{tabular}}\vspace*{-2mm}
	\end{table}
	
	\begin{table*}[t]
		\centering
		\caption{Common quantum algorithms}\label{tab:benchmarks}
		\resizebox{0.9\linewidth}{!}{\scriptsize
			\begin{tabular}[h]{l r r r @{\hskip 0.4in} r r @{\hskip 0.4in} r r @{\hskip 0.4in} r r}
				\toprule
				\multicolumn{4}{c}{Benchmark}    &  \multicolumn{2}{c}{Equivalent} &  \multicolumn{2}{c@{\hskip 0.4in}}{1 Gate
					Missing} & \multicolumn{2}{c}{Flipped CNOT}  \\ \hline                                                                    
				Name                 & $n$ & $|G|$ & $|G^\prime|$   & $t_{\text{zx}}$[\si{s}]   & $t_{\text{qcec}}$[\si{s}] &
				$t_{\text{zx}}$[\si{s}]
				& $t_{\text{qcec}}$[\si{s}] & $t_{\text{zx}}$[\si{s}] &
				$t_{\text{qcec}}$[\si{s}]
				\smallskip\\ 
				\multicolumn{10}{c}{Compiled Circuits} \\ \midrule
				Grover            & 6  & 1606   & 2803   &   0.39     &   3.40    &  0.31    &   0.04      & 0.47      & 0.04    \\
				Grover            & 7  & 4732   & 8476   &   1.24   &   0.30    &  3.03    &   0.14      & 4.16     & 0.14    \\
				Grover            & 8  & 12482  & 22860  &   12.15  &   0.91    &  5.11   &   0.42      & 189.61    & 0.39    \\
				QFT               & 23 & 1311   & 3741   &   0.06    &   2.00    &  0.05    &   $>$3600   & 0.05      & 902.99  \\
				QFT               & 38 & 3591   & 10449  &   0.32   &   $>$3600 &  0.20    &   $>$3600   &  0.21    & $>$3600 \\
				Random-Walk       & 7  & 6523   & 8955   &   150.36  &   0.24    &  9.35   &   0.14      & 55.44     & 0.16    \\
				Random-Walk       & 8  & 14084  & 19755  &   1289.13 &   0.57    &  455.58  &   0.33      & 687.95    & 0.31    \\
				Random-Walk       & 9  & 29325  & 41942  &   $>$3600 &   1.31    &  1001.93 &   0.59      & 2477.31   & 0.50    \\
				QPE-Exact         & 22 & 1217   & 3006   &   0.83    &   0.10    &  0.78    &   $>$3600   & 0.79 & 0.82 \\
				QPE-Exact         & 39 & 3823   & 11552  &   3.19    &   $>$3600 &  2.89    &   $>$3600   & 2.92      & $>$3600 \\
				GHZ & 65 & 130 & 493 & 0.06 & $<$0.01 & 0.06 & $<$0.01 & 0.59 & 0.01 \\
				Graph State & 62 & 403 & 2041 & 0.36 & 0.17 & 0.43 & 0.17 & 0.35 & 0.17  \smallskip\\ 
				\multicolumn{10}{c}{Optimized Circuits}                                                                        \\ \midrule
				Grover            & 8  & 12479  & 12287  &   8.00 &   0.04    &  2.11   &   0.24      & 9.837    & 0.04    \\
				Grover            & 9  & 37193  & 36881  &   82.9446 &   0.14    &  3.02  &   129.56    & 145.574   & 0.17    \\
				Grover            & 10 & 104977 & 104501 &   779.291 &   0.42    &  72.05 &   $>3600$   & 90.68   & 41.24   \\
				Grover            & 11 & 308074 & 307322 &   2178.13 &   0.07    &  2316.00 &   588.62   & 2264.67   & 281.j04   \\
				QFT & 32 & 2544 & 2482  & 1.43 & 0.04    & 1.49 & 3.57    & 1.89 & 14.53   \\
				QFT & 43 & 4601 & 4502  & 2.86 & 10.837  & 2.86 & 17.78   & 2.81 & 1.02    \\
				QFT & 44 & 4818 & 4702  & 3.01 & $>$3600 & 3.05 & 1.27    & 2.90 & 1.21    \\
				QFT & 75 & 14136 & 11013 & 1.23 & $>$3600 & 1.30 & $>$3600 & 1.26 & $>$3600 \\
				Random-Walk       & 7  & 2351   & 1906   &   140.90  &   0.02    &  11.86   &   0.10      & 212.527     & 0.02    \\
				Random-Walk       & 8  & 4648  & 3925  &   2175.39 &  0.04    &  78.91  &   62.15      & 149.49    & 0.11    \\
				Random-Walk       & 9  & 9249  & 7987  &   $>$3600 &   0.09    &  245.39&   0.11      & $>$3600   & 0.11    \\
				\bottomrule
		\end{tabular}}
		\vspace*{-4mm}
	\end{table*}

        \autoref{fig:results_clifford} shows runtimes for the proposed ZX-calculus checker and Feynver for the set of random Clifford benchmarks. \autoref{fig:random_clifford_gates} shows runtimes with respect to the number of gates in the original circuit for fixed numbers of qubits and \autoref{fig:random_clifford_gates} shows runtimes with respect to the number of qubits for fixed numbers of gates.
        The similarity between the plots for both methods suggests that both methods scale somewhat similarly with respect to the size of the circuits, however.
        Indeed, the two methods exhibit the same asymptotic behavior.
        However, considering the scaling of the vertical axis in the plots, one can clearly see that the proposed implementation outperforms Feynver by orders of magnitude on all benchmarks.
        
	In the comparison with QCEC, both methods managed to prove the correct result for all considered circuits where a result is obtained within the given
	time frame. 
	As discussed before, this is not guaranteed by the theory of the \mbox{ZX-calculus}. 
	On the other hand, the question of completeness for the decision diagram
	based approach is trivial. Decision diagrams are a canonical representation of a matrix. Thus, if the combined circuit
	$G^\dagger G^\prime$ has the identity system matrix, the decision diagram for $G^\dagger G^\prime$ has to be the
	identity decision diagram as well.
	
        For the set of reversible benchmarks (\autoref{tab:rev1} and \autoref{tab:rev2}), the two methods finished within \SI{10}{\second} of each other for
	\SI{92}{\percent} of benchmark instances in the case of equivalent instances for both optimization levels. 
	The remaining reversible benchmarks and circuits containing large reversible parts in their high-level description (such as Grover's algorithm and the Quantum Random Walk) favor the decision diagram based approach.
	These circuits can be \emph{exactly} compiled to polynomially-sized quantum circuits comprised only of Clifford+T gates, i.e., circuits only using Hadamard ($H$), Phase ($S$), CNOT ($CX$), and $T$ gates.
	As a consequence, the respective functionalities (i.e., the system matrices) possess lots of structure that can be exploited by decision diagrams and, additionally, only feature a very limited set of complex numbers which limits the effect of numerical instabilities.
	In contrast, the ZX-calculus based approach does not benefit from this structure very much. 
	
        In the case of proper quantum circuits (\autoref{tab:benchmarks}) the story looks a bit different. For circuits containing no or smaller reversible parts (such as the QFT or Quantum Phase Estimation), the ZX-calculus approach fairs much better in comparison to decision diagrams. 
	The main obstacle in these cases is that the considered algorithms feature many rotation gates with arbitrarily small rotation angles.
	Due to numerical instabilities and rounding errors, it might happen that two decision diagram nodes that should be identical in theory, differ by a small margin in practice.
	As a consequence, inherent redundancies in the underlying representations cannot be captured accurately anymore.
	Thus, while the resulting decision diagram is very close to the identity with respect to the Hilbert-Schmidt norm, it might grow exponentially large in the worst case.
	In contrast, ZX-diagrams are not susceptible to such exponential growth under numerical errors.
	
	The above observations are similar in the case of non-equivalent instances. Although runtimes for both methods are generally lower, the relative performances are still similar. 
	Since the resulting decision diagram is almost guaranteed to not be very close to the identity during the equivalence check, the alternating scheme discussed cannot be as efficient as in the equivalent case.
	Due to this, QCEC resorts to simulations of the circuit with random inputs which, as shown in~\cite{burgholzerAdvancedEquivalenceChecking2021}, are expected to show the non-equivalence within a few simulations.
	Yet, the complexity of decision diagram based simulation is still exponential in the worst case.
	The rewriting approach of the ZX-calculus is less volatile to errors in the circuit. 
	During the equivalence check, the combined circuit diagram is simplified as much as possible until no more rules can be applied. 
	Depending on the severity and kind of error, the procedure stops sooner or later.
        Of course, the ZX-calculus checker cannot prove non-equivalence of circuits, but the experiments show that inability to show equivalence of circuits with the ZX-calculus at least gives a strong indication that two circuits are indeed non-equivalent since the ZX-calculus checker managed to prove equivalence in all equivalent benchmarks.
        \begin{figure*}[t]
          \centering

	\begin{subfigure}[b]{0.49\linewidth}
		\centering
		\begin{minipage}[b]{.49\linewidth}
                  \includegraphics[width=\linewidth]{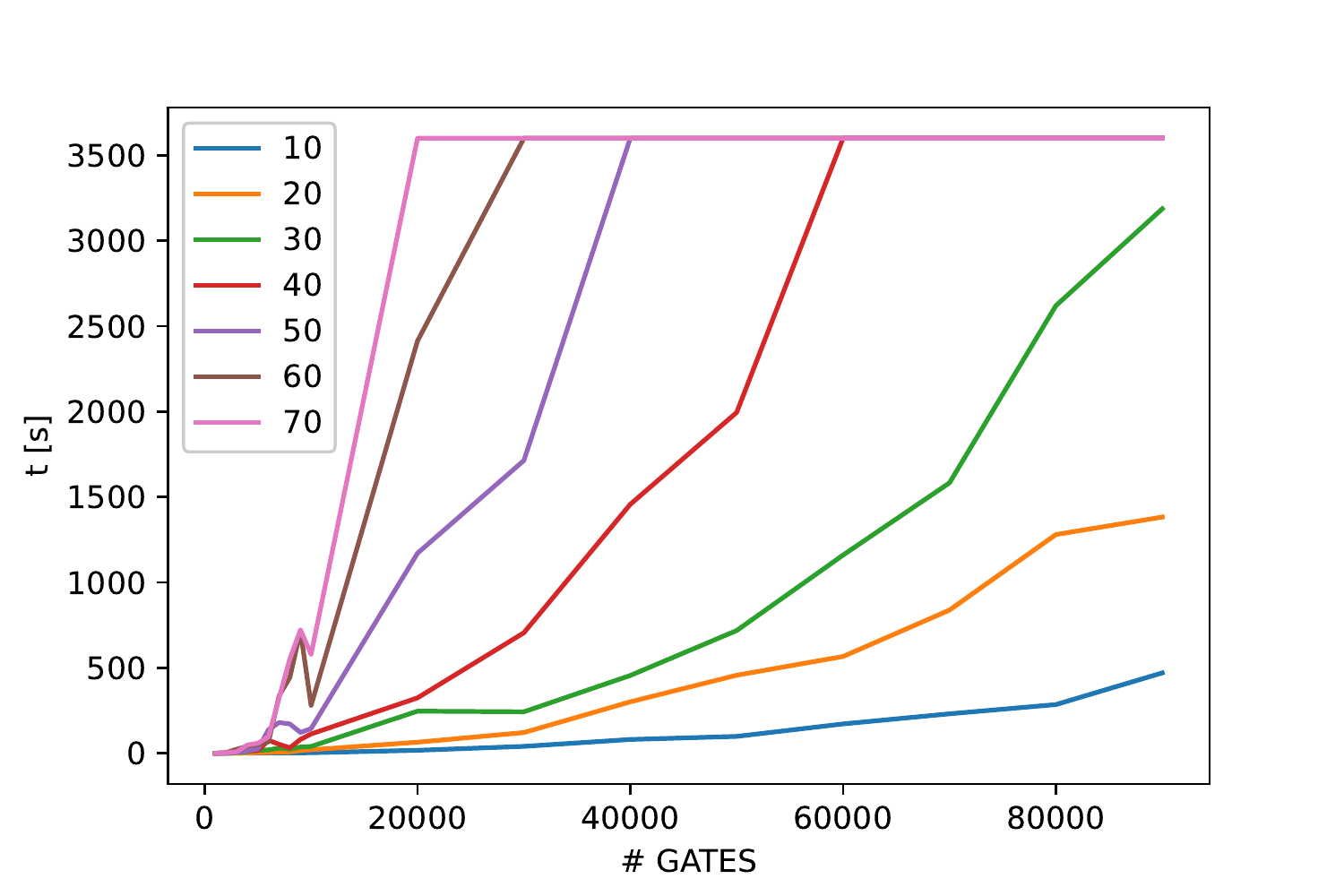}
                  \captionsetup{labelformat=empty}
			\caption{ZX-calculus checker $O1$}
		\end{minipage}
		\begin{minipage}[b]{.49\linewidth}
                  \includegraphics[width=\linewidth]{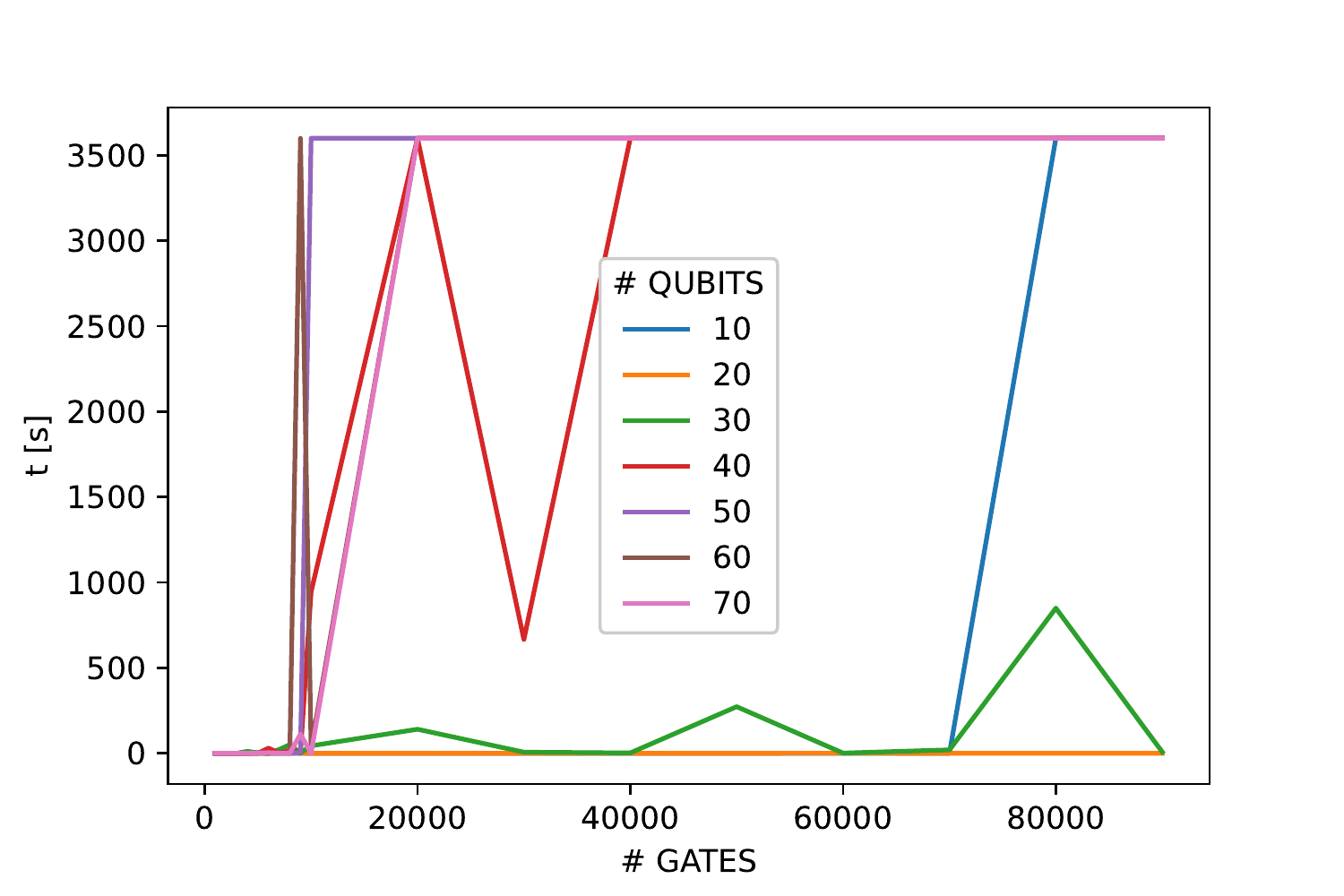}
                  \captionsetup{labelformat=empty}
			\caption{QCEC $O1$}
		\end{minipage}
		
		\begin{minipage}[b]{.49\linewidth}
                  \includegraphics[width=\linewidth]{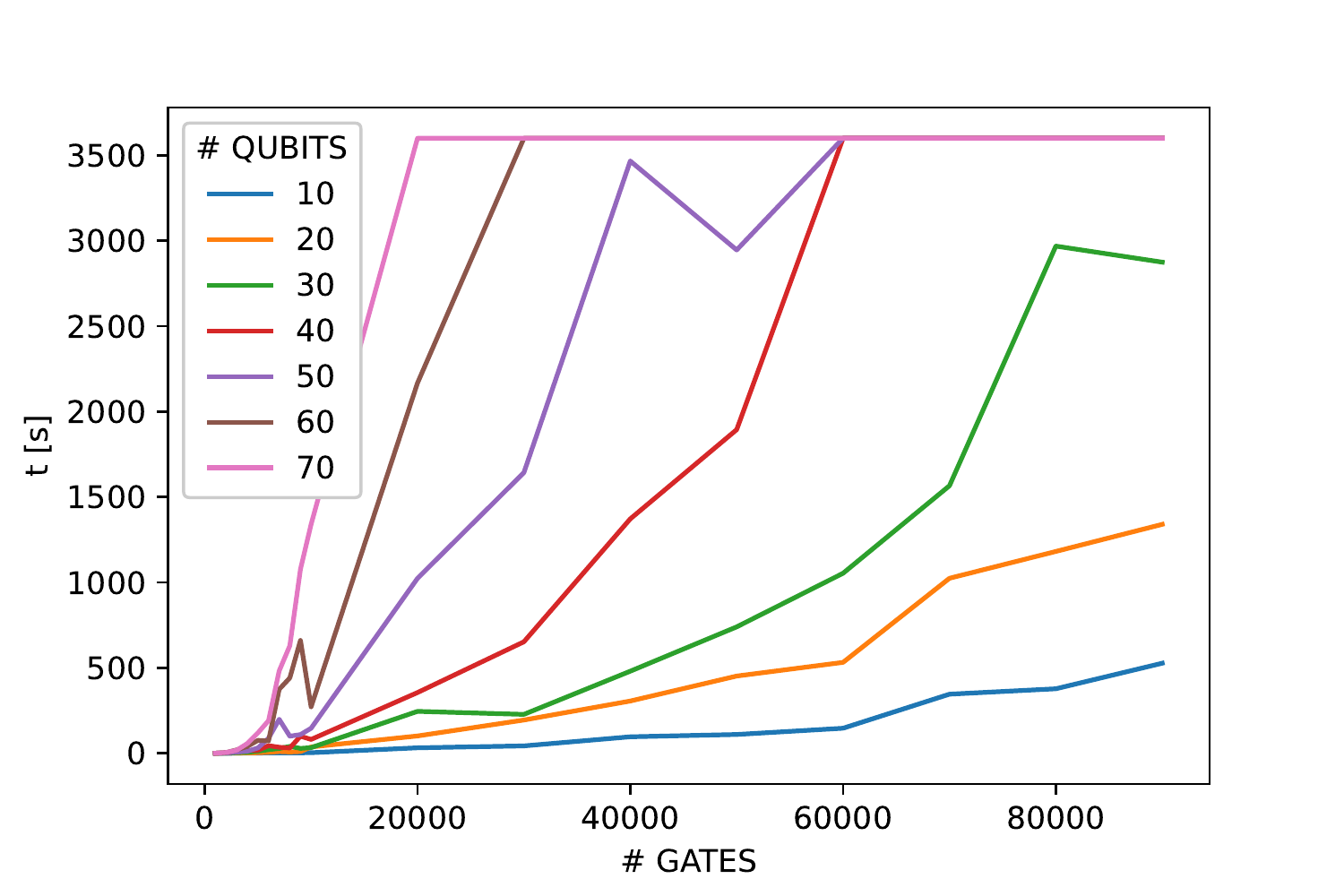}
                  \captionsetup{labelformat=empty}
			\caption{ZX-calculus checker $O2$}
		\end{minipage}
		\begin{minipage}[b]{.49\linewidth}
                  \includegraphics[width=\linewidth]{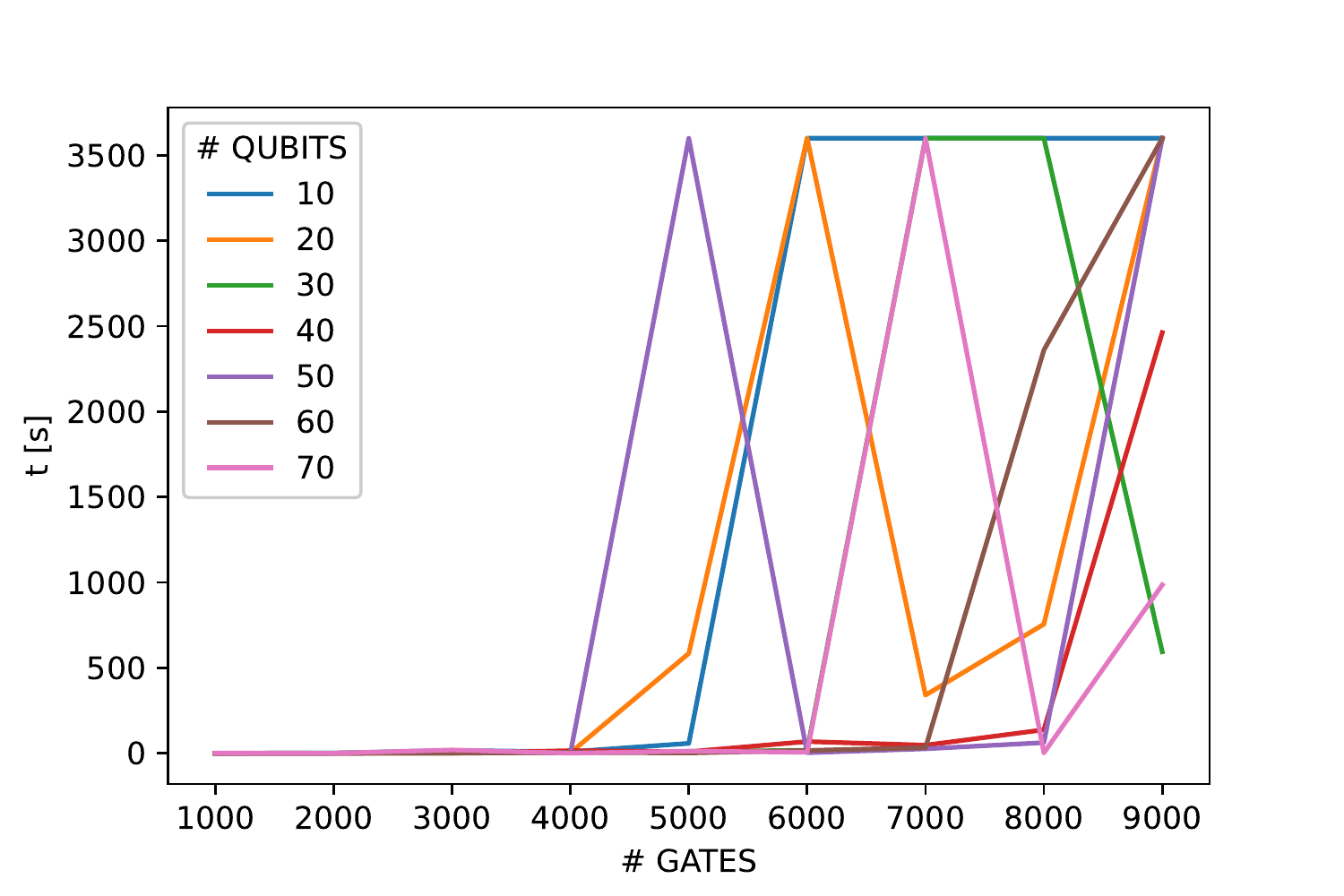}
                  \captionsetup{labelformat=empty}
			\caption{QCEC $O2$}
                      \end{minipage}
                      \setcounter{subfigure}{0}\vspace*{-6mm}
		\caption{Runtimes for differing gate count}\label{fig:random_gates}\vspace*{-1mm}
	\end{subfigure}	
	\begin{subfigure}[b]{0.49\linewidth}
		\centering
		\begin{minipage}[b]{.49\linewidth}
                  \includegraphics[width=\linewidth]{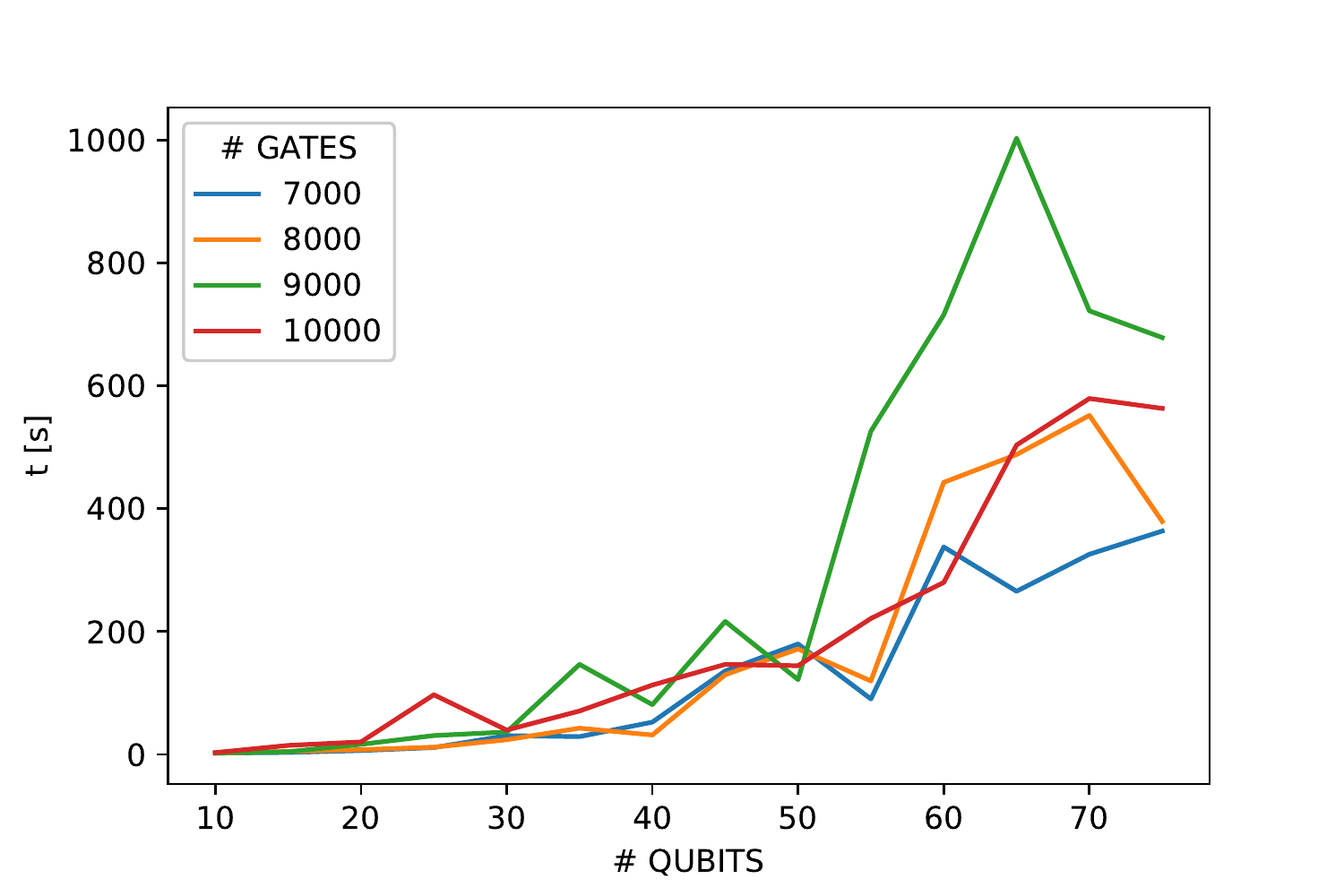}
                  \captionsetup{labelformat=empty}
                  \caption{ZX-calculus checker $O1$}
		\end{minipage}
		\begin{minipage}[b]{.49\linewidth}
                  \includegraphics[width=\linewidth]{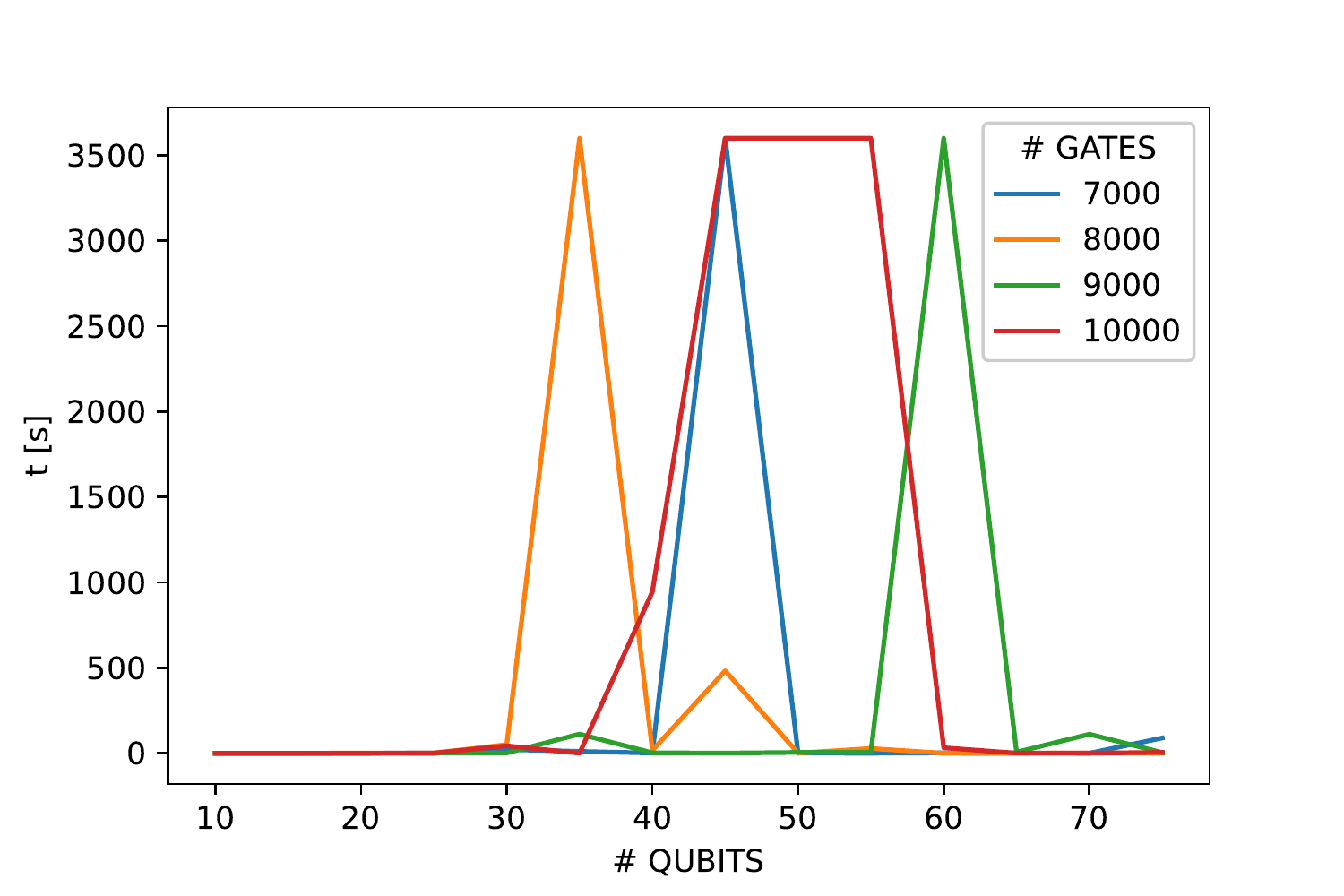}
                  \captionsetup{labelformat=empty}
                  \caption{QCEC $O1$}
		\end{minipage}
		
		\begin{minipage}[b]{.49\linewidth}
                  \includegraphics[width=\linewidth]{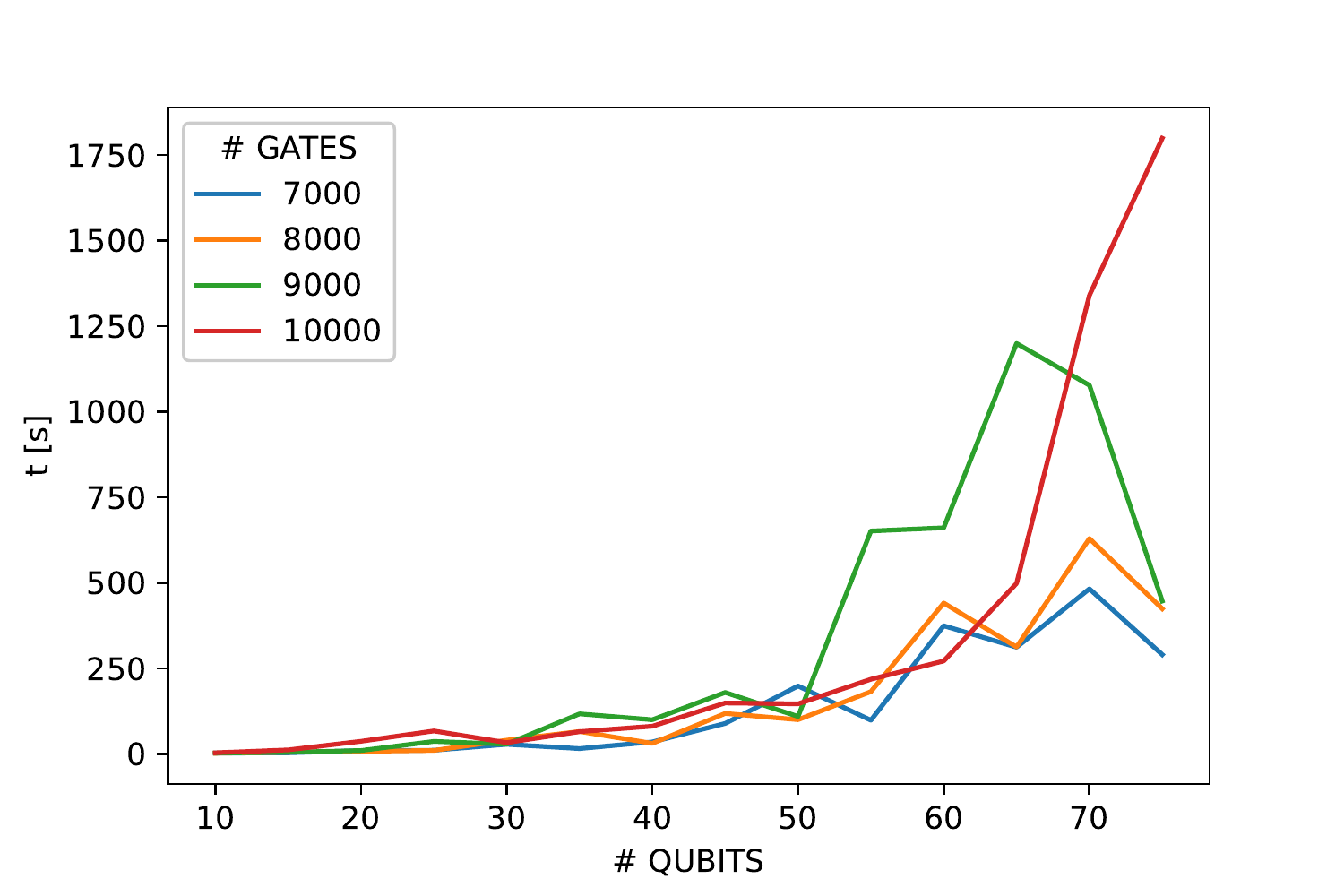}
                  \captionsetup{labelformat=empty}
                  \caption{ZX-calculus checker $O2$}
		\end{minipage}
		\begin{minipage}[b]{.49\linewidth}
                  \includegraphics[width=\linewidth]{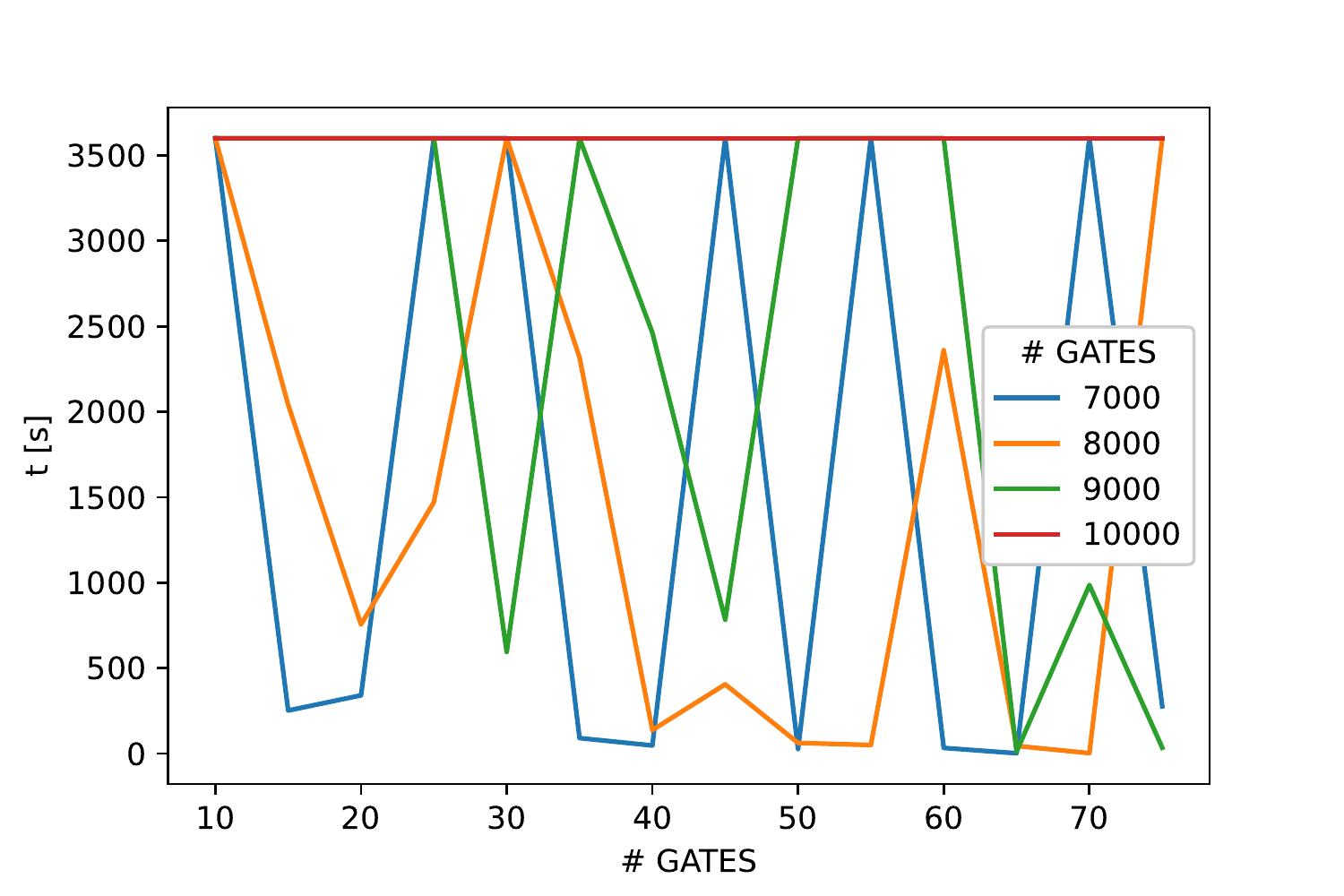}
                  \captionsetup{labelformat=empty}
			\caption{QCEC $O2$}
                      \end{minipage}
                      \setcounter{subfigure}{1}\vspace*{-6mm}
		\caption{Runtimes for differing qubit count}\label{fig:random_qubits}\vspace*{-1mm}
              \end{subfigure}                
          \caption{Equivalence checking random Clifford+T benchmarks}
          \label{fig:results}\vspace*{-4mm}
        \end{figure*}
	
	What \autoref{tab:benchmarks} also shows is the volatility of the decision diagram based
	approach. Runtimes can increase dramatically for the same type of circuit with a differing number of qubits. As
	soon as the intermediate decision diagram does not admit a compact representation, applying gates to it is a costly
	operation. This volatility is shown in more detail in the case of random Clifford+T
	benchmarks. \autoref{fig:random_gates} and \autoref{fig:random_qubits} show the runtimes of equivalence checking random
	Clifford+T circuits with increasing gate count and number of qubits. 
	
	In \autoref{fig:random_gates} the different lines correspond to benchmarks with differing qubit counts. If a data point
	is at $3600\si{s}$ in the graph this indicates that the equivalence check took longer than the timeout limit of
	$3600\si{s}$ or---in the case of decision diagrams--- it means that the memory limit has been exceeded. In \autoref{fig:random_qubits} different lines correspond to benchmarks with
	differing gate counts in the original circuit. With $O1$ an average of $0.9\%$ of gates were optimized away. With $O2$
	an average of $6.1\%$ of gates were optimized away.
	
	Both \autoref{fig:random_gates} and \autoref{fig:random_qubits} clearly show that the equivalence checking routine based
	on the ZX-calculus has a clear correlation between runtime and the size of the circuits, whether that size is due to the
	number of gates or qubits. Although there are some simpler instances where the runtime decreases, the general trend can
	be clearly seen. \autoref{fig:random_gates} also further supports the claim in~\cite{kissingerReducingTcountZXcalculus2020} that the
	complexity of reducing a ZX-diagram to reduced gadget form is between $O(n)$ and $O(n^2)$ where $n$ is the number of gates in the circuit.
	
	On the other hand, the plots for QCEC show no such correlation. Whether QCEC manages to prove equivalence for a circuit
	only depends on the specific circuit in question after a certain circuit size and complexity has been reached. This is
	not too surprising---decision diagrams can blow up to exponential size with respect to the number of qubits. This
	volatility can hardly be held against decision diagrams though. After all, \autoref{fig:random_gates} and
	\autoref{fig:random_qubits} show runtimes for \emph{random} benchmarks, i.e.\ circuits that do not exhibit much
	structure. This volatility
	is actually a positive feature of the decision diagram based approach. If this method yields a result at all it usually
	does so using significantly less time than the ZX-calculus based method and is, therefore, able to prove equivalence of
	some very large circuits where the worst case complexity of $O(n^3)$ of the ZX-calculus based approach leads to long runtimes.
	
	This shows that the two methods are complementary and are best used in tandem, especially for more optimized
	circuits. For circuits with many qubits but a smaller number of gates, the ZX-calculus based approach performs more
	favorably. For even larger circuits the problem itself is too complex to be solved even in polynomial time for the
	ZX-calculus based approach. In this case, the decision diagram based method might still be able to show equivalence by
	keeping the intermediate decision diagrams small. Because the size of the ZX-diagram during the equivalence check is bounded by the size of the original circuit (the number of spiders is strictly decreasing) the
	ZX-calculus based equivalence checking method has a low memory footprint. It can therefore easily be used in parallel
	with the decision diagram based method without using too many resources.

        \vspace*{-3mm}
	\section{Conclusion}\label{cha:conclusion}
	
	In this work, we examined the viability and effectiveness of the ZX-calculus for equivalence checking of quantum
	circuits. By improving the state of the art to be able to handle inaccurate representations of complex numbers,
	permutations of the input, and output layout of a circuit and ancillary qubits, we can now verify the results of
	compilation flows with the \mbox{ZX-calculus}. We have also discussed the limitations of the \mbox{ZX-calculus} based approach which
	prevents it from being a general equivalence checking method. 
	
	To give empirical results on the practicality of the \mbox{ZX-calculus} in equivalence checking, we conducted a case study
	comparing the ZX-calculus equivalence checker with one based on path-sums and one based on decision diagrams.
        
        Empirical results show that path-sums and ZX-calculus exhibit similar scaling when checking the equivalence of Clifford circuits but the ZX-calculus based approach is still orders of magnitude faster on average.
        Also, the ZX-calculus and decision diagram show similar performance in many cases: but they differ in key aspects.
	Decision diagrams show significant benefits for circuits containing large reversible parts, such as oracles or adders.
	The sensibility of decision diagrams to numerical imprecision makes them hard to use on quantum algorithms that cannot
	be exactly represented using floating points, such as algorithms relying on arbitrary rotation angles, due to the
	potential blow-up of the intermediate representation.  
	The ZX-calculus based equivalence checking procedure is less sensitive to this and is useful in showing equivalence in
	these cases. However, the ZX-calculus tends to be more suitable for verifying smaller building blocks than whole quantum
	algorithms due to the large number of involved gates. In conclusion, we can see that decision diagrams and the
	\mbox{ZX-calculus} can serve as complementary approaches for the equivalence checking problem. 

\subsection*{Acknowledgements}
This work received funding from the European Research Council (ERC) under the European Union’s Horizon 2020 research and innovation program (grant agreement No. $101001318$), was part of the Munich Quantum Valley, which is supported by the Bavarian state government with funds from the Hightech Agenda Bayern Plus, and has been supported by the BMWK on the basis of a decision by the German Bundestag through project QuaST.

        \balance
	\printbibliography

\begin{IEEEbiography}
	[{\includegraphics[width=1in,height=1.25in,clip, keepaspectratio]{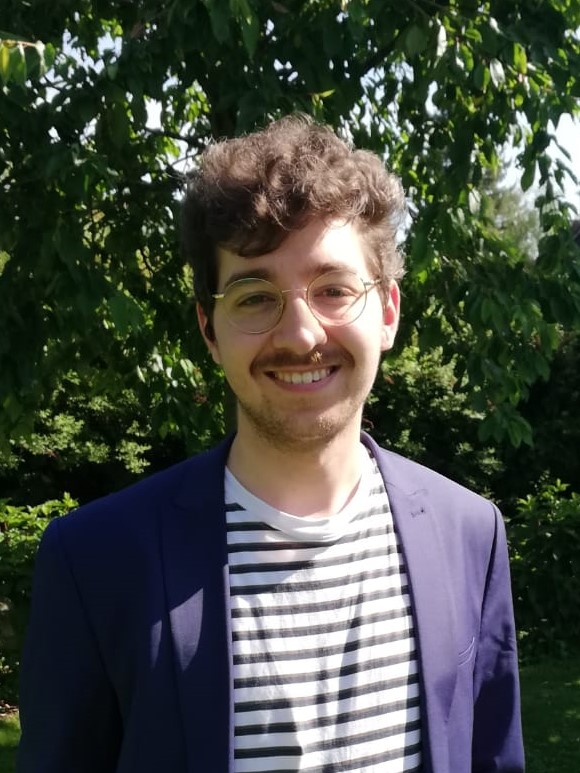}}]{Tom Peham}
  Tom Peham received his Master's degree in computer science (2022) from the Johannes Kepler University Linz, Austria.
  He is currently a Ph.D. student at the Chair for Design Automation at the Technical University of Munich, Germany. 
  His research interests include design automation for quantum computing---currently focusing on applications of the ZX-calculus in this domain. 
\end{IEEEbiography}

\begin{IEEEbiography}
  [{\includegraphics[width=1in,height=1.25in,clip,keepaspectratio]{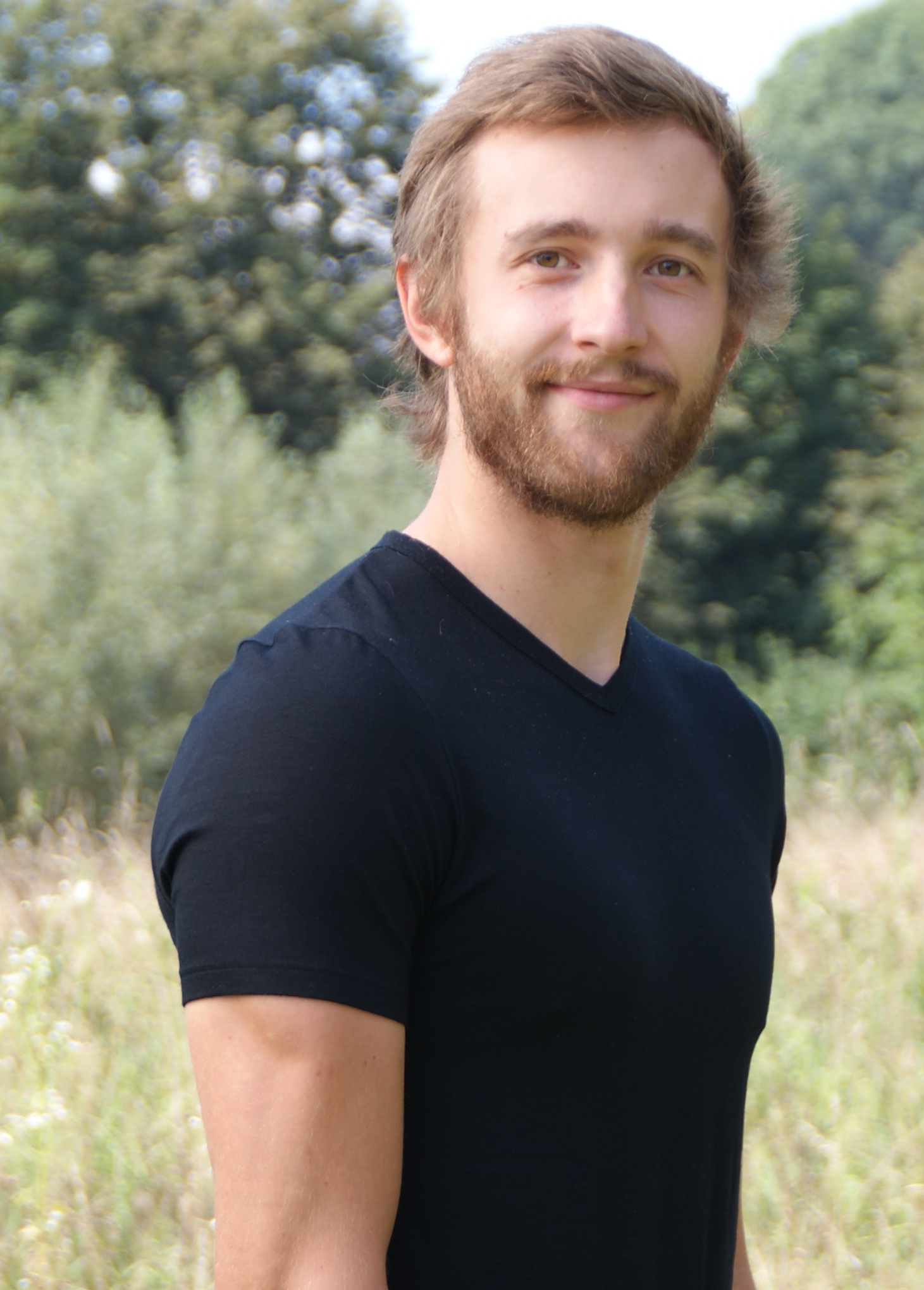}}]{Lukas Burgholzer}
  Lukas Burgholzer (S’19) received his Master's degree in industrial mathematics (2018) and Bachelor's degree in computer science (2019) from the Johannes Kepler University Linz, Austria.
  He is currently a Ph.D. student at the Institute for Integrated Circuits at the Johannes Kepler University Linz, Austria. 
  His research focuses on design automation and software for quantum computing. In these areas, he has published several papers on international conferences such as ASP-DAC, DAC, ICCAD, DATE, and QCE.
\end{IEEEbiography}

\begin{IEEEbiography}
	[{\includegraphics[width=1in,height=1.25in,clip,keepaspectratio]{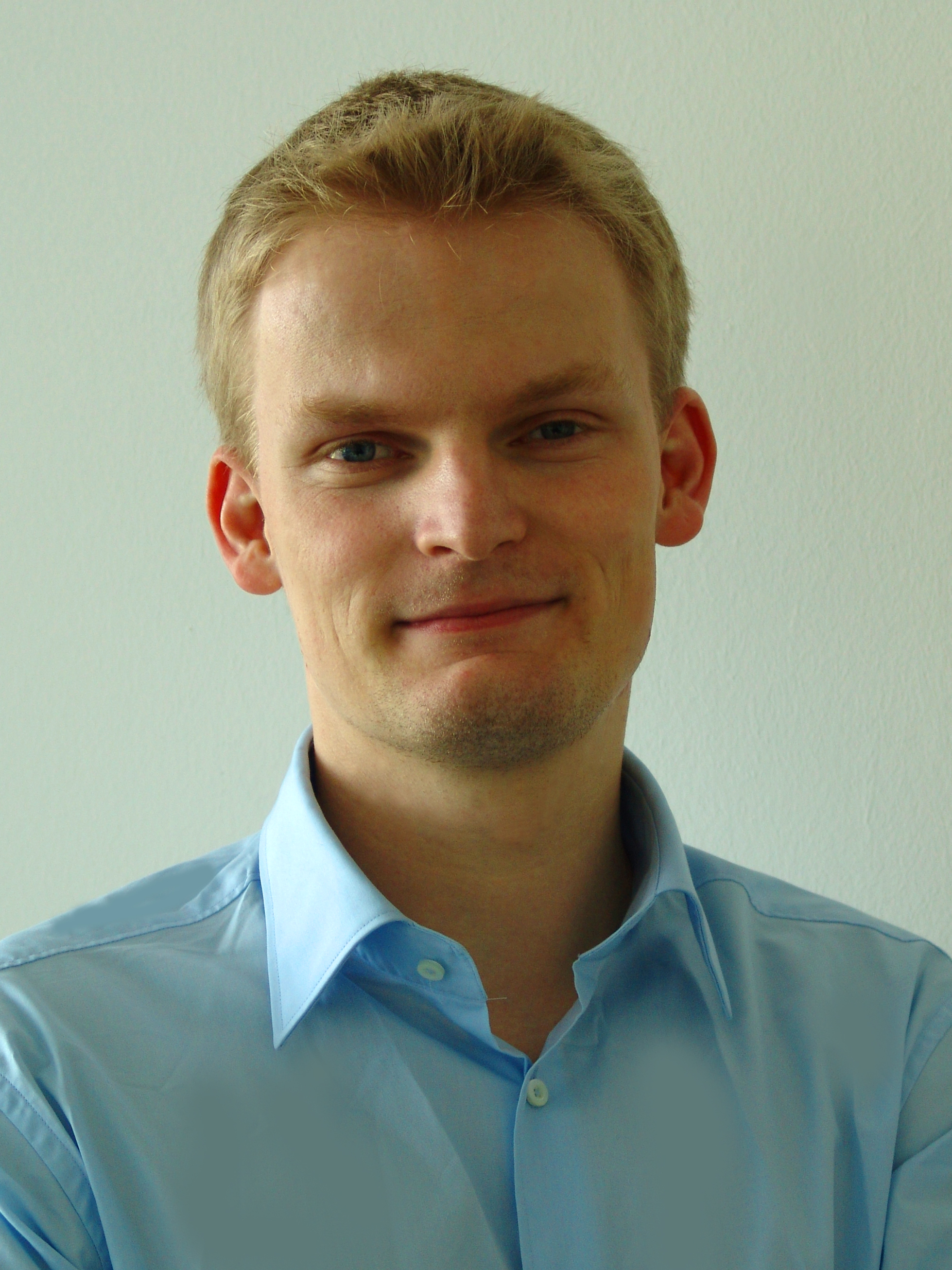}}]{Robert Wille}
        Robert Wille is a Full and Distinguished Professor at the Technical University of Munich, Germany, and Chief Scientific Officer at the Software Competence Center Hagenberg, Austria. He received the Diploma and Dr.-Ing. degrees in Computer Science from the University of Bremen, Germany, in 2006 and 2009, respectively. Since then, he worked at the University of Bremen, the German Research Center for Artificial Intelligence (DFKI), the University of Applied Science of Bremen, the University of Potsdam, and the Technical University Dresden. From 2015 until 2022, he was Full Professor at the Johannes Kepler University Linz, Austria, until he moved to Munich. His research interests are in the design of circuits and systems for both conventional and emerging technologies. In these areas, he published more than 400 papers and served in editorial boards as well as program committees of numerous journals/conferences such as TCAD, ASP-DAC, DAC, DATE, and ICCAD. For his research, he was awarded, e.g., with Best Paper Awards, e.g., at TCAD and ICCAD, an ERC Consolidator Grant, a Distinguished and a Lighthouse Professor appointment, a Google Research Award, and more.
\end{IEEEbiography}

\end{document}